\documentclass[sigconf]{acmart}
\AtBeginDocument{%
  }

\copyrightyear{2026}
\acmYear{2026}
\setcopyright{cc}
\setcctype{by}
\acmConference[CHI '26]{Proceedings of the 2026 CHI Conference on Human Factors in ComputingSystems}{April 13--17, 2026}{Barcelona, Spain}
\acmBooktitle{Proceedings of the 2026 CHI Conference on Human Factors in Computing Systems (CHI'26), April 13--17, 2026, Barcelona, Spain}
\acmDOI{10.1145/3772318.3791379}
\acmISBN{979-8-4007-2278-3/2026/04}

\usepackage{graphicx}  
\usepackage{makecell}  
\usepackage{multirow}

\begin{document}

\title["I Was Told to Come Back and Share This"]{“I Was Told to Come Back and Share This”: Social Media-Based Near-Death Experience Disclosures as Expressions of Spiritual Beliefs}


\author{Yifan Zhao}
\email{yzhao597-c@my.cityu.edu.hk}
\orcid{0009-0001-0487-785X}
\affiliation{
\institution{City University of Hong Kong}
\city{Hong Kong}
\country{China}}

\author{Yuxin Fang}
\email{fangyuxin226@gmail.com}
\orcid{0009-0007-7127-7962}
\affiliation{
\institution{Gannan University of Science and Technology\\Ganzhou Key Laboratory of Digital Cultural Preservation and Intelligent Innovation  }
\city{Gan Zhou}
\country{China}}

\author{Yihuan Chen}
\email{yihuachen6-c@my.cityu.edu.hk}
\orcid{0009-0004-2723-7398}
\affiliation{
\institution{City University of Hong Kong}
\city{Hong Kong}
\country{China}}

\author{RAY LC}
\authornote{Correspondences can be addressed to ray.lc@cityu.edu.hk.}
\email{ray.lc@cityu.edu.hk}
\orcid{0000-0001-7310-8790}
\affiliation{
\institution{City University of Hong Kong\\Studio for Narrative Spaces}
\city{Hong Kong, SAR}
\country{China}}

\renewcommand{\shortauthors}{Zhao, et al.}

\begin{abstract}

People who experienced near-death events often turn to personal expression as a way of processing trauma and articulating beliefs. While scholars have examined how individuals share near-death experiences (NDEs), limited research has explored how these narratives are communicated collaboratively on today’s social media platforms. We analyzed 200 randomly sampled TikTok videos tagged with \#nde and related hashtags. Content analysis revealed that individuals often use NDE narratives to articulate personal meaning, with spiritual and religious themes appearing in the majority of posts and serving as a means of exploring and making sense of personal spiritual perspectives. Consistent with this, analyses of comment sections reveal that videos containing spiritual themes tend to attract more engagement and foster deeper conversations around faith and meaning. Our findings offer insights into how online platforms facilitate community-level engagement with spirituality, and suggest implications for design of spaces that support shared expression and connection in specialized communities.


\end{abstract}



\maketitle

\section{Introduction}\label{sec:Introduction}

Near-death experiences (NDEs) are profound psychological experiences at the threshold of life and death \cite{moody_semantic_1975}. They often involve sensations of leaving the body, entering realms beyond reality, or encountering spiritual beings \cite{ring_near-death_1997,blackmore_dying_1993}. Research has shown that about 20\% of Near-Death Experiencers (NDErs) retain memories related to transcendental realms after the event \cite{zingrone_pleasurable_2009} and subsequently go through stages of sharing and disclosure \cite{hoffman_disclosure_1995}. For these experiencers, self-disclosure is a common and necessary part of dealing with what happened \cite{hoffman_disclosure_1995-1}. Describing their experiences helps them to understand the event and can influence their sense of self and worldview \cite{charland-verville_near-death_2015}.



While NDErs can find recounting their experiences helpful, doing so is difficult due to NDEs’ unreal and supernatural qualities \cite{hoffman_disclosure_1995}. Many experiencers hesitate in deciding whether and how to disclose their accounts, and some ultimately decide to remain silent. Social taboos around death, often shaped by cultural expectations and sensitivities around mortality, help enforce silence and can leave those who disclose their experiences feeling exposed \cite{staudt_covering_2009}. When individuals are willing to share, creating an open and empathetic environment that supports conversations about death and spirituality is essential \cite{erika_domeika_pratte_wellbeing_2021}.  



As social media platforms have grown, NDErs have increasingly shared about their NDEs on those platforms, which facilitate narrating one’s experiences and enable forms of interaction that foster intimacy, solidarity, and emotional support while easing the social pressures that often discourage talking openly about death. TikTok has become an important venue for sharing NDE content. Many creators make their experiences visible through videos, and viewers express support and share similar stories in the comment sections, collectively fostering a highly interactive and participatory digital community \cite{walther_attraction_2002}.

Studies have shown that the disclosure of NDEs can affect individuals’ lives and worldviews \cite{belanti_phenomenology_2008}. Yet, despite the growing prevalence of such disclosures on social media sites, few studies have examined how they unfold online, particularly regarding expression and their cultural or religious dimensions. Moreover, little research has been conducted on the interactive aspects of NDE disclosure; researchers have paid limited attention to how audience responses to NDE videos shape storytelling or shared understandings of spirituality.


To study how NDErs express themselves on social media platforms and the societal implications of such expressions, we conducted a mixed-method analysis of 200 TikTok videos in which users disclose their NDEs. Drawing from grounded theory and multimodal discourse analysis, we examined both the narrative content of the videos and the patterns of interaction within their comment sections. Specifically, we categorized the videos into thematic types, analyzed how religious and spiritual elements were framed, and coded comments to understand viewer engagement styles. This approach enables us to identify not only how creators narrate their NDEs, but also how audiences receive, interpret, and co-construct meaning around these disclosures in the digital environment. Accordingly, we propose the following research questions (RQs):

RQ1: How do creators express NDE content through video? What narrative forms and perspectives are typically used in these portrayals?

RQ2: How do people express their NDE as reflective of their cultural and religious background?

RQ3: How do creators connect with and express to viewers their NDE content and beliefs through comments on the videos?

By analyzing the narrative techniques and viewer engagement patterns, we aim to provide insights that could foster design practices supporting clearer testimonial expression, reducing context collapse, and encouraging a healthier, more open, and empathetic online environment for sensitive disclosure, while also offering directions for supporting thanatosensitivity.

As researchers, we approach NDE content from a sociotechnical rather than a theological or experiential perspective. Although none of us are NDErs, one team member participates in NDE-related communities and works in death education, offering familiarity with relevant norms. We acknowledge that our positional limitations may introduce interpretive gaps when analyzing NDE content online \cite{andalibi_social_2018} and remain attentive to our own identities and limited firsthand understanding of these experiences.

\section{Background}\label{sec:Background}

\subsection{Online Disclosure and NDE Expression}

\subsubsection{Self-Disclosure and Expression}

Self-disclosure and expression are fundamental practices in online environments that shape how users share about personal experiences and interact with others \cite{zhang_image_2025,zhou_eternagram_2024}. Various disclosure forms exist, including text-based \cite{ma_self-disclosure_2017} and video-based practices \cite{mckim_investigating_2021}, and they can convey such emotions as gratitude, humor, and distress \cite{gong_if_2025}.

Factors such as linguistic habits \cite{tang_tale_2011}, gender \cite{de_choudhury_gender_2017}, and sociocultural background \cite{punuru_cultural_2020} influence people’s disclosure behaviors. For instance, women are generally more willing to disclose than men \cite{wang_modeling_2016}, and men often express higher levels of negative emotions and show less willingness to seek social support \cite{de_choudhury_gender_2017}. Users from different cultural contexts also display distinct levels of expressiveness \cite{de_choudhury_gender_2017}; individuals from developing regions tend to be more reserved \cite{de_choudhury_gender_2017}. Temporal and contextual variables also influence engagement; the reception of a creator’s self-disclosure depends on how active their community members are at the time the creator posts \cite{lu_when_2015}, and disclosures’ perceived publicity significantly shapes recipients’ impressions \cite{bazarova_contents_2012}. Investigating these dynamics helps to understand how users form trust and build community engagement online.  

Designing online communities that foster healing, resilience, and recovery is critical for well-being \cite{semaan_military_2017, zhou_retrochat_2025}. Members of stigmatized or vulnerable groups frequently use digital platforms to circumvent constraints and connect with supportive audiences. Online spaces have become important venues for disclosure among LGBTQ+ individuals \cite{punuru_cultural_2020}, people recovering from addiction \cite{mckim_investigating_2021}, and people sharing illness- or trauma-related experiences \cite{bak_examining_2024, wang_weaving_2023}. Among these disclosures, the most prevalent forms include deep personal self-narratives\cite{mckim_investigating_2021}, expressions of negative emotions\cite{andalibi_self-disclosure_2017}, and explicit requests for help\cite{wang_weaving_2023}. These disclosures often elicit empathy and support while minimizing harmful behaviors \cite{andalibi_sensitive_2017}.

In disclosing sensitive experiences online, creators need to make strategic choices to balance their need for support against the risks of exposure \cite{pinch_its_2021}. Groups, including stay-at-home fathers \cite{ammari_thanks_2016}, veterans  \cite{haque_perceptions_2023}, and immigrants \cite{lingel_city_2014}, have varied reasons for disclosing. NDErs have unique motivations for and needs in narrating their experiences, such as seeking support to cope with lingering aftereffects and expressing a strong desire to talk through what they encountered\cite{pehlivanova_support_2025}, yet little is known about how they engage in online discourse. Studying their online disclosure behaviors will help clarify how platforms can support their self-expression and connections.  

\subsubsection{NDE Characteristics and Disclosure}

NDEs are psychological events with transcendent and mysterious elements, typically occurring when individuals are near death or facing extreme physical or emotional danger \cite{greyson_defining_1999}.They are widely regarded as a prototypical core experience that transcends cultural, social, and religious boundaries \cite{belanti_phenomenology_2008,blackmore_dying_1993}, as descriptions from diverse sources consistently reveal shared structural and phenomenological features. Over the past decades, researchers have developed multiple instruments for identifying NDErs \cite{greyson_near-death_1983,ring_life_1982} and have attempted to measure and evaluate the intensity, components, and psychological correlates of NDEs \cite{ring_life_1982,jr_experience_1972}. The Greyson NDE Scale operationalizes NDEs by grouping the experience into four measurable clusters, namely cognitive, affective, transcendental, and paranormal features, thereby providing a structured definition of the phenomenon \cite{greyson_near-death_1983}. More recent tools such as the NDE-C \cite{martial_near-death_2020} further refine these dimensions using contemporary psychometric validation and factor-analytic approaches.

Individuals who have experienced NDEs frequently provide detailed narratives of their experiences \cite{thonnard_characteristics_2013, palmieri_reality_2014}. Their narratives often contain cognitive phenomena such as time distortion, emotions such as love or feelings of unity, paranormal features, and transcendental aspects such as mystical encounters or perceptions of a boundary \cite{greyson_near-death_1983}. Accounts of NDEs commonly include sensations of peace, out-of-body experiences, viewing one’s surroundings from above, passing through darkness or a void, or entering areas of exceptional brightness or beauty \cite{ring_life_1982, zingrone_pleasurable_2009}. Some individuals report encounters with deceased relatives, divine figures, or other beings. While most NDEs are described as positive and filled with peace, joy, and love, a small proportion involve distressing emotions such as fear, terror, and guilt \cite{bush_distressing_2009}.

Many NDErs face integration challenges—long-lasting psychological and emotional aftereffects of an NDE \cite{foster_practical_2009}. Integrating the experience can take years and reshape a person’s worldview \cite{ sutherland_very_1991}. Successful integration is often associated with positive changes in empathy and spiritual awareness as well as reduced fear of death and diminished focus on materialism or competition \cite{khanna_daily_2014, greyson_western_2015}. The intensity of an NDE is strongly correlated with both the number and severity of the aftereffects \cite{morris_nature_2003}. 

Disclosing an NDE, particularly the initial act of sharing, plays a crucial role in helping individuals process and integrate what they have gone through. However, many experiencers hesitate to share their accounts due to fears of stigma, misdiagnosis, or dismissal \cite{foster_practical_2009}. Consequently, NDErs need supportive environments where they can disclose their experiences safely and meaningfully.

\subsection{Religious Interpretation of NDEs}

Studying NDE narratives without considering religion or spirituality would entail overlooking a crucial dimension of many experiencers’ meaning-making. NDErs frequently mention mystical experiences that involve conscious encounters with ultimate truth or divinity in their narratives \cite{greyson_congruence_2014}, and NDEs often resemble religious experiences reported in various faith traditions \cite{badham_religious_1997}. For instance, NDEs and the experiences of medieval Catholic mystics share such features as ecstatic out-of-body journeys, visions of divine beings, and the alleviation of death anxiety \cite{pahnke_psychedelic_1969, pennachio_near-death_1986}. Moreover, the lasting changes some NDErs report resemble descriptions of mystical consciousness, which highlight awareness of sacredness, preternatural insight, transience, and lasting transformation \cite{greyson_congruence_2014}. Many NDErs interpret their experiences through a spiritual lens \cite{zingrone_pleasurable_2009, liu_many_2009}. Additionally, surveys indicate that large segments of the public believe in or report having encountered supernatural phenomena such as reincarnation, spiritual energy, or contact with the dead \cite{liu_many_2009}. 

Since many religions address fundamental human concerns about death and the afterlife, religion plays a central role in shaping how individuals interpret NDEs. People who frequently engage in religious activities often show lower levels of death anxiety than those who do not \cite{martin_relationship_1965}, and people with deeply internalized religious beliefs tend to adopt more hopeful views of death and the afterlife than people with no or less firmly held religious beliefs \cite{spilka_death_1977}. Preexisting religious background sometimes influences how people interpret their NDEs \cite{badham_religious_1997, ring_life_1982}; religiously inflected NDEs are more common among individuals who regularly engage in religious practices than among those who do not \cite{mclaughlin_near-death_1984}. Conversely, NDEs can reshape or strengthen people’s spiritual outlook. Many experiencers describe becoming more devout \cite{sabom_recollections_1982}, and group-level observations suggest that NDEs often spark increased religious sentiment and greater focus on faith \cite{mclaughlin_near-death_1984}.

\subsection{Death and HCI Research}

In HCI research, work on death has primarily focused on three themes: digital remains, remembrance, and coping \cite{albers_dying_2023}. The term thanatosensitivity has been introduced to describe technological designs that address the experiences of dying individuals or bereaved groups \cite{massimi_dying_2009}. However, most existing studies concentrate on the latter and pay limited attention to how people perceive and reflect on death \cite{eum_how_2021,massimi_dealing_2011}. Yet, for many individuals, reflecting on mortality can help to “die well." \cite{meier_defining_2016} Dying well refers to consciously shaping one’s own end-of-life process to make it more meaningful and more bearable \cite{glaser_awareness_2017,kellehear_social_2007}. This process not only involves creating supportive external conditions but also encompasses inner efforts to face and cope with death positively \cite{fairbairn_good_2002}. 

Scholars have recently begun exploring how technology might enhance people’s awareness of death. Early attempts include investigating whether chatbots that engage users in conversations about mortality reduce death-related fear and support end-of-life preparation \cite{albers_lets_2024}. Similarly, games have been designed to connect and comfort people by having them share their death-related fears \cite{luo_between_2025}. Such work suggests that technology can play a meaningful role in facilitating understanding and reflecting on death \cite{albers_dying_2023,chen_what_2021,bahng_reflexive_2020}. 

Nevertheless, several potentially important dimensions remain underexplored. One such dimension is the spiritual construction of meaning \cite{albers_dying_2023}, which remains insufficiently theorized, including the unexplored question of whether such spiritual meaning is co-constructed. By examining the perceptions and expressions of individuals who have undergone NDEs, we hope to provide new perspectives for future thanatosensitive design.

\subsection{Platform Characteristics of TikTok}

TikTok differs from earlier social media platforms in how personal experiences are surfaced and circulated \cite{bhandari_whys_2022}. Its For You Page (FYP) algorithm distributes content based on engagement-driven signals such as watch time, comments, and shares \cite{xu_research_2019}. Unlike traditional networked publics, which are structured around interpersonal connections, this algorithmic logic enables TikTok to function as clustered publics formed through shared interests and content affinities \cite{gerbaudo_tiktok_2024}. As a result, compared with platforms such as Instagram Reels and YouTube Shorts, TikTok’s more fine-grained recommendation system allows users to encounter highly personalized video streams \cite{roberts_technology_2025}, with content exhibiting strong characteristics of virality and personalization.

Users’ behaviors and strategies are also shaped by these platform characteristics. In response to TikTok’s opaque recommendation system, creators often develop “folk theories” about how the algorithm operates and adjust their content production accordingly, such as optimizing interaction metrics and maintaining rapid posting rhythms to sustain algorithmic visibility \cite{karizat_algorithmic_2021}. TikTok’s platform affordances further influence creators’ practices: social and creative features such as Duet, Stitch, and video-based comment replies not only incentivize audience participation by rewarding interaction and responsiveness \cite{kaye_jazztok_2023}, but also shape how creators plan, extend, and adapt their posting strategies in order to maintain ongoing visibility and engagement on the platform \cite{herman_for_2023}.

\section{Methods}\label{sec:Methods}
\subsection{Data Collection}

To identify hashtags and keywords related to NDE self-disclosures, we first referenced \textit{Life After Life} \cite{moody_life_1976}, which is widely regarded as the first systematic account of modern NDE reports, and searched for directly related terms such as “near-death experience" and “near-death." During this initial stage, we followed our IRB-approved protocol, which restricted data collection to publicly accessible content and prohibited gathering any personal identifiers beyond what was required to retrieve the videos. We then identified frequently occurring hashtags in the posts we found and used those as new search terms, continuing this iterative process until no new terms emerged. This process resulted in a list of keywords that included “afterdeath," “afterlife," “lifeafterdeath," “neardeathexperience," “nde," and “neardeath". 

On December 21 and 22, 2024, we collected posts using the TikTok data API. Our search encompassed posts uploaded over a three-year timeframe (December 20, 2021–December 20, 2024). We chose this period because it covers the recent rise of NDE-related sharing on TikTok, as reflected in the noticeable increase of NDE-related hashtags and posting frequency in our preliminary scans, and it also provides a broad enough time span to  include different kinds of first-person NDE narratives as well as related explanatory or commentary posts. Consistent with guidance for ethical social media research \cite{rsumovic_social_2016,fielding_sage_2016}, we did not circumvent any technical barriers and collected only data that TikTok makes publicly available through its API, avoiding any access to private accounts or restricted content. We gathered a total of 12,097 posts. The average video length was 109.5 seconds. To facilitate analysis and minimize outliers, we excluded image posts and videos shorter than 5 seconds or longer than 690 seconds (three standard deviations above the mean). We considered very short clips insufficient to contain meaningful narratives, while unusually long videos were often full-length programs or non-typical content, which could distort the analysis. This filtering process resulted in a final dataset of 9,868 videos.

Our keyword-based search method yielded a large number of irrelevant results. For example, searching for “afterlife" surfaced posts related to a concert tour of that name. To address this, we manually filtered out posts that were clearly unrelated to NDEs based on their titles and descriptions. During this initial stage, we obtained institutionally approved IRB protocol for publicly available social media research, which restricted data collection to publicly accessible content and prohibited the collection of any personal identifiers beyond what was necessary to retrieve the videos. To ensure a consistent and equitable workflow across all videos, we initially relied on machine translation to process non-English titles and text. However, automated translation can distort culturally embedded expressions and introduce errors that undermine coding reliability. To maintain analytic consistency and minimize misinterpretation risks, we restricted our final dataset to videos expressed in English. We then performed a second round of screening to further ensure our sample consisted of videos related to NDE self-disclosure. We briefly viewed each video to confirm whether it met all of the following criteria: directly related to an NDE, an original creation (not a clip or reposted segment from a documentary or television show), and produced by individual video bloggers who had experienced an NDE. Through this screening process, we identified 1017 NDE self-disclosure videos.

\subsection{Data Analysis}

\subsubsection{Video Narrative Analysis: Content and Function}
We adopted a hierarchical mixed approach to examine NDE-related videos at two levels: narrative content (what was said) and narrative function (why it was said). For the content level, we drew on the established NDE framework \cite{hashemi_explanation_2023}(Table 1) , which integrates elements from classic NDE models \cite{greyson_near-death_1983,ring_life_1982,moody_semantic_1975} and organizes NDE accounts into four recurring experiential domains: supernatural experiences, spiritual or religious experiences, cognitive experiences, and emotional experiences. Two researchers independently applied a preliminary set of codes to a shared subset of 20 videos and reached substantial agreement (Cohen’s $\kappa$ = 0.83). Building on this pilot, we randomly sampled an additional 180 videos, producing a total of 200  (follower counts: approximately 50–2.4M), which represented about 20\% of the dataset of 1017 videos.  Before coding, we fully transcribed all videos in the dataset. Three researchers independently viewed the videos, segmented them into units of around five seconds that usually corresponded to a complete spoken sentence or a coherent narrative moment, and wrote frame-specific notes.We also made frame-specific notes while at the same time checking the transcripts so that their judgments were based on both the video content and the written text. We judged that thematic saturation had been reached when new videos no longer generated additional codes or categories. Through iterative comparison and affinity diagramming, the open codes were refined during axial and selective coding and organized into four major frames: emotional, cognitive, religious or spiritual, and supernatural. Each frame was further divided into subcategories such as positive and negative affect, out-of-body experiences, life review, and encounters with the dead. The finalized content codebook is presented in Table 2 and was used to annotate the full dataset. The complete workflow of data collection, preprocessing, and coding is documented in the appendix figures for reference.

Guided by the four domains of the framework, we first distinguished accounts that focused solely on physical danger from those that contained cognitive, emotional, or supernatural components. During open coding, we also observed that spiritual and religious references consistently formed a coherent narrative pattern. In contrast, cognitive, emotional, and supernatural elements frequently co-occurred and were difficult to isolate as independent categories. For this reason, we structured our coding system around three types of disclosures: Physiological Crisis Experiences (PCEs), Classical NDEs that contained cognitive or supernatural elements but did not feature spiritual or religious components (NS-NDEs), and a distinct subset of Classical NDEs featuring prominent spiritual or religious themes (S-NDEs), as shown in Table 2. This coding structure informed the subsequent content analysis.

\begin{table}[t]
\centering
\renewcommand{\arraystretch}{1.3}
\begin{tabular}{|
  >{\raggedright\arraybackslash}m{0.28\linewidth}|
  m{0.68\linewidth}|
}
\hline
\textbf{Frame} & \textbf{Definition} \\
\hline
\textbf{Supernatural experiences} 
& Experiences involving separation from the body and metaphysical perceptions beyond normal physical boundaries. \\
\hline
\textbf{Spiritual and religious experiences} 
& Encounters with spiritual beings or deceased individuals and visions of religious figures or realms. \\ 
\hline
\textbf{Cognitive experiences} 
& Mental clarity and perceptions such as enhanced senses, time distortion, and deep insight, often while unconscious. \\
\hline
\textbf{Emotional experiences} 
& Strong emotional states, usually peace and bliss, but sometimes distress, occurring during or after the experience. \\
\hline
\end{tabular}
\caption{\textbf{Definitions of frames in the NDE framework identified in \cite{hashemi_explanation_2023}}}
\end{table}

\begin{table}[htbp]
\centering
\renewcommand{\arraystretch}{1.3}
\begin{tabular}{|
  >{\centering\arraybackslash}m{0.12\linewidth}|
  >{\centering\arraybackslash}m{0.14\linewidth}|
  m{0.60\linewidth}|
}
\hline
\multicolumn{2}{|c|}{\textbf{Category}} & \textbf{Definition} \\
\hline
\multicolumn{2}{|c|}{\textbf{PCEs}} &
Depicting physical-level near-death events (e.g., trapped in rapids). These experiences do not qualify as ``spiritual NDEs'' (e.g., entering a tunnel of light or meeting deceased loved ones), which are not considered typical or borderline NDEs. \\
\hline
\multirow{2}{*}{\textbf{NDEs}}
& NS-NDEs &
Classical NDEs that do not include spiritual or religious content. These may still feature elements like out-of-body experiences, life reviews, or heightened senses. \\
\cline{2-3}
& S-NDEs &
Classical NDEs that include spiritual and religious elements such as encounters with religious figures, oneness with the universe, or visions of the afterlife. \\
\hline
\end{tabular}
\caption{\textbf{Narrative Typology of Near-Death Experience Disclosures}}
\end{table}

\begin{table*}[t]
\centering
\renewcommand{\arraystretch}{1.5}
\begin{tabular}{|>{\raggedright\arraybackslash}p{1cm}|>{\raggedright\arraybackslash}p{4cm}|>{\raggedright\arraybackslash}p{7.5cm}|}
\hline
 & \textbf{Concept} & \textbf{Description} \\
\hline
\multirow{3}{*}{\rotatebox{90}{\makecell{\textbf{Emotional}\\\textbf{experiences}}}} 
 & \textbf{Positive experiences} & Feeling emotions such as peace, love, or joy. \\
\cline{2-3}
 & \textbf{Negative experiences} & Experiencing fear, anxiety, or distress. \\
\cline{2-3}
 & \textbf{Absence of experience} & No emotional response or feeling reported. \\
\hline

\multirow{9}{*}{\rotatebox{90}{\makecell{\textbf{Supernatural}\\\textbf{experiences}}}} 
 & \textbf{Out of body experiences} & Perceiving oneself as separate from the physical body. \\
\cline{2-3}
 & \textbf{Seeing your body} & Observing one's physical body from an external perspective. \\
\cline{2-3}
 & \textbf{Tunnel experience} & Traveling through a dark or lighted tunnel. \\
\cline{2-3}
 & \textbf{Awareness of the surrounding environment} & Perceiving details of the nearby space or setting. \\
\cline{2-3}
 & \textbf{Move to the ceiling} & Floating upward and viewing the scene from above. \\
\cline{2-3}
 & \textbf{Presence in new places (paradise)} & Being in a different or heavenly location. \\
\cline{2-3}
 & \textbf{Floating} & Feeling weightless or drifting without physical support. \\
\cline{2-3}
 & \textbf{Understanding the presence of another} & Perceiving another being nearby, without necessarily seeing them. \\
\cline{2-3}
 & \textbf{Seeing the dark} & Being surrounded by or moving through darkness. \\
\hline

\multirow{4}{*}{\rotatebox{90}{\makecell{\textbf{Cognitive}\\\textbf{experiences}}}} 
 & \textbf{Heightened senses} & Intensified perception, such as sharper vision or hearing. \\
\cline{2-3}
 & \textbf{Life review} & Recalling and re-experiencing past life events. \\
\cline{2-3}
 & \textbf{Changing the nature of time} & Experiencing time as faster, slower, or non-linear. \\
\cline{2-3}
 & \textbf{Awareness of the future} & Perceiving or understanding future events. \\
\hline

\multirow{4}{*}{\rotatebox{90}{\makecell{\textbf{Spiritual and}\\\textbf{religious}\\\textbf{experiences}}}} 
 & \textbf{Meeting with the dead} & Encountering deceased individuals. \\
\cline{2-3}
 & \textbf{Meeting with religious figures} & Interacting with spiritual or divine beings. \\
\cline{2-3}
 & \textbf{Oneness with the world} & Feeling deeply connected with the universe or all existence. \\
\cline{2-3}
 & \textbf{Observing punishment and reward} & Witnessing judgment, consequences, or moral outcomes. \\
\hline

\end{tabular}
\caption{\textbf{The concepts and definitions of sub-frames in NDE framing from}}
\label{tab:nde-compact}
\end{table*}

After establishing the content categories, we turned to the functional or motivational dimension of the narratives. Using the same set of 200 videos, the research team conducted another round of open coding to examine the communicative purposes underlying each segment. For example, the statement “I was crying because I was really scared" was coded as a negative emotional experience at the content level and as self-expression at the functional level. Similarly, “I met my deceased mother and she told me to come back and keep living" was categorized as meeting the dead in terms of content and as identity clarification in terms of function. Through affinity diagramming and selective coding, we clustered these functional codes and identified six recurring themes. To assign each video to a predominant theme, we considered the distribution and relative weight of functional codes, and reached decisions through team discussion. When videos contained multiple codes, we prioritized both their frequency and interpretive salience in determining which of the six themes was most representative. The six functional themes and their descriptions are summarized in Table 3.

\begin{table}[t]
\centering
\renewcommand{\arraystretch}{1.3}
\begin{tabular}{|
  >{\raggedright\arraybackslash}m{0.30\linewidth}|
  m{0.65\linewidth}|
}
\hline
\textbf{Theme} & \textbf{Description} \\
\hline
\textbf{Identity clarification} 
& To increase personal clarification and convey one's personal identity \\
\hline
\textbf{Self-expression} 
& Express feelings and thoughts; release feelings \\
\hline
\textbf{Information sharing} 
& To benefit other(s) by sharing information or a personal experience \\
\hline
\textbf{Resource gain} 
& Obtain benefits and information from others; to seek for help \\
\hline
\textbf{Entertainment} 
& Personal enjoyment, future use (storage), and pleasure \\
\hline
\textbf{Emotional Support} 
& Provide comfort, encouragement, or empathy to foster a sense of connection \\
\hline
\end{tabular}
\caption{\textbf{Six Functional Themes Identified in NDE-Related Video Content}}
\end{table}

\subsubsection{Visual Analysis}
We also conducted a separate round of coding that focused on the visual presentation of the videos alongside their titles and descriptions. This stage aimed to capture how experiencers framed and sustained their accounts through visual and paratextual strategies. We examined features such as imagery, editing patterns, and sequencing, and noted whether a video appeared as an independent story, as a direct reply to viewers’ comments, or as part of a continuing series. Self-initiated video postings were often the creator’s first or only NDE-related video and visually centered solely on the narrator without additional elements. Posting in response to comments typically displayed the question being answered on screen and mentioned it in the title or description. Updating a video series was marked by notations such as “p2.PT.3" within the video. When a video simultaneously included both a p2 notation and a displayed viewer question, we categorized it as responding to comments, reasoning that the continuation was primarily triggered by audience interaction. Throughout this process, our visual analysis focused strictly on publicly available content and avoided any inferential judgments about creators’ identities, intentions, or personal circumstances. Through iterative viewing and discussion, we consolidated these observations into three disclosure patterns on TikTok: self-initiated video posting, responding to comments, and updating video series. Table 4 provides an overview of these patterns and their defining visual and paratextual features.

\begin{table}[t]
\centering
\renewcommand{\arraystretch}{1.2}
\begin{tabular}{|c|}
\hline
\textbf{Self-Initiated Video Posting} \\
\includegraphics[width=0.9\columnwidth]{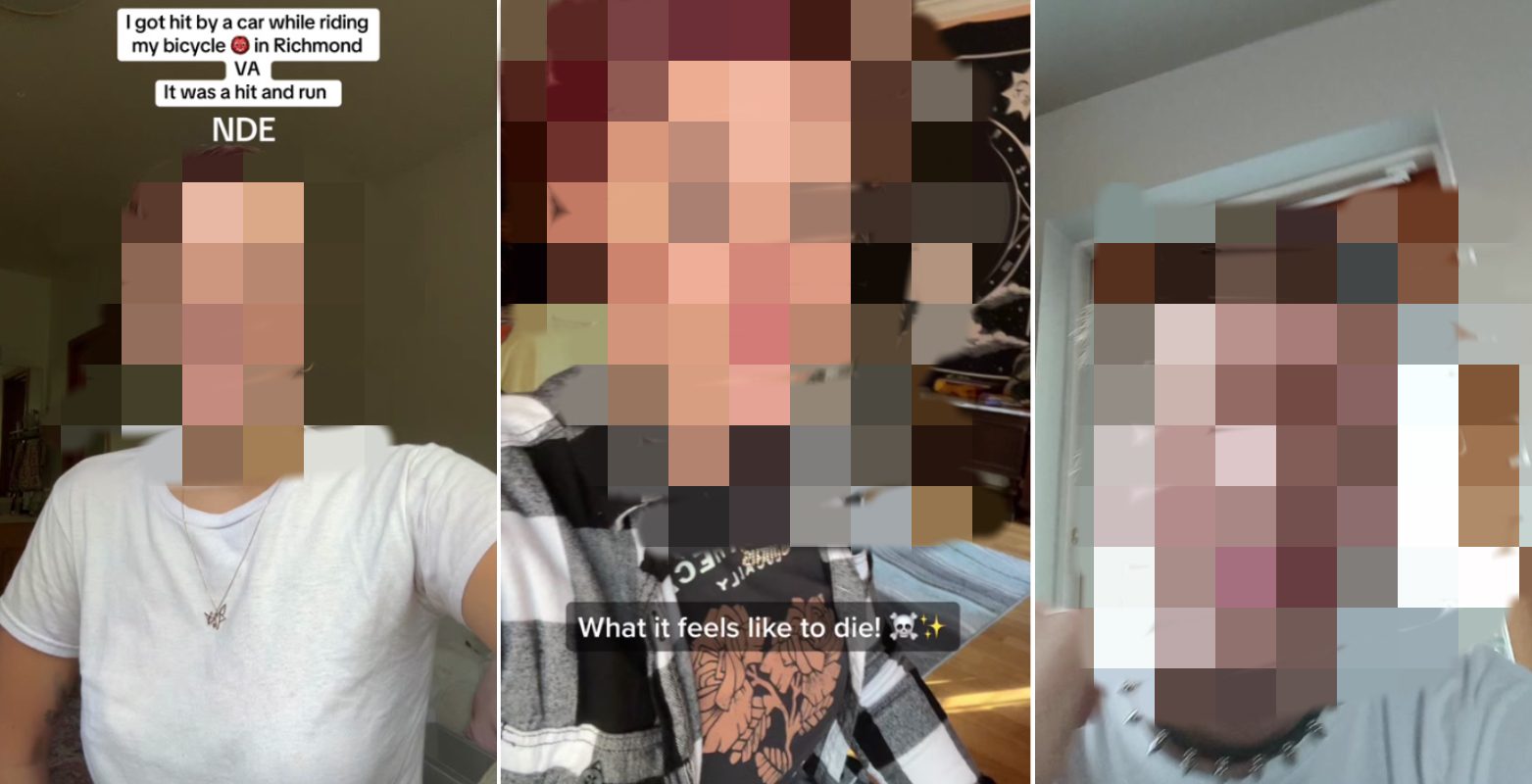} \\
\hline
\textbf{Responding to Comments} \\
\includegraphics[width=0.9\columnwidth]{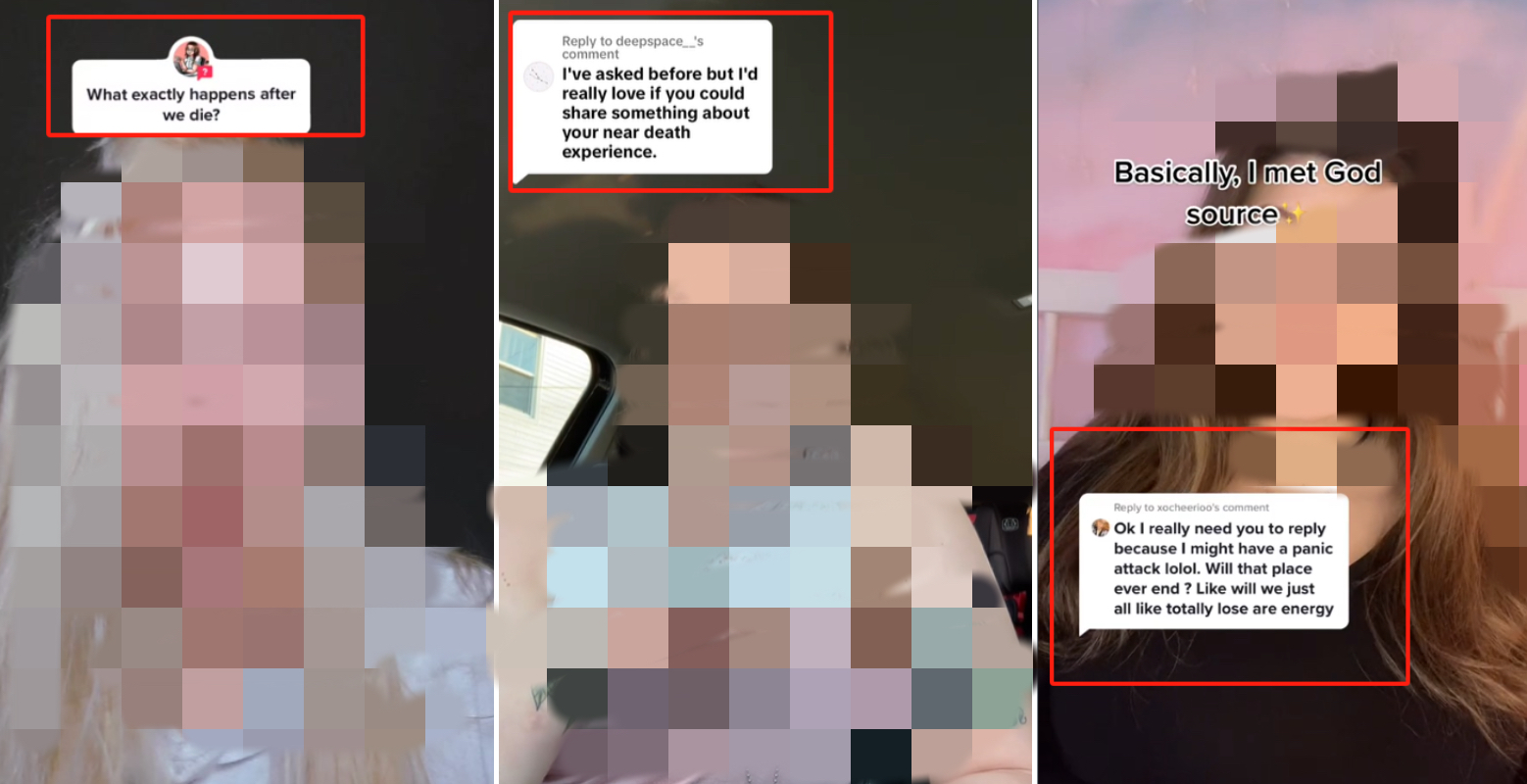} \\
\hline
\textbf{Updating Video Series} \\
\includegraphics[width=0.9\columnwidth]{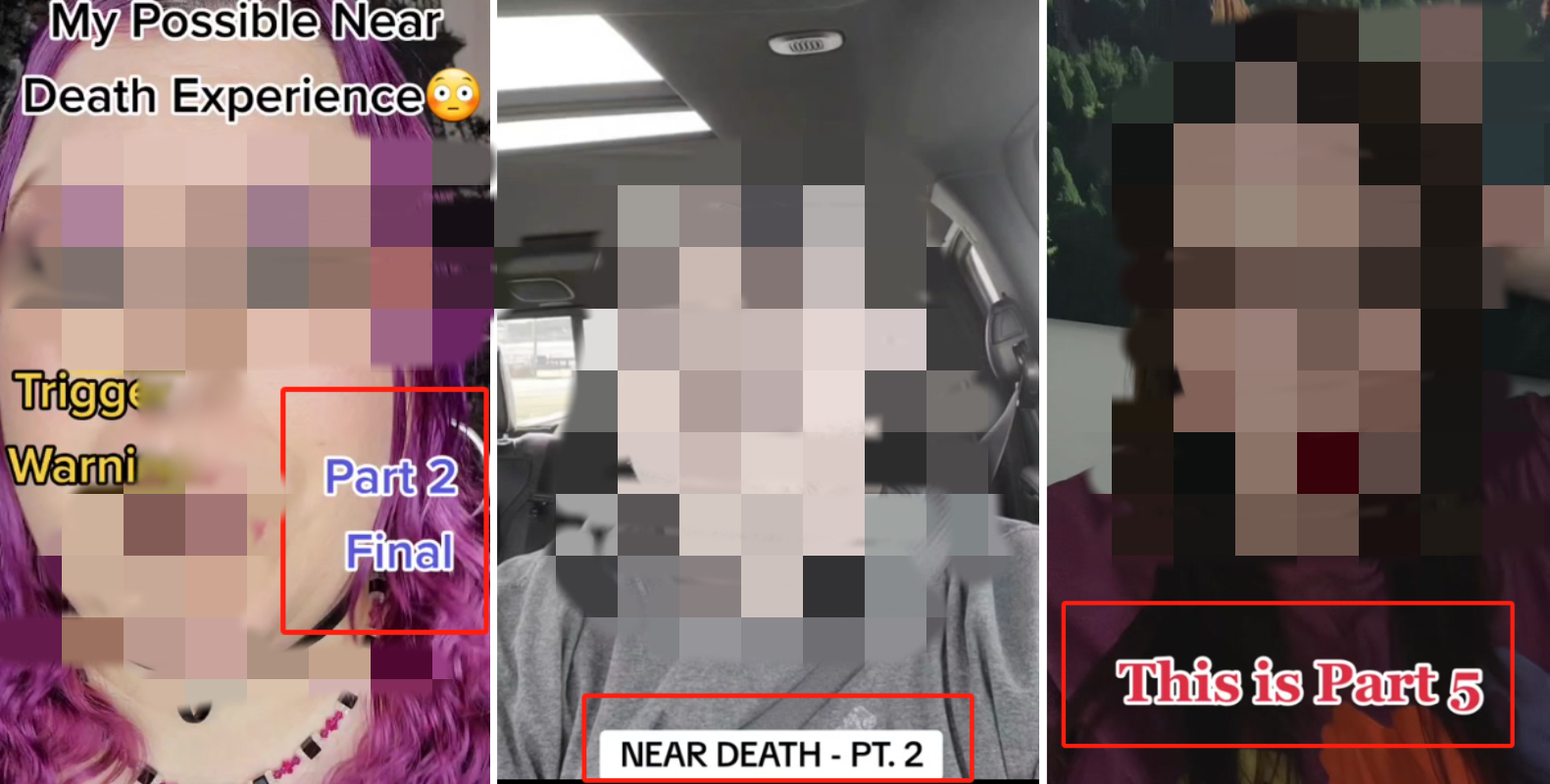} \\
\hline
\end{tabular}
\caption{\textbf{Three Patterns of NDE Video Disclosure on TikTok}}
\end{table}

\subsubsection{Comment Sampling and Coding}
To examine audience engagement, we sampled TikTok comments at the level of individual videos. Each of the 200 videos in the dataset was treated as a separate unit. For each video, we drew up to 10\% of its comments, capping the maximum at 20 to prevent extremely popular videos from dominating the analysis. Videos with fewer than three comments were included in full, and those without comments were excluded. This procedure resulted in a final corpus of 1,943 comments, with per-video samples ranging from 0 to 20. We used a fixed random seed and reservoir sampling to ensure reproducibility. Following pre-registered criteria, we removed duplicates, spam, and empty @-mentions. Because the 20-comment cap lowered the inclusion probability for heavily commented videos, we calculated sampling weights for each comment and used these to compute all proportions and confidence intervals.

\begin{table*}
\centering
\renewcommand{\arraystretch}{1.4}
\begin{tabular}{|p{6cm}|p{11cm}|}
\hline
Topic & Description \\
\hline
Topic 1 (Personal Experience) & Sharing of one's own or others’ NDE or similar experiences. \\
\hline
Topic 2 (Discussion and Thoughts) & Discussion of the video content, including explanations and interpretations. \\
\hline
Topic 3 (Questioning) & Expressions of doubt, questions or skepticism about the content. \\
\hline
Topic 4 (Memorial) & Comments expressing condolences to those who have passed. \\
\hline
Topic 5 (Faith and Religious) & Comments grounded in religious or spiritual beliefs. \\
\hline
Topic 6 (Gratitude) & Viewers express thanks or appreciation for the content shared in the video. \\
\hline
Topic 7 (Support and Empathy) & Affirmations, emotional support, and expressions of empathy \\
\hline
Topic 8 (Humor and Interaction) & Casual or social interaction that may or may not directly relate to the video’s topic \\
\hline

\end{tabular}
\caption{\textbf{Comment topics for 200 TikTok videos}}
\end{table*}

To address potential biases associated with platform ordering \cite{lupinacci_phenomenal_2024}, all comments were collected using TikTok’s “Newest” ordering rather than the algorithmically ranked “Top” view. This ensured that our sample reflected the chronological flow of audience engagement rather than TikTok’s relevance-based ranking. Pinned comments, which always appear at the top in “Top” view, do not automatically rise to the top in “Newest” mode and were therefore sampled in the same manner as other comments. If a pinned comment fell within the randomly selected range, it was included; otherwise it had no special influence on the sample. Comments deleted prior to data collection were not available to us and could not be recovered, which may introduce a minor survivorship bias; however, because our scraping window occurred after most videos had accumulated stable engagement, we expect the missingness to have minimal impact on thematic distribution.

We developed the coding scheme inductively. Two coders began by closely reading and open-coding a subset of comments, generating tentative labels for the main ideas expressed. Through iterative comparison and discussion, we consolidated these labels into eight themes(table 5): personal experience, discussion of the video, questioning, memorial expression, religious belief, gratitude, support and empathy, and humor. To validate the scheme, two coders independently annotated 10–15\% of the data and reached an inter-rater reliability of (Cohen’s $\kappa = 0.79$). Disagreements were reviewed, and definitions were refined before one lead coder completed the remaining annotations with periodic double-coding checks. Each comment could be assigned multiple labels, but coders also designated a primary theme to capture its main orientation. All subsequent analyses were based on this refined codebook. For ethical considerations, all quoted comments were lightly paraphrased to remove identifiable details while preserving their original meaning.

\subsubsection{Ethical Considerations}

This study followed an institutionally approved IRB protocol for research involving publicly available social media data. The approved use includes personal account metadata and publicly accessible TikTok videos and specified procedures for handling, storing, and analyzing creators’ posts without attempting to identify, contact, or intervene with users. In conducting the study, we adhered to strict protection of user anonymity, the exclusion of non-public or restricted content, and careful consideration of ethical risks associated with studying sensitive, death-related disclosures.

We acknowledge that the public availability of data does not automatically render its use ethically unproblematic. Given that users may not fully leverage privacy settings \cite{sarikakis_social_2017, lenhart_i_2025}, and that even publicly posted content is often shared with contextual privacy expectations \cite{fiesler_participant_2018,klassen_this_2022}, and recognizing that TikTok creators may not anticipate researchers as part of their imagined audience or that videos may be repurposed beyond the uploader’s intended context, we took steps to minimize risks to individuals whose content we analyzed.

First, we collected only API-accessible data, stored no usernames or profile images, and conducted all screening procedures within secure workstations. We excluded reposted content, minors, and third-party accounts to avoid analyzing involuntary disclosures or material shared without full personal agency. Second, we avoided verbatim quotations that could be reverse-searched and reported all content in aggregated or paraphrased form \cite{roberts_ethical_2015}, ensuring that creators’ posts remained embedded within a larger dataset \cite{fiesler_participant_2018}. Following recommended practices that treat anonymized citations with user consent as the safest way to reference sensitive content \cite{vitak_ethics_2017}, we contacted the relevant TikTok users and obtained permission for images from videos. In addition, to maximize community benefit \cite{fiesler_participant_2018,gilbert_when_2023}, when requesting permission from creators to use their video content, we also promised to share our research findings with them. Finally, we considered how our work might affect individuals who have already posted sensitive content \cite{fernandez_informing_2003}. Since social media users may have concerns about how their data are used in research and may even experience harm as a result \cite{fiesler_participant_2018}, we respected their actions and dignity \cite{vitak_beyond_2016}, refrained from making evaluative claims about the authenticity of NDEs, and avoided highlighting specific creators or identifiable cases.

If readers are concerned about their data appearing in this study, they may contact the corresponding author for clarification. If any misuse or identification of creators is brought to our attention, we will report such cases to our institutional ethics committee and, if necessary, contact the SIGCHI Research Ethics Committee and ACM. We hope these procedures help prevent potential misuse and ensure that our analysis remains respectful of individuals whose content is represented in the dataset.

\section{Results}\label{sec:Results}

In this section, we examine how TikTok creators and commenters engage with the topic of NDEs, using spiritual and religious references to share their thoughts, feelings, and personal stories. We begin by introducing the dataset and summarizing how the videos were coded into different NDE types to identify the motivations behind disclosure. We then describe the ways spiritual and religious themes are presented in the videos and how these themes are echoed or supported in the comments. 

\subsection{Expression Patterns and Thematic Distribution of NDE Self-Disclosure}

\subsubsection{Types of NDE}

Our analysis identified two overarching categories of accounts. Physiological Crisis Experiences (n = 57) described physical danger without cognitive, spiritual, or supernatural elements. These cases focused on moments of acute threat to physical survival, such as drowning incidents or near-accidents.

Classical NDEs (n = 143) included experiential components that extended beyond the physical domain, such as out-of-body sensations, heightened perception, life reviews, and encounters with realms beyond ordinary experience. Within these Classical NDEs, S-NDEs, which contain spiritual and religious references, formed a distinct subset. This pattern appeared in 113 cases, representing 79\% of all Classical NDEs, and accounted for a large portion of NDE narratives in our dataset.

\subsubsection{Distribution of Narrative Motivations }

We further examined the distribution of narrative motivations across different types of near-death experiences. As shown in Figure 1, the most frequent motivations are identity clarification, self-expression, and information sharing, with information sharing being the most common. Resource gain appears only rarely, with very few cases. Entertainment is concentrated primarily in PCE videos, where it accounts for a substantial share, while NDE videos almost never involve entertainment. Self-expression appears in both PCEs and NDEs, but PCEs show a relatively higher proportion, often using lighthearted or humorous approaches to convey emotions after the experience. In contrast, NDE videos are more often associated with identity clarification and information sharing, and they also include a notable proportion of emotional support. A closer look shows that a large number of identity clarification videos belong to the S-NDE category, while no S-NDEs appear in the entertainment category. 

\begin{figure}
    \centering
    \includegraphics[width=1\linewidth]{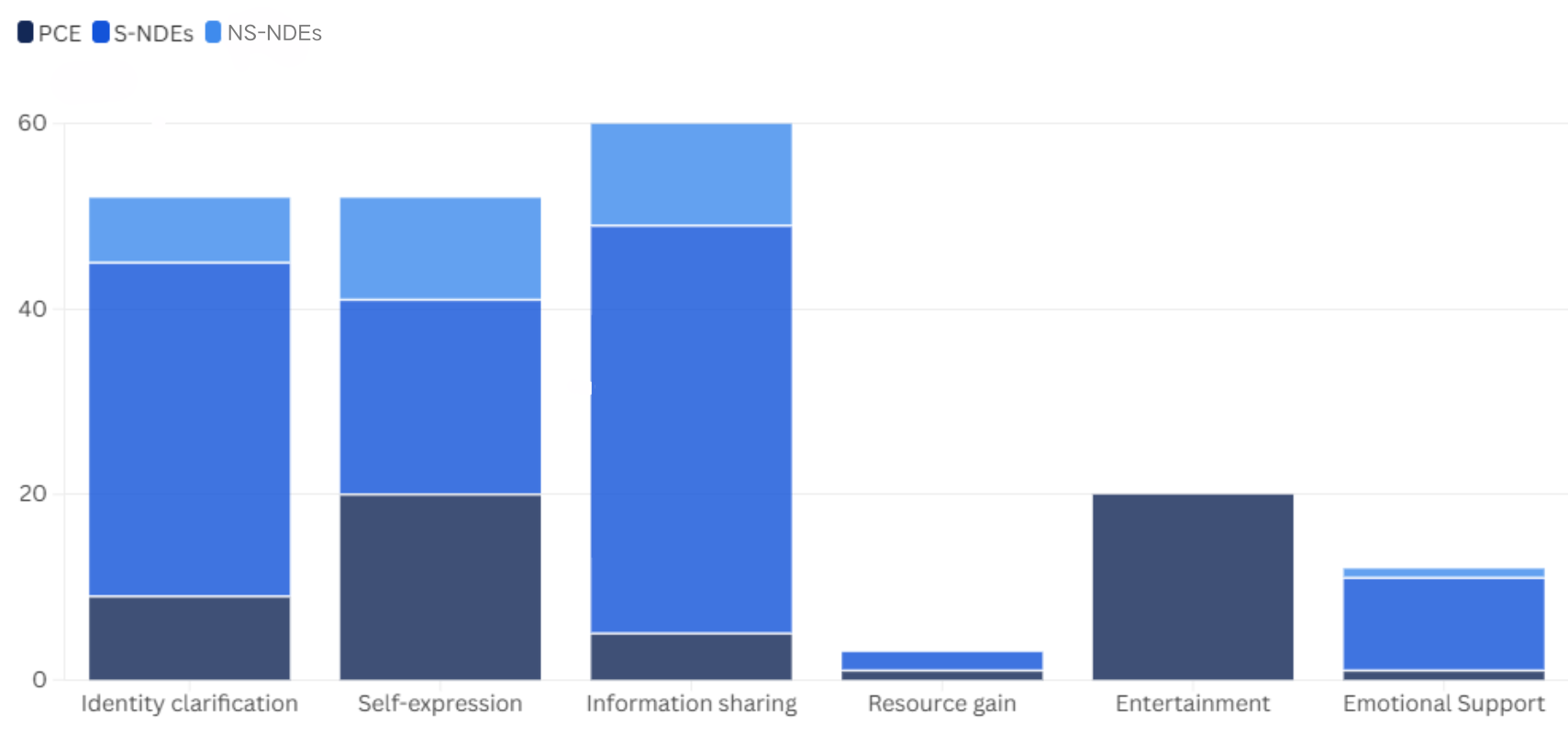}
    \caption{\textbf{Motivations of NDE TikTok Videos by Category}
}
    \label{fig:enter-label}
\end{figure}

\subsubsection{Forms of NDE Disclosure }

In addition to identifying overarching narrative motivations such as self-expression and identity clarification, we also observed three distinct forms of NDE disclosure on TikTok (Table 4), identified through the visual presentation of the videos together with their titles and descriptions: initial self-disclosure, comment-driven interaction, and ongoing serial updates. 

Initial self-disclosure (n = 103) was the most common. These videos typically represent a creator’s first attempt to publicly recount a near-death experience. Narratives usually begin directly with a description of the event, without references to earlier posts or external information. Creators in this category often focus on the experience itself and its impact on their thoughts or perspectives, rather than explicitly explaining their motivation for posting. Comment-driven disclosures (n = 45) were produced in response to audience comments or questions. These videos frequently open with phrases such as “answering a question" or “replying to a comment" and often display comment excerpts on screen. In 21 cases, creators explicitly mentioned that they chose to share their NDE after being encouraged or requested to do so by friends or followers. Typical viewer prompts included “Can you share your story with us?", “Have you ever made a video about your NDE?", and “You should post this on TikTok." Some videos were labeled as “Part 2" or “Part 3" while also addressing viewer questions; we placed these in the comment-driven category, since they were shaped more by audience requests than by the creators’ own initiative to continue narrating. Ongoing serial updates (n = 52) involved creators posting follow-up videos to elaborate on earlier accounts. These often carried titles such as “Part 2" or “Follow-up," and introduced new reflections, details, or clarifications. In some cases, creators revisited their initial stories to provide more context or to respond to misunderstandings raised in the comments. Together, these patterns show that many creators who first shared their NDEs later returned with additional posts, either to extend their narratives, add new insights, or engage with the reactions of their audiences.


\subsection{ \textbf{Religious / Spiritual Interpretation of NDE Experiences }}

To better understand how individuals express spiritual and religious experiences in NDEs, we conducted a focused analysis of S-NDEs videos. The three most prevalent motivations were selected for detailed examination. 

\subsubsection{Emotional Self-Expression } Self-expression content in S-NDE videos primarily consisted of narratives depicting personal feelings, emotional responses, and encounters with religious figures. Emotionally rich language (such as \textit{“beautiful," “terrifying," “unexplainable"}) was prevalent, and expressive conversational phrases used to describe one’s own emotional state were frequently used to convey the emotional intensity of the experience. For example, one user expressed that after seeing Jesus, they cried and felt an overwhelming desire to share their story. In addition, speakers in these videos often displayed visible signs of emotional breakdown, such as crying, trembling, or pausing to collect themselves. These behaviors helped to reinforce the authenticity and sincerity of their storytelling. As one experiencer repeatedly emphasized these events were beyond words, and they were only trying their best to convey what had happened. Others adopted more creative forms of expression, such as drawing, singing, or poetry, to visually represent their spiritual experiences. Unlike information-sharing or identity-clarification videos, those focused on self-expression did not explain mechanisms or provide factual knowledge. Instead, they relied on the emotional dimension of storytelling to convey spirituality and faith. The intensity of their disclosure heightened its emotional resonance, and the latter, rather than rational argumentation, lent credibility to the spiritual or religious elements reported. 

\subsubsection{Affirmation of Identity}
 Identity clarification content in S-NDE videos centered on the way that individuals come to understand, affirm, or reconstruct their sense of self through the experience. The most common NDE features in this category included Meeting with Religious Figures and Oneness With the World. Speakers often described how they felt \textit{“seen"} or \textit{“understood"} during the experience, or how they finally knew who they were, indicating that their self-perception was transformed. For example, one individual shared that they were explicitly told to return with a responsibility to help and heal others, while another described being asked to assist people and speak on behalf of a divine figure. These constructions of identity were often tied to a sense of mission that arose after an encounter with a religious figure. Some videos emphasized an increase in abilities or a sense of awakening following the NDE. One speaker described feeling as though they had become all-knowing, possessing answers to every question on earth, while another explained that they returned with a clear personal mission to bring a higher sense of “home” or spiritual resonance back to the world. The emotional tone of these videos was generally calm and reflective, with most focusing more on internal realization than on external expression. Compared to other categories of video content, identity clarification videos were more likely to use spiritual and religious experiences (e.g., encounters with religious figures or divine presence) as catalysts for transformation. These experiences prompted the individual to make changes in their life based on insight into personal identity, core values, or spiritual purpose. 

\subsubsection{Dissemination of Spiritual Beliefs}
 Information-sharing content in S-NDE videos mainly appeared in descriptions of the NDE that were intended to help others. The most common experiential features in this category were Oneness With the World, Positive Experiences, and Out-of-Body Experiences. Characteristics across the spiritual, cognitive, emotional, and supernatural dimensions were mentioned with relatively equal frequency. These videos often adopted an instructive tone, with speakers summarizing what they had learned or explaining why they believed people exist for a purpose. Moreover, speakers tended to describe the situations and feelings they experienced during the NDE in an objective manner, frequently delivering lessons about love, judgment, the afterlife, or the nature of the soul. For example, one person shared the view that all individuals are expressions of God and are formed from the same unified energy, while another explained the idea they received that concepts such as time and space do not truly exist. The primary aim in these videos was not emotional catharsis or self-discovery but rather the transmission of knowledge, insight, or spiritual and religious truths. In many videos, speakers responded directly to viewers’ comments and engaged in Q\&A. For instance, one person responded to a comment asking about the origin of time and how it came into being. A few also included diagrams, hand-drawn visuals, quotations from religious texts and cosmological drawings to help structure their accounts.

\begin{table*}[t]
  \centering
  \begingroup
  \captionsetup{aboveskip=2pt, belowskip=2pt}
  \setlength{\textfloatsep}{6pt}
  \setlength{\floatsep}{6pt}
  \setlength{\intextsep}{6pt}
  \setlength{\tabcolsep}{3pt}\renewcommand{\arraystretch}{1.03}

  \begin{tabular}{
    >{\centering\arraybackslash}m{0.24\linewidth}
    >{\centering\arraybackslash}m{0.24\linewidth}
    >{\centering\arraybackslash}m{0.24\linewidth}
    >{\centering\arraybackslash}m{0.24\linewidth}
  }
    \includegraphics[width=\linewidth]{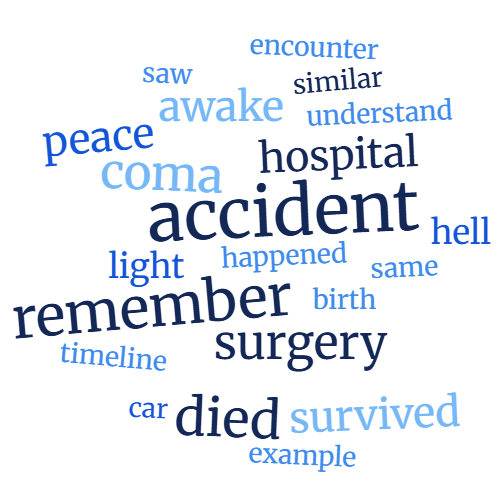} &
    \includegraphics[width=\linewidth]{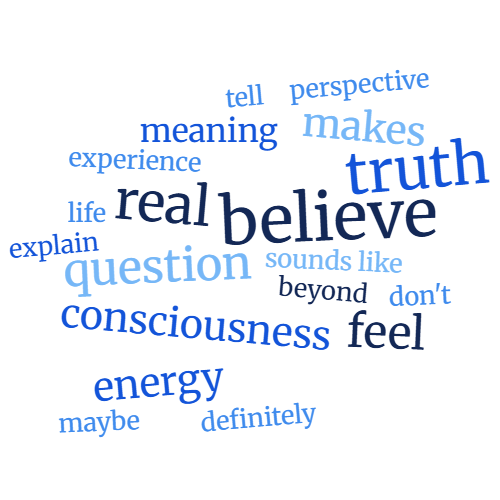} &
    \includegraphics[width=\linewidth]{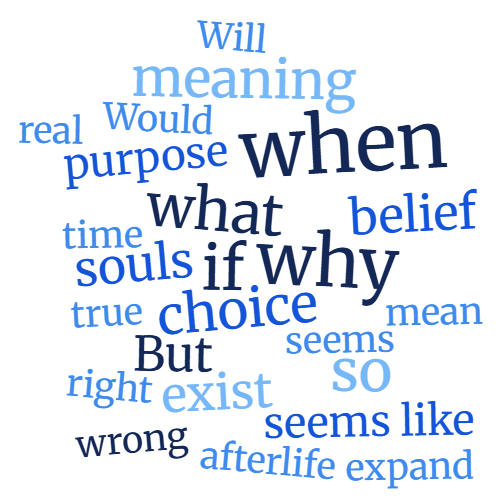} &
    \includegraphics[width=\linewidth]{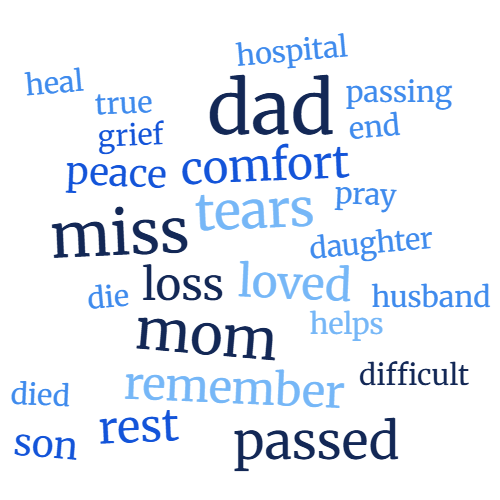} \\
    \makecell{Topic 1\\Personal Experience} &
    \makecell{Topic 2\\Discussion and Thoughts} &
    \makecell{Topic 3\\Questioning} &
    \makecell{Topic 4\\Memorial} \\
    \includegraphics[width=\linewidth]{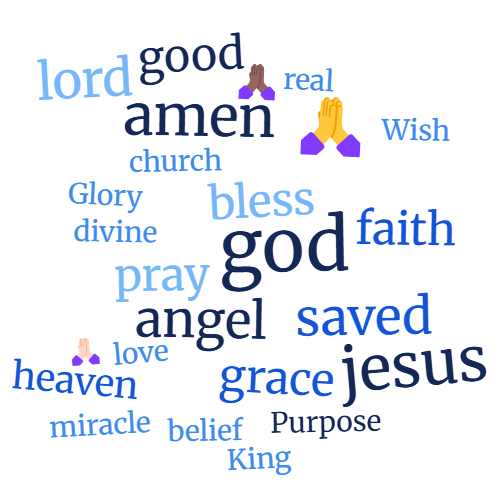} &
    \includegraphics[width=\linewidth]{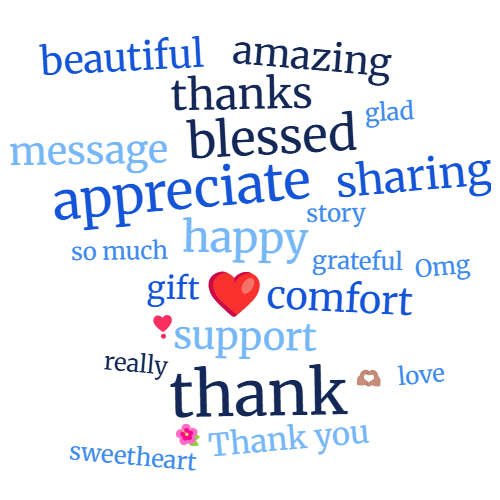} &
    \includegraphics[width=\linewidth]{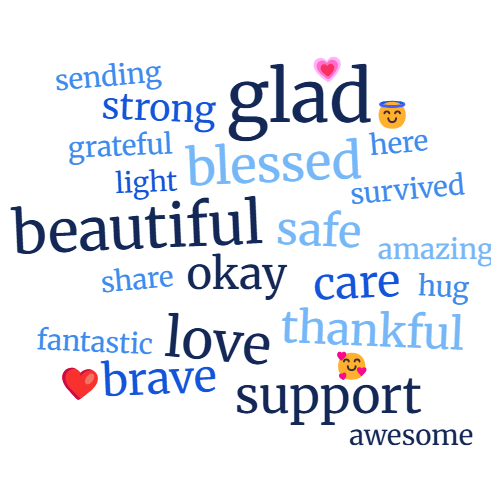} &
    \includegraphics[width=\linewidth]{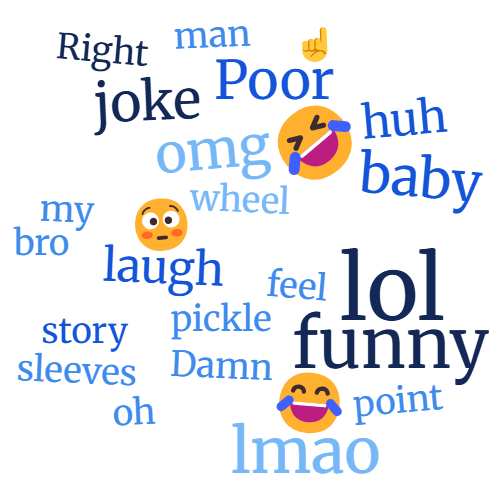} \\
    \makecell{Topic 5\\Faith and Religious} &
    \makecell{Topic 6\\Gratitude} &
    \makecell{Topic 7\\Support and Empathy} &
    \makecell{Topic 8\\Humor and Interaction}
  \end{tabular}

  \caption{Comment Themes (Topics 1–8) with Representative Thumbnails}
  \label{tab:comment-themes}
  \endgroup
\end{table*}

\subsection{Connections Through Comments}

Finally, we analyzed the comments sections of the 200 videos. The word cloud visualizations in Table 7 show the keywords in each topic. Figure 2 illustrates the distribution of comment topics across the three categories of NDE content—PCEs, S-NDEs and NS-NDEs. 

\begin{figure}
    \centering
    \includegraphics[width=1\linewidth]{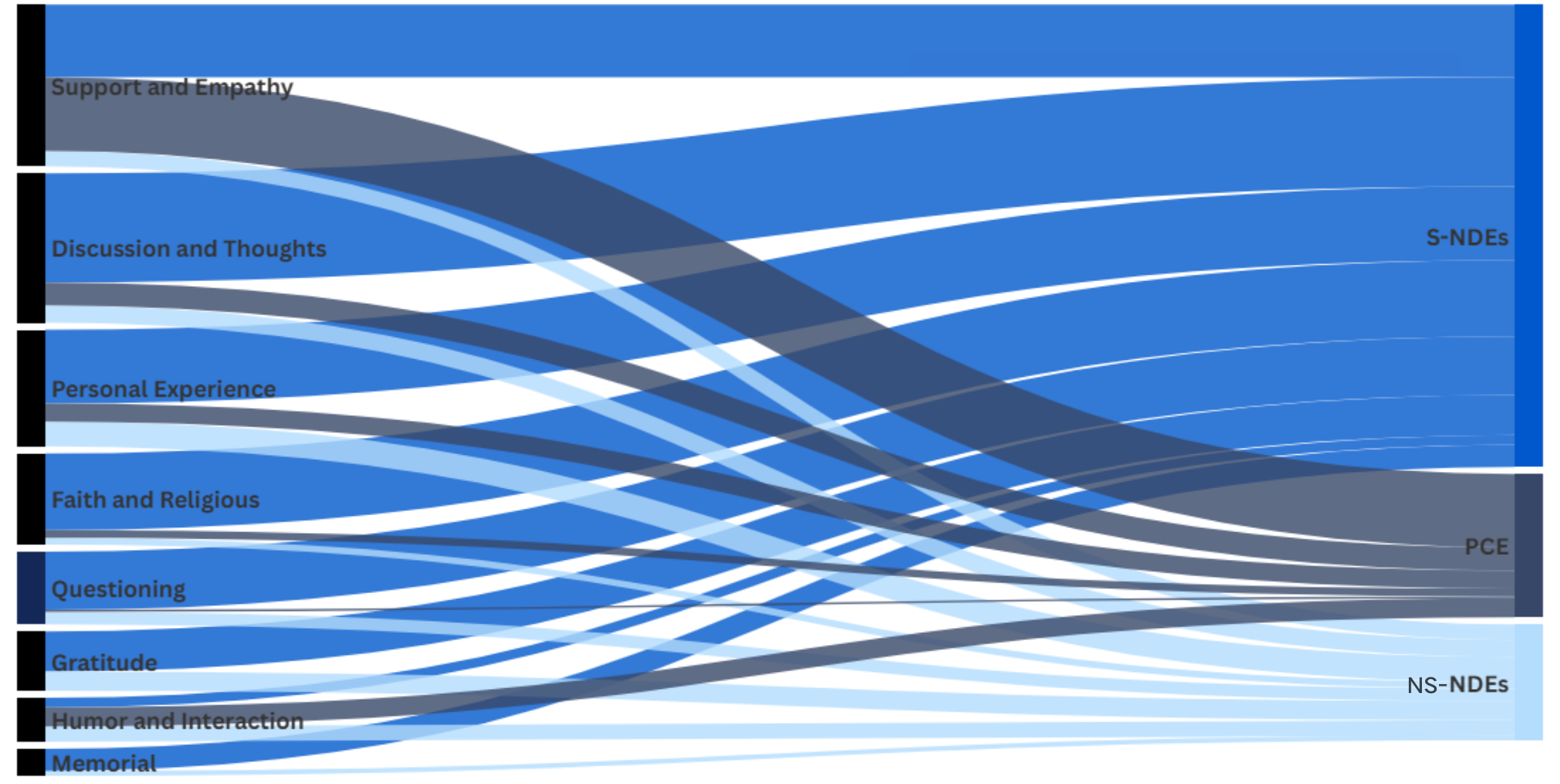}
    \caption{\textbf{The associations between comment topics and video themes. Topics are in descending order of comment count.}
}
    \label{fig:enter-label}
\end{figure}

\subsubsection{Support, Empathy, and Gratitude}

Across all videos except PCEs, Support and Empathy emerged as the most dominant comment types. Viewers frequently left emotionally affirming messages, expressing encouragement, solidarity, or compassion. These expressions often included emotive language and emoji cues(e.g., “\includegraphics[height=1em]{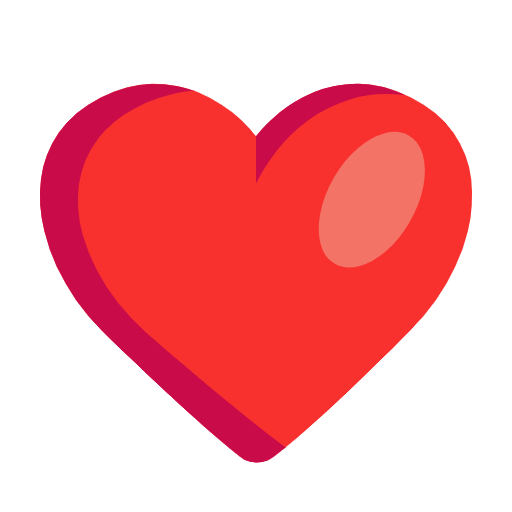},"  “\includegraphics[height=1em]{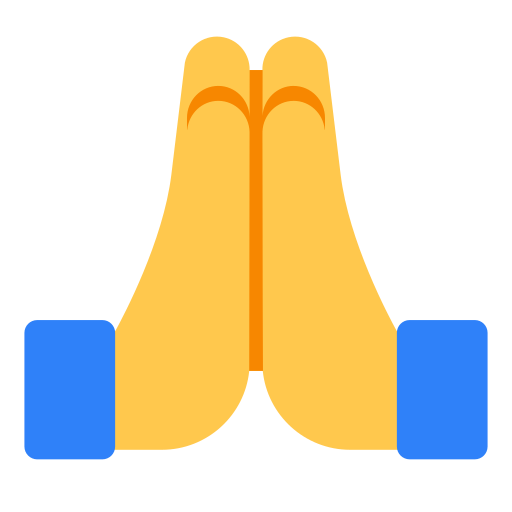}") Many comments also praised the creators, acknowledging their strength, courage, or effort, and explicitly offering comfort and companionship.

Again, in most videos except PCEs, viewers also expressed Gratitude. Comments often thanked the creator for posting the video or for sharing their experience, sometimes accompanied by supportive emojis such as “\includegraphics[height=1em]{emojis/red-heart_2764-fe0f.png}"and“\includegraphics[height=1em]{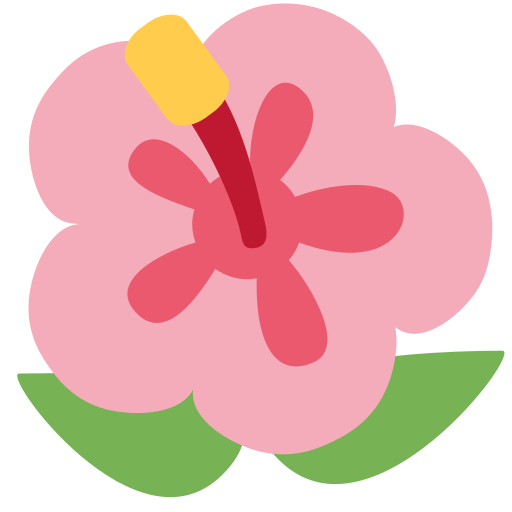}" In contrast, such appreciative comments were far less common in PCE videos, where the tone shifted more often toward humor. Some viewers described the content as frightening but admitted that the storyteller’s delivery made them laugh, reacting with crying-laughing emojis like “\includegraphics[height=1em]{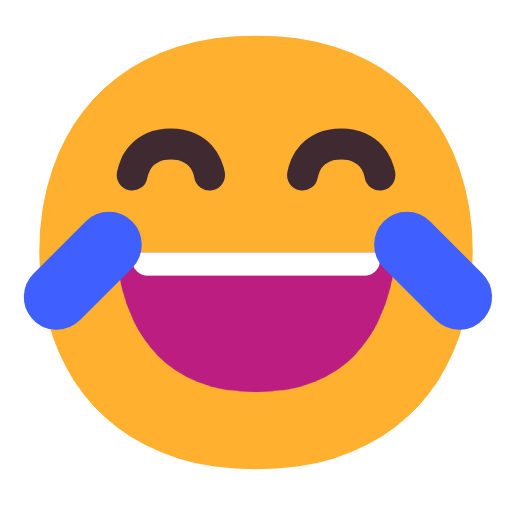}". This light-hearted emotional tone was significantly less common in S-NDEs, which tended to elicit more emotionally supportive engagement.

\subsubsection{ Faith-Based Expressions and Religious Affirmations}

Comments under S-NDE videos often contained a high frequency of Faith-Based Expressions and Religious Affirmations. These included theological reflections and personal testimonials of spiritual encounters. Many comments were direct responses to the video content, rearticulating references to divine figures such as God, Jesus, and angels, and expressing beliefs about spiritual protection or divine intervention. For example, one viewer explained to the creator that she was in the process of receiving a calling from God, while another offered a similar interpretation, saying that the creator had been protected by God and therefore needed to continue praying and placing trust in divine guidance. These expressions were often accompanied by prayer-related emojis such like \includegraphics[height=1em]{emojis/folded-hands_1f64f.png} Some viewers cited religious texts or phrases drawn from Christian teachings, such as statements affirming that Jesus represents truth, light, and the path forward. Others responded by sharing their own experiences of healing or divine presence during critical moments. For example, some described past recoveries they believed were aided by God, and others recalled moments when they saw a divine figure while facing life-threatening situations. These types of disclosures served to build rapport with the video creators and other commenters, reinforcing a shared belief system through testimony.

In addition to responding directly to the video content, commenters also employed a range of expressive formats to communicate their religious sentiments. Emotional shout-outs using emojis and religious language, such as \textit{“Amen," “Praise God," “Thank you Lord,"} and “\includegraphics[height=1em]{emojis/folded-hands_1f64f.png}" were a common format that served to create a communal atmosphere of gratitude and reverence. Another format was more directive: Viewers encouraged others to embrace faith through phrases like\textit{ “choose Jesus," “surrender,"} and \textit{“trust God."} These comments often took on an evangelistic tone and emphasized religious commitment as a solution or path forward. 

Although most of the religious content was Christian, some comments referenced Buddhism and other Eastern philosophies as meaningful ways to interpret NDEs. For example, one viewer remarked that studying Eastern religions can be beneficial, and another noted that Buddhist teachings can offer deeper insight. Overall, the comments on S-NDE videos suggest that public sharing of spiritual and religious NDEs can invite a range of faith-based reactions. Viewers may use the comment space not only to affirm the experiences shared in the videos but also to express their own religious identities, offer encouragement, or promote spiritual worldviews. 

Overall, comments in this category mainly involved affirming religious meanings, sharing personal spiritual experiences, and expressing faith-based interpretations of the videos.

\subsubsection{Discussion and Reflection}

Comments categorized as Discussion and Reflection appeared across all video types but were particularly prominent in S-NDE and NS-NDE videos. These comments typically involved viewers responding to the video content by interpreting it, providing supplementary information, or commenting reflectively on it. Rather than primarily expressing emotional responses, viewers in this category engaged cognitively, often seeking to deepen their understanding or exchange perspectives. 

One common format involved sharing personal insights as a way of interpreting the themes raised in the videos. For example, when a video mentioned consciousness, some viewers used the comments to express their own understanding of what it is and how it works.

Another response format consisted of direct responses to prompts by the video creators. In one video, the creator asked viewers to share their thoughts on death, which led to comments expressing fear or a reluctance to leave family members behind. In these cases, viewers were expressing personal views that related closely to a given topic.  

Some comments attempted to resolve questions raised in the video. For instance, when the creator wondered about a language they believed they had learned in heaven, one viewer tried to offer an explanation and shared information they knew about similar dialects. Other comments provided evaluations of the ideas discussed, expressing reflections on whether life has a purpose or lessons to take away.

Lastly, some commenters interpreted the video content from a neuroscientific perspective, offering explanations grounded in rational frameworks rather than metaphysical belief. Of all the comments under the category of Discussion and Reflection, these tended to involve the most cognitive participation and the least emotional resonance. Viewers shared their own viewpoints and engaged in rational inquiry related to NDE topics.

Overall, comments in this category focused on interpreting the themes in the videos, exchanging viewpoints, and exploring possible explanations, showing a primarily cognitive mode of engagement.

\subsubsection{ Questioning and Doubt}

A number of comments characterized by questioning, critical reflection, and inquiry appeared under S-NDE videos. These comments took on various forms, ranging from open-ended questions to direct challenges of the content presented in these videos. 

Some comments consisted of non-confrontational questions expressing personal confusion or curiosity. These were typically framed as open inquiries rather than skeptical challenges. For example, some commenters posed follow-up questions related to the video narrative, including whether the afterlife might resemble a dreamlike realm and whether the body and consciousness continue to exist in some form after death. Such comments sought clarification or elaboration and reflected viewers’ efforts to make sense of the content. Other comments exhibited critical responses to specific claims made in the videos. For instance, when a creator described the soul’s purpose as learning lessons, some viewers raised concerns about how this would apply to people who die by suicide or why suffering and hardship exist in everyday life. In another instance, when the creator suggested that moral binaries do not exist beyond the physical world, a viewer questioned why people should still strive to be good if concepts such as right and wrong are absent. These comments demonstrated viewers’ dialectical thinking and their willingness to critically examine the concepts raised in the videos.

A portion of responses centered on religious interpretation and doctrinal consistency. For example, when a speaker mentioned not encountering God during their NDE, one viewer countered by listing other cases in which people reported seeing God and used these examples to question the creator’s account. Another viewer went further by directly asking whether the creator was denying the existence of God, Jesus, or the Holy Spirit. These comments were critical of the video and introduced comparative accounts to challenge its implications. 

A smaller subset of commenters expressed skepticism toward the authenticity of NDEs as a phenomenon. These included doubts about whether the person had truly died and claims that no one can return from actual death, suggesting that the creator was conflating different concepts. Some viewers stated that they did not consider personal testimonies of NDEs to be reliable or credible. In other cases, commenters offered scientific explanations as counterarguments. For example, one viewer suggested that what the creator saw was merely a hallucination caused by chemicals in the brain. These contributors positioned NDEs as physiological events and questioned their reality. 

Overall, comments in this category focused on cognitive and existential engagement with NDE narratives. Many tried to make sense of the creator’s account by asking questions or examining its internal logic, and some approached the videos as a form of public storytelling that could be evaluated or challenged. Instead of affirming or empathizing with the content, the commenters used the platform to ask questions, seek clarification, or offer alternative ways of explaining ideas about death, consciousness, and spirituality.

\subsubsection{Personal Narratives}

 Lastly, Personal Narratives offered as commentary were frequent across all video types. Many viewers described having gone through similar experiences in the past, sharing their own near-death events or moments when they felt close to dying. Others shared stories they had heard from family members or close acquaintances, often recalling what relatives said during critical or end-of-life moments. In these comments, viewers shared either firsthand or secondhand accounts of NDEs, often aligning their own narratives with those presented in the original videos. 

Another common occurrence in NS-NDE and S-NDE videos that was absent in PCE videos was to express remembrance and grief for deceased loved ones. Some viewers used these videos as a source of emotional consolation, drawing comfort from the idea that their loved ones may have experienced peace in their final moments. For example, one viewer shared that a relative had passed away many years ago and that learning about the beauty of the near-death state brought them a sense of relief. Comments such as this suggest that engagement with NDE content may provide viewers with a framework for coping with personal loss, particularly in the context of bereavement. 

Overall, our findings indicate that NDErs employ multiple approaches to online self-disclosure, including storytelling, opinion expression, and stance articulation. In addition, most NDE narratives on TikTok contain religious and spiritual content, with comment sections reflecting support, religious affirmation, and cognitive engagement. This form of disclosure may influence how creators continue narrating their experiences and how viewers respond to them. Building upon and adapting the disclosure motivation framework discussed earlier (i.e., self-expression, identity clarification, and information sharing), we analyze in the following sections the modes of NDE disclosure, the associated methods of community interaction, and the opportunities for providing support to these online exchanges.  

\section{Discussion}\label{sec:Discussion}

\subsection{User Motivations, Narrative Strategies, and Community Interaction in NDE Disclosure}

\subsubsection{Personalized, Affective, and Identity-Driven Motivations for NDE Disclosure}

On TikTok, NDE-related self-disclosures tend to foreground personalized, affective, and identity-assertive motivations rather than scientific or entertainment-oriented intentions. Although TikTok is widely known as an entertainment-driven platform \cite{ivana_stamenkovic_motivation_nodate}, we found that NDE videos rarely adopt humorous or playful narrative styles, and most disclosures maintain a serious and restrained tone. Analytically, drawing on the established NDE dimensions, emotional and spiritual dimensions were most strongly associated with affective self-expression and identity clarification, while cognitive dimensions more often surfaced in information sharing and comment-based negotiation.

A primary motivation underlying these disclosures is information sharing. TikTok’s platform logic systematically pushes private experiences toward semi-knowledgeable forms of public content \cite{al-megren_dementia_2021}. Because the platform conducts relatively limited moderation, removal, or fact-checking of experiential videos \cite{sharevski_debunk-it-yourself_2025}, users frequently default to trusting short-form content and exhibit potentially inaccurate trust in influencers or accessible (non-)experts \cite{stephenson_sharenting_2024}. Consequently, users who present themselves as having interpretive authority over highly subjective NDEs often leverage information sharing as an opportunity to educate or guide others. Many NDErs describe themselves as motivated by a sense of self-efficacy, a belief that they can offer valuable insight to audiences, and emphasize detailed descriptions alongside reflections or lessons learned \cite{hsu_acceptance_2008,lin_why_2013,lou_contributing_2013}. Entertainment is rarely foregrounded in these narratives.

A second common motivation involves affective storytelling. First, TikTok’s platform mechanisms reduce psychological distance \cite{nguyen_tiktok_2023}, and the online disinhibition effect \cite{suler_online_2004} further enables experiencers to share emotions more openly. Many NDE videos integrate descriptions of death-related encounters with visible emotional expressions such as crying, trembling, or joy. Second, emotional expression functions as a coping strategy for grief and trauma, allowing experiencers to seek connection through affective articulation \cite{olson_distance_2000,barak_fostering_2008}. Such displays of vulnerability often elicit strong empathic responses from audiences \cite{andalibi_sensitive_2017}. Third, TikTok’s broader expressive norms, emphasizing “authenticity” and just-be-you styles of emotional openness \cite{milton_i_2023}, frame vulnerability as a marker of genuineness. Finally, affective expression reduces the perceived need for logical or scientific justification. Following the classical triad of ethos, logos, and pathos \cite{miceli_emotional_2006}, emotional force often surpasses logical reasoning in shaping audience reception \cite{jorgensen_chapter_1996}. Given that NDEs commonly involve supernatural or non-scientific content, experients frequently embed emotion to enhance narrative credibility when rational explanation is insufficient or contested.

The third motivation centers on identity construction. In online environments, individuals selectively construct and present facets of the self \cite{kim_examining_2011,huang_gender_2018}, often driven by self-expression \cite{hogan_presentation_2010} and the desire for short-term social recognition \cite{oudshoorn_how_2005}. Given the extraordinary nature of NDEs, narrators frequently draw on spiritual or religious elements to position themselves as individuals with unique and meaningful experiences. TikTok’s content ecosystem further reinforces this process through nichification, where creators become increasingly “typed” into a single narrative category \cite{schafer_i_2025}. Through identity flattening \cite{devito_how_2022}, creators are compressed into a “single-content persona,” producing narratives that are easily consumable within the platform’s recommendation logic \cite{lutz_were_2024}. As a result, experiencers may emphasize or amplify certain spiritual or religious dimensions to sustain their identity, often elaborating through additional videos or comment replies that reinforce the coherence of their self-presentation over time.

\subsubsection{Religious Features in NDE Narratives}

Religious and non-ordinary NDE narratives on TikTok illustrate how the spiritual dimensions of NDEs are not only narrated but actively negotiated through platform-mediated interaction, often inviting both attention and debate. Such narratives more readily elicit questions and doubt; however, these interactions are not simple rejections. Instead, they function as important processes through which users make sense of, interpret, and construct meaning around supernatural narratives.

In videos related to spiritual themes, NDE self-disclosers often narrate their personal experiences through religious and spiritual concepts such as the soul and consciousness. Prior offline NDE research has also shown that such testimonies frequently include religious or faith-oriented interpretations \cite{badham_religious_1997,greyson_implications_2010}, and in our dataset, 79\% of classical NDE disclosures mentioned content related to faith and religion. Many experiencers interpret their encounters as sacred events or spiritual missions and use them to reconnect with their religious backgrounds. Much of identity clarification centers on the idea of a religious mission. Many disclosers frame their NDEs as “mission-bestowing" stories to reconstruct their personal identities and life purposes \cite{greyson_near-death_2000}, thereby enhancing social comprehensibility and achieving self-legitimation \cite{yaden_varieties_2017}. Given the ineffable nature of religious experiences \cite{james_varieties_2003}, disclosers rely on affective storytelling techniques to compensate for the limits of rational articulation. They attempt to reinforce experiential authority through emotional and persuasive language \cite{carr_reason_2004}. Audience interactions generate a large number of religious comments, producing an echo chamber effect \cite{cinelli_echo_2021}, which further facilitates the formation of communities centered on shared faith \cite{byun_effect_2022,moller_online_2021}.

Compared with non-religious content, religious narratives in NDE more easily trigger broader discussions and cognitive conflict. First, because social media lowers social pressure and provides anonymity-based protection \cite{christopherson_positive_2007}, people are more willing to openly express doubt, critique, and dissent \cite{suler_online_2004}. Second, NDErs often reassess their religious beliefs after the experience \cite{greyson_near-death_2006}; some recount episodes that contradict their prior faith and raise questions about existing doctrines. Since religious understanding is closely tied to culturally embedded schemas \cite{dimaggio_culture_1997}, narratives that deviate from traditional teachings often generate opposing views. In such cases, some viewers respond by defending their own religious beliefs, while others challenge the supernatural elements of the videos through scientific or medical explanations. These varied experiences and cultural backgrounds make it less likely for viewers to converge on a single shared interpretation.

Another point worth noting is that, compared with PCEs, spiritual or non-ordinary NDEs and S-NDEs more often triggered questions and doubt in the comment sections. First, experiences involving extrasensory perception or spiritual elements can come with interpretive pressure and potential stigma \cite{mayer_extraordinary_2008}, and doubt is therefore a common reaction \cite{arrowood_existential_2022}. In our data, these comments often appeared as questions about the narrative’s internal logic and mainly sought further clarification of the concepts mentioned by the creators. Second, “doubt” is also understood as a religious or spiritual struggle and is considered part of religious experience \cite{exline_doubt_2025}. Faith is not simply the acceptance of religious or spiritual ideas, and uncertainty is often expressed and negotiated through interaction and discussion \cite{lamine_i_2014}. For this reason, most questions directed at spiritual NDEs in the comment sections were not simple rejection but attempts to address uncertainty when encountering supernatural narratives. Viewers’ questions can be seen as a way of exploring the content, and doubt and inquiry function as their means of understanding these narratives.

\subsubsection{Peer Support, Empathy, and NDE Community Building via Comments
}

We observed three primary comment cultures within NDE content: supportive empathy, content discussion, and experiential sharing. Across all NDE videos, the most common phenomenon was viewers sharing their own stories and experiences in the comment sections. In sensitive contexts, viewers openly disclose similar experiences in response to content creators, engaging in empathic mirroring \cite{osler_taking_2024}. This reciprocal disclosure model, which involves sharing vulnerabilities to build communities, has also been observed in other contexts such as trauma-related discussions \cite{sprecher_taking_2013,trepte_reciprocal_2013}. Through shared vulnerability \cite{lagerkvist_grand_2017}, commenters not only strengthen their sense of belonging and identity affirmation but also regulate trauma and emotional distress. In addition, comment sections often provide emotional support and esteem support, helping strengthen emotional connections and social affirmation \cite{roland_affective_2017}. Facilitated by supportive platform affordances, such as commenting and lightweight affective feedback (e.g., likes), users frequently offer encouragement and comfort to NDE storytellers.

Compared with S-NDEs and classical NDEs, PCEs include a larger proportion of encouraging and supportive comments. In online spaces, disclosures about health crises and illness tend to elicit a consistent pattern of supportive responses, such as encouragement, blessings, and a strong survivor-support culture \cite{yang_channel_2019,eijk_using_2013}. At the same time, PCEs may be perceived as more realistic and verifiable, and are therefore more likely to receive emotional support rather than doubt \cite{phillips_emotional_2025}. PCEs in NDE contexts often involve concrete medical events (such as cardiac arrest or miscarriage) or traffic accidents, making it easier for users to understand the associated risks and fear, which naturally triggers empathic responses. By contrast, S-NDEs involve elements such as the soul, consciousness, and divine encounters, which more readily evoke interpretive pressure \cite{jones_can_2015} and religious doubt \cite{schellenberg_wisdom_2007}. Thus, while supportive responses are prevalent, NDE comment spaces also display varying degrees of questioning, depending on the thematic content of the NDE videos.

However, unlike typical supportive comment spaces, NDE-related videos also create atmospheres of mourning and remembrance. Bereaved viewers use NDE content to access new perspectives on understanding and confronting death, thereby obtaining comfort and meaning and alleviating their grief. Online support similarly offers healing effects for those coping with loss \cite{baglione_modern_2018}. They treat NDE videos as emotional resources, using the comment space to express mourning and remembrance, thereby facilitating the social integration and healing of their bereavement experiences \cite{walter_new_2015}. Through these interactions, NDE comment sections also exhibit the potential for digital mourning \cite{giaxoglou_mediatization_2018}.

\subsection{Platform-Driven Influences on NDE Disclosure}

Beyond personal motivations for disclosure, NDE narratives may also be shaped by platform-oriented dynamics. To adapt to the content ecology of short-video platforms \cite{pera_shifting_2024, baumann_dynamics_2025}, NDE disclosures typically take the form of personalized and first-person storytelling. TikTok’s platform structure encourages users to “stage” their personal experiences and continuously perform and extend their online identities \cite{biggs_tiktok_2023}. Many creators continue producing additional segments after their initial NDE video. This pattern is not only motivated by narrative needs but also influenced by TikTok’s visibility logic \cite{simpson_rethinking_2023}, which pushes creators toward content nichification and rewards the sustained production of similar material \cite{lee_algorithmic_2022}. As a result, creators repeatedly publish related content \cite{losh_are_2023} , gradually shifting from “expressing a unique personal experience” to “maintaining the visibility of their account” \cite{devito_how_2022}. At the same time, because TikTok’s For You Page prioritizes content with clear symbolic cues that are easy to classify \cite{simpson_for_2021}, religious NDE videos appear more frequently, as they more readily form algorithmically recognizable identity categories. In addition, many NDErs update their narratives through Q\&A responses, often prompting viewers to comment at the end of their videos. These dialogic interaction structures become key drivers of the creators’ continued content production \cite{le_compte_its_2021}. This may relate to creators’ “folk theories” of the algorithm \cite{klug_trick_2021,wang_weaving_2023}—specifically, the belief that comments significantly increase the likelihood that a video will be promoted—encouraging them to extend their NDE storytelling through Q\&A-style videos.

The three primary comment cultures surrounding NDE content may likewise be influenced by the platform’s affordances. TikTok’s comment sections often function as sociotechnical spaces of mutual support, belonging, and emotional connection \cite{biggs_tiktok_2023}. Through “algorithmic community-building,” the platform directs users toward content aligned with their identity needs \cite{moolenijzer_they_2023}, gathers individuals with similar identities into communities \cite{devito_how_2022}, and continuously pushes material closely related to their emotional or experiential states \cite{de_los_santos_tiktok_2022}. Moreover, because TikTok emphasizes call-to-action dynamics and interaction-driven engagement \cite{pera_shifting_2024}, audiences are prompted to participate actively. Viewers often recognize “a part of themselves” through algorithmically surfaced content, forming brief yet genuine emotional connections \cite{lee_algorithmic_2022}. As a result, prayer-like responses and supportive behaviors are widespread in NDE comment sections, where users frequently reply to creators with disclosures of their own experiences.

\subsection{Design Implications}

\subsubsection{Testimony-Oriented Design}

 NDEs belong to extraordinary experiences or sacred testimonies, and most creators share their encounters through a first-person testimonial mode. However, TikTok’s entertainment-oriented mechanisms weaken the seriousness of these disclosures. The hashtags associated with these videos show substantial overlap, which may obscure the heterogeneity of the content and lead to potential misunderstandings regarding the nature of the videos \cite{zhang_pragmatic_2023, zeng_whatieatinaday_2025}. Unlike platforms such as Instagram, which allow users to control audience access through features such as Stories or private group functions \cite{chiu_last_2021}, TikTok lacks effective tools for contextual separation. Conflicts and personal attacks may appear in the comment sections due to divergent religious stances, producing a case of context collapse \cite{marwick_i_2011}, in which heterogeneous audiences are brought together by TikTok’s shared comment space and respond to near-death experience testimonies with incompatible interpretive expectations.

We suggest that platforms adopt more fine-grained zoning that supports respectful expression of sacred or extraordinary experiences while reducing entertainment-driven or conflict-prone interpretations caused by context collapse. First, Testimony Labeling can provide optional testimony frames for narratives involving sacred, religious, or end-of-life experiences. Without mechanisms that help viewers understand the nature of the content, emotional conflict in the comment sections may intensify \cite{mori_design_2012}. By offering a testimony frame, platforms can help differentiate these narratives from entertainment-oriented content and guide viewers’ expectations. Second, Contextual Credibility Support can allow creators to attach optional background prompts, such as “personal experience, not medical advice,” which would help viewers interpret the content more accurately. Third, Soft-Gated Exposure can provide lightweight filtering options for content involving death, bereavement, psychological crisis, or religious reevaluation. This approach is particularly relevant because TikTok’s novelty-driven and watch-time optimized structure tends to amplify religious or emotional content \cite{karizat_algorithmic_2021,boeker_empirical_2022}. Soft-gated mechanisms preserve user autonomy while reducing unwanted exposure among vulnerable audiences. We recognize that TikTok’s underlying architecture prioritizes novelty and watch-time optimization. Our suggestions do not reject engagement-based platform logics but propose refinements within these constraints; rather than claiming that TikTok systematically trivializes NDE testimonies, we note that its predominantly entertainment-oriented content framing may create interpretive ambiguity when sacred narratives circulate alongside recreational content.

\subsubsection{The Potential Supportive Role of NDE Content for Thanatosensitivity}

Based on our dataset, we find that NDE-related disclosures may provide meaningful support for thanatosensitive populations, including individuals who are sensitive to death, those experiencing bereavement, and those facing existential anxiety \cite{massimi_dying_2009}. Online narrative sharing can create alternative pathways for processing grief and can help users maintain connections with deceased loved ones \cite{robinson_online_2019,massimi_dealing_2011,massimi_death_2010}. In many religious NDE comment sections, bereaved viewers express grief, longing, and sorrow. They often describe emotional comfort when encountering NDE narratives that portray peaceful transitions or continued bonds with the deceased. Bereaved individuals often sustain continuing bonds through processes of making sense of death and constructing personal meaning \cite{klass_continuing_1996}, and NDE disclosures may therefore offer a channel that helps them reduce distress and cultivate a sense of meaning.

NDE narratives also have the potential to provide a gentle entry point for viewers who experience death anxiety or are interested in fundamental questions about human existence. For thanatosensitive populations, this type of meaning-focused content may help reduce anxiety related to uncertainty and promote a more constructive understanding of death \cite{albers_dying_2023,albers_lets_2024,bahng_reflexive_2020}. Digital content allows users to encounter death-related narratives in an indirect and mediated way, which can strengthen their psychological preparedness and reduce the fear associated with confronting death \cite{gulotta_engaging_2016}. Under religious or philosophical NDE videos, we observe frequent questions regarding the continuation of consciousness or the existence of an afterlife. These interactions indicate that NDE narratives function as a space for exploring life and death in a non-threatening environment. In this space, users can express uncertainty, raise questions, and seek explanations.

Building on these observations, these patterns suggest that NDE-related spaces could be further supported through thoughtful design considerations within HCI. As users turn to online platforms to process uncertainty and explore questions about death, systems can play a role in shaping how these interactions unfold. Rather than directing users toward specific interpretations, platforms could offer gentle structures that help them navigate diverse narratives, engage in reflective discussion, and encounter sensitive content at a comfortable pace. Such approaches may help sustain the supportive qualities already emerging in these communities while providing clearer pathways for users who seek understanding, connection, or reassurance in moments of vulnerability.

\subsection{Limitation and Future Work}

\subsubsection{Linguistic and Regional Bias.}

Our dataset excluded non-English videos, with most content originating from the United States and other Western regions. Consequently, the religious and cultural contexts were largely confined to Christian or broadly Western frameworks, leaving out non-Western systems such as Buddhism or Chinese and Indian folk religions. Since religious experience and self-expression are culturally embedded, this bias may have shaped both narratives and audience responses. Future work should adopt cross-cultural and multilingual sampling to capture a broader range of near-death spiritual narratives.

\subsubsection{Limits of Data Collection Scope.}
This study used keyword- and hashtag-based search rather than personalized FYP exposure. The resulting dataset represents a topic-centered collection of NDE-related content, not users’ routine, serendipitous encounters via algorithmic recommendation. Accordingly, our findings do not claim how or how often TikTok surfaces NDE content to general audiences, but reflect interaction and community formation within an NDE-focused content enclave.

\subsubsection{Absence of Creator Interviews.}

By focusing only on NDE disclosures and comment interactions, we lacked direct input from creators. This reliance on publicly available content introduces risks of researcher and confirmation bias \cite{onwuegbuzie_validity_2007}, and limits our insight into creators’ intentions, emotional responses, and reflections. Future studies could incorporate interviews or longitudinal follow-ups to better understand how online disclosure informs identity construction and spiritual integration.

\subsubsection{Platform-Specific Constraints.}

Certain platform-specific constraints on 
TikTok’s design emphasizes short-form, high-velocity content circulation, which may shape how NDE disclosures emerge, gain visibility, and accumulate interaction. Our study prioritized an in-depth qualitative analysis of narrative and interactional structures, and therefore focused on the experiential and expressive aspects of NDE videos rather than platform-level performance indicators. While follower counts, view rates, and other engagement metrics could further contextualize how narratives travel within TikTok’s algorithmic ecosystem \cite{boeker_empirical_2022}, such metrics were beyond the scope of our qualitative inquiry. Future work that integrates quantitative platform analytics could complement our findings by offering a fuller account of how algorithmic visibility intersects with disclosure practices.

\subsubsection{Limited Visibility of Negative Interactions.}

We observed relatively few instances of overt conflict, argumentation, or explicitly negative comments (e.g., bullying) in our dataset. One possible explanation is that some creators actively delete negative comments or block specific users \cite{thomas_its_2022,hodl_social_2024}, as many feel a sense of responsibility to manage discussions surrounding their content \cite{zeng_content_2022}. However, the precise motivations and practices underlying such moderation remain unclear. Future work should include creator interviews to better understand how moderation behaviors shape the visibility and tone of NDE-related interactions.

\section{Conclusion}\label{sec:Conclusion}

In this paper, we examined how TikTok users disclose near-death experiences and how audiences engage with these narratives through comments. We identified three main motivations behind NDE disclosures, namely self-expression, identity clarification, and information sharing, and observed that spiritual or religious content appeared in the majority of videos. Comments under these videos commonly expressed empathy, shared similar experiences, or discussed religious beliefs, with S-NDE videos drawing more emotionally intense and theologically focused interactions. We also found that some creators used serial videos to revisit and reinterpret their experiences, suggesting an ongoing process of identity work. Based on these findings, we propose design strategies to support respectful testimony sharing and design potential for thanatosensitivity. While this study focused on publicly available content in English, future work could incorporate more diverse cultural contexts and engage creators directly to deepen the understanding of NDE disclosure practices online.

\bibliographystyle{ACM-Reference-Format}
\bibliography{references}

@article{karizat_algorithmic_2021,
	title = {Algorithmic {Folk} {Theories} and {Identity}: {How} {TikTok} {Users} {Co}-{Produce} {Knowledge} of {Identity} and {Engage} in {Algorithmic} {Resistance}},
	volume = {5},
	shorttitle = {Algorithmic {Folk} {Theories} and {Identity}},
	url = {https://dl.acm.org/doi/10.1145/3476046},
	doi = {10.1145/3476046},
	number = {CSCW2},
	urldate = {2026-01-24},
	journal = {Proc. ACM Hum.-Comput. Interact.},
	author = {Karizat, Nadia and Delmonaco, Dan and Eslami, Motahhare and Andalibi, Nazanin},
	year = {2021},
	pages = {305:1--305:44},
}

@article{roberts_technology_2025,
	title = {Technology {Affordances}, {Social} {Media} {Engagement}, and {Social} {Media} {Addiction}: {An} {Investigation} of {TikTok}, {Instagram} {Reels}, and {YouTube} {Shorts}},
	volume = {28},
	issn = {2152-2715},
	shorttitle = {Technology {Affordances}, {Social} {Media} {Engagement}, and {Social} {Media} {Addiction}},
	url = {https://www.liebertpub.com/doi/abs/10.1089/cyber.2024.0338},
	doi = {10.1089/cyber.2024.0338},
	number = {5},
	urldate = {2026-01-24},
	journal = {Cyberpsychology, Behavior, and Social Networking},
	author = {Roberts, James A. and David, Meredith E.},
	month = may,
	year = {2025},
	note = {Publisher: Mary Ann Liebert, Inc., publishers},
	pages = {318--325},
}

@article{herman_for_2023,
	title = {For who page? {TikTok} creators’ algorithmic dependencies},
	shorttitle = {For who page?},
	url = {https://dl.designresearchsociety.org/iasdr/iasdr2023/fullpapers/224},
	journal = {IASDR Conference Series},
	author = {Herman, Laura},
	month = oct,
	year = {2023},
}

@article{kaye_jazztok_2023,
	title = {{JazzTok}: {Creativity}, {Community}, and {Improvisation} on {TikTok}},
	volume = {6},
	issn = {2578-4765},
	shorttitle = {{JazzTok}},
	url = {https://doi.org/10.5406/25784773.6.2.05},
	doi = {10.5406/25784773.6.2.05},
	number = {2},
	urldate = {2026-01-24},
	journal = {Jazz and Culture},
	author = {Kaye, D. Bondy Valdovinos},
	month = dec,
	year = {2023},
	pages = {92--116},
}

@article{gerbaudo_tiktok_2024,
	title = {{TikTok} and the algorithmic transformation of social media publics: {From} social networks to social interest clusters},
	issn = {1461-4448},
	shorttitle = {{TikTok} and the algorithmic transformation of social media publics},
	url = {https://doi.org/10.1177/14614448241304106},
	doi = {10.1177/14614448241304106},
	language = {EN},
	urldate = {2026-01-24},
	journal = {New Media \& Society},
	author = {Gerbaudo, Paolo},
	month = dec,
	year = {2024},
	note = {Publisher: SAGE Publications},
	pages = {14614448241304106},
}

@article{xu_research_2019,
	title = {Research on the {Causes} of the “{Tik} {Tok}” {App} {Becoming} {Popular} and the {Existing} {Problems}},
	issn = {21680787},
	url = {http://www.joams.com/index.php?m=content&c=index&a=show&catid=76&id=477},
	doi = {10.18178/joams.7.2.59-63},
	urldate = {2026-01-24},
	journal = {Journal of Advanced Management Science},
	author = {Xu, Li and Yan, Xiaohui and Zhang, Zhengwu},
	year = {2019},
	pages = {59--63},
}

@article{bhandari_whys_2022,
	title = {Why’s {Everyone} on {TikTok} {Now}? {The} {Algorithmized} {Self} and the {Future} of {Self}-{Making} on {Social} {Media}},
	volume = {8},
	issn = {2056-3051},
	shorttitle = {Why’s {Everyone} on {TikTok} {Now}?},
	url = {https://doi.org/10.1177/20563051221086241},
	doi = {10.1177/20563051221086241},
	language = {EN},
	number = {1},
	urldate = {2026-01-24},
	journal = {Social Media + Society},
	author = {Bhandari, Aparajita and Bimo, Sara},
	month = jan,
	year = {2022},
	note = {Publisher: SAGE Publications Ltd},
	pages = {20563051221086241},
}

@inproceedings{vitak_beyond_2016,
	address = {New York, NY, USA},
	series = {{CSCW} '16},
	title = {Beyond the {Belmont} {Principles}: {Ethical} {Challenges}, {Practices}, and {Beliefs} in the {Online} {Data} {Research} {Community}},
	isbn = {978-1-4503-3592-8},
	shorttitle = {Beyond the {Belmont} {Principles}},
	url = {https://dl.acm.org/doi/10.1145/2818048.2820078},
	doi = {10.1145/2818048.2820078},
	urldate = {2025-12-04},
	booktitle = {Proceedings of the 19th {ACM} {Conference} on {Computer}-{Supported} {Cooperative} {Work} \& {Social} {Computing}},
	publisher = {Association for Computing Machinery},
	author = {Vitak, Jessica and Shilton, Katie and Ashktorab, Zahra},
	year = {2016},
	pages = {941--953},
}

@article{gilbert_when_2023,
	title = {When research is the context: {Cross}-platform user expectations for social media data reuse},
	volume = {10},
	issn = {2053-9517},
	shorttitle = {When research is the context},
	url = {https://doi.org/10.1177/20539517231164108},
	doi = {10.1177/20539517231164108},
	language = {EN},
	number = {1},
	urldate = {2025-12-05},
	journal = {Big Data \& Society},
	author = {Gilbert, Sarah and Shilton, Katie and Vitak, Jessica},
	month = jan,
	year = {2023},
	note = {Publisher: SAGE Publications Ltd},
	pages = {20539517231164108},
}

@article{fiesler_participant_2018,
	title = {“{Participant}” {Perceptions} of {Twitter} {Research} {Ethics}},
	volume = {4},
	issn = {2056-3051},
	url = {https://doi.org/10.1177/2056305118763366},
	doi = {10.1177/2056305118763366},
	language = {EN},
	number = {1},
	urldate = {2025-12-05},
	journal = {Social Media + Society},
	author = {Fiesler, Casey and Proferes, Nicholas},
	month = jan,
	year = {2018},
	note = {Publisher: SAGE Publications Ltd},
	pages = {2056305118763366},
}

@article{andalibi_social_2018,
	title = {Social {Support}, {Reciprocity}, and {Anonymity} in {Responses} to {Sexual} {Abuse} {Disclosures} on {Social} {Media}},
	volume = {25},
	issn = {1073-0516},
	url = {https://dl.acm.org/doi/10.1145/3234942},
	doi = {10.1145/3234942},
	number = {5},
	urldate = {2025-12-04},
	journal = {ACM Trans. Comput.-Hum. Interact.},
	author = {Andalibi, Nazanin and Haimson, Oliver L. and Choudhury, Munmun De and Forte, Andrea},
	year = {2018},
	pages = {28:1--28:35},
}

@article{vitak_ethics_2017,
	title = {Ethics {Regulation} in {Social} {Computing} {Research}: {Examining} the {Role} of {Institutional} {Review} {Boards}},
	volume = {12},
	issn = {1556-2646},
	shorttitle = {Ethics {Regulation} in {Social} {Computing} {Research}},
	url = {https://doi.org/10.1177/1556264617725200},
	doi = {10.1177/1556264617725200},
	language = {EN},
	number = {5},
	urldate = {2025-12-04},
	journal = {Journal of Empirical Research on Human Research Ethics},
	author = {Vitak, Jessica and Proferes, Nicholas and Shilton, Katie and Ashktorab, Zahra},
	month = dec,
	year = {2017},
	note = {Publisher: SAGE Publications Inc},
	pages = {372--382},
}

@article{fernandez_informing_2003,
	title = {Informing {Study} {Participants} of {Research} {Results}: {An} {Ethical} {Imperative}},
	volume = {25},
	issn = {0193-7758},
	shorttitle = {Informing {Study} {Participants} of {Research} {Results}},
	url = {https://www.jstor.org/stable/3564300},
	doi = {10.2307/3564300},
	number = {3},
	urldate = {2025-12-04},
	journal = {IRB: Ethics \& Human Research},
	author = {Fernandez, Conrad V. and Kodish, Eric and Weijer, Charles},
	year = {2003},
	note = {Publisher: Hastings Center},
	pages = {12--19},
}

@article{roberts_ethical_2015,
	title = {Ethical {Issues} in {Conducting} {Qualitative} {Research} in {Online} {Communities}},
	volume = {12},
	issn = {1478-0887},
	url = {https://doi.org/10.1080/14780887.2015.1008909},
	doi = {10.1080/14780887.2015.1008909},
	number = {3},
	urldate = {2025-12-04},
	journal = {Qualitative Research in Psychology},
	author = {Roberts, Lynne D.},
	month = jul,
	year = {2015},
	note = {Publisher: Routledge
\_eprint: https://doi.org/10.1080/14780887.2015.1008909},
	pages = {314--325},
}

@article{klassen_this_2022,
	title = {“{This} {Isn}’t {Your} {Data}, {Friend}”: {Black} {Twitter} as a {Case} {Study} on {Research} {Ethics} for {Public} {Data}},
	volume = {8},
	issn = {2056-3051},
	shorttitle = {“{This} {Isn}’t {Your} {Data}, {Friend}”},
	url = {https://doi.org/10.1177/20563051221144317},
	doi = {10.1177/20563051221144317},
	language = {EN},
	number = {4},
	urldate = {2025-12-04},
	journal = {Social Media + Society},
	author = {Klassen, Shamika and Fiesler, Casey},
	month = oct,
	year = {2022},
	note = {Publisher: SAGE Publications Ltd},
	pages = {20563051221144317},
}

@article{sarikakis_social_2017,
	title = {Social {Media} {Users}’ {Legal} {Consciousness} {About} {Privacy}},
	volume = {3},
	issn = {2056-3051},
	url = {https://doi.org/10.1177/2056305117695325},
	doi = {10.1177/2056305117695325},
	language = {EN},
	number = {1},
	urldate = {2025-12-04},
	journal = {Social Media + Society},
	author = {Sarikakis, Katharine and Winter, Lisa},
	month = jan,
	year = {2017},
	note = {Publisher: SAGE Publications Ltd},
	pages = {2056305117695325},
}

@article{lupinacci_phenomenal_2024,
	title = {Phenomenal algorhythms: {The} sensorial orchestration of “real-time” in the social media manifold},
	volume = {26},
	issn = {1461-4448},
	shorttitle = {Phenomenal algorhythms},
	url = {https://doi.org/10.1177/14614448221109952},
	doi = {10.1177/14614448221109952},
	language = {EN},
	number = {7},
	urldate = {2025-12-04},
	journal = {New Media \& Society},
	author = {Lupinacci, Ludmila},
	month = jul,
	year = {2024},
	note = {Publisher: SAGE Publications},
	pages = {4078--4098},
}

@book{jones_can_2015,
	title = {Can {Science} {Explain} {Religion}?: {The} {Cognitive} {Science} {Debate}},
	isbn = {978-0-19-024939-7},
	shorttitle = {Can {Science} {Explain} {Religion}?},
	language = {en},
	publisher = {Oxford University Press},
	author = {Jones, James W.},
	month = sep,
	year = {2015},
	note = {Google-Books-ID: lgE7CgAAQBAJ},
}

@book{schellenberg_wisdom_2007,
	title = {The {Wisdom} to {Doubt}: {A} {Justification} of {Religious} {Skepticism}},
	isbn = {978-0-8014-4554-5},
	shorttitle = {The {Wisdom} to {Doubt}},
	language = {en},
	publisher = {Cornell University Press},
	author = {Schellenberg, J. L.},
	year = {2007},
}

@article{phillips_emotional_2025,
	title = {Emotional language reduces belief in false claims},
	volume = {20},
	issn = {1930-2975},
	url = {https://www.cambridge.org/core/journals/judgment-and-decision-making/article/emotional-language-reduces-belief-in-false-claims/3EF5128ED9C9CE3813B7C6BDBD64A48A},
	doi = {10.1017/jdm.2025.10019},
	language = {en},
	urldate = {2025-12-03},
	journal = {Judgment and Decision Making},
	author = {Phillips, Samantha C. and Wang, Sze Yuh Nina and Carley, Kathleen M. and Rand, David G. and Pennycook, Gordon},
	month = jan,
	year = {2025},
	pages = {e43},
}

@article{eijk_using_2013,
	title = {Using {Online} {Health} {Communities} to {Deliver} {Patient}-{Centered} {Care} to {People} {With} {Chronic} {Conditions}},
	volume = {15},
	url = {https://www.jmir.org/2013/6/e115},
	doi = {10.2196/jmir.2476},
	language = {EN},
	number = {6},
	urldate = {2025-12-03},
	journal = {Journal of Medical Internet Research},
	author = {Eijk, Martijn van der and Faber, Marjan J. and Aarts, Johanna WM and Kremer, Jan AM and Munneke, Marten and Bloem, Bastiaan R.},
	month = jun,
	year = {2013},
	note = {Company: Journal of Medical Internet Research
Distributor: Journal of Medical Internet Research
Institution: Journal of Medical Internet Research
Label: Journal of Medical Internet Research
Publisher: JMIR Publications Inc., Toronto, Canada},
	pages = {e2476},
}

@inproceedings{yang_channel_2019,
	address = {New York, NY, USA},
	series = {{CHI} '19},
	title = {The {Channel} {Matters}: {Self}-disclosure, {Reciprocity} and {Social} {Support} in {Online} {Cancer} {Support} {Groups}},
	isbn = {978-1-4503-5970-2},
	shorttitle = {The {Channel} {Matters}},
	url = {https://dl.acm.org/doi/10.1145/3290605.3300261},
	doi = {10.1145/3290605.3300261},
	urldate = {2025-12-03},
	booktitle = {Proceedings of the 2019 {CHI} {Conference} on {Human} {Factors} in {Computing} {Systems}},
	publisher = {Association for Computing Machinery},
	author = {Yang, Diyi and Yao, Zheng and Seering, Joseph and Kraut, Robert},
	year = {2019},
	pages = {1--15},
}

@article{arrowood_existential_2022,
	title = {The {Existential} {Quest}: {Doubt}, {Openness}, and the {Exploration} of {Religious} {Uncertainty}},
	volume = {32},
	issn = {1050-8619},
	shorttitle = {The {Existential} {Quest}},
	url = {https://doi.org/10.1080/10508619.2021.1902647},
	doi = {10.1080/10508619.2021.1902647},
	number = {2},
	urldate = {2025-12-03},
	journal = {The International Journal for the Psychology of Religion},
	author = {Arrowood, Robert B. and Vail III, Kenneth E. and Cox, Cathy R.},
	month = apr,
	year = {2022},
	note = {Publisher: Routledge
\_eprint: https://doi.org/10.1080/10508619.2021.1902647},
	pages = {89--126},
}

@incollection{lamine_i_2014,
	title = {“{I} {Doubt}. {Therefore}, {I} {Believe}”: {Facing} {Uncertainty} and {Belief} in the {Making}},
	shorttitle = {“{I} {Doubt}. {Therefore}, {I} {Believe}”},
	url = {https://brill.com/display/book/9789004277793/B9789004277793_006.xml},
	language = {zh},
	urldate = {2025-12-03},
	publisher = {Brill},
	author = {Lamine, Anne-Sophie},
	month = jan,
	year = {2014},
	doi = {10.1163/9789004277793_006},
	note = {Section: Religion in Times of Crisis},
}

@book{mayer_extraordinary_2008,
	title = {Extraordinary {Knowing}: {Science}, {Skepticism}, and the {Inexplicable} {Powers} of the {Human} {Mind}},
	isbn = {978-0-553-38223-5},
	shorttitle = {Extraordinary {Knowing}},
	language = {en},
	publisher = {Random House Publishing Group},
	author = {Mayer, Elizabeth Lloyd},
	month = feb,
	year = {2008},
	note = {Google-Books-ID: mxgghclkCg8C},
}

@incollection{exline_doubt_2025,
	edition = {2},
	title = {Doubt as a {Form} of {Spiritual} {Struggle}},
	booktitle = {The {Routledge} {Handbook} of the {Uncertain} {Self}},
	publisher = {Routledge},
	author = {Exline, Julie J. and Keller, Yehudis and Moffitt, Andrew C. and Pargament, Kenneth I.},
	year = {2025},
	note = {Num Pages: 17},
}

@misc{rsumovic_social_2016,
	title = {Social {Media} {Research}: {A} {Guide} to {Ethics}},
	shorttitle = {Social {Media} {Research}},
	url = {https://seenpm.org/social-media-research-guide-ethics/},
	language = {en-US},
	urldate = {2025-12-03},
	journal = {SEENPM},
	author = {Ršumović, Nevena},
	month = jul,
	year = {2016},
}

@article{fielding_sage_2016,
	title = {The {SAGE} {Handbook} of {Online} {Research} {Methods}},
	url = {https://www.torrossa.com/en/resources/an/5019467},
	language = {en},
	urldate = {2025-12-03},
	author = {Fielding, Nigel G. and Blank, Grant and Lee, Raymond M.},
	year = {2016},
	note = {Publisher: SAGE Publications Ltd},
	pages = {1--684},
}

@article{lenhart_i_2025,
	title = {I {Feel} {Like} {All} of {This} {Is} {Already} {Happening} {Anyways}?: {Context} {Import} and {Young} {Adults}' {Perspectives} on {Researcher} {Access} to {TikTok} {Data}},
	volume = {9},
	shorttitle = {I {Feel} {Like} {All} of {This} {Is} {Already} {Happening} {Anyways}?},
	url = {https://dl.acm.org/doi/10.1145/3757535},
	doi = {10.1145/3757535},
	number = {7},
	urldate = {2025-12-03},
	journal = {Proc. ACM Hum.-Comput. Interact.},
	author = {Lenhart, Anna and Shilton, Katie},
	year = {2025},
	pages = {CSCW354:1--CSCW354:44},
}

@article{hodl_social_2024,
	title = {Social {Media} {Feedback} {Dynamics}: {The} {Influence} of {Community} and {Technology} on {Content} {Creators}},
	shorttitle = {Social {Media} {Feedback} {Dynamics}},
	url = {https://aisel.aisnet.org/icis2024/digtech_fow/digtech_fow/1},
	journal = {ICIS 2024 Proceedings},
	author = {Hödl, Tatjana},
	month = dec,
	year = {2024},
}

@inproceedings{thomas_its_2022,
	address = {New York, NY, USA},
	series = {{CHI} '22},
	title = {“{It}’s common and a part of being a content creator”: {Understanding} {How} {Creators} {Experience} and {Cope} with {Hate} and {Harassment} {Online}},
	isbn = {978-1-4503-9157-3},
	shorttitle = {“{It}’s common and a part of being a content creator”},
	url = {https://dl.acm.org/doi/10.1145/3491102.3501879},
	doi = {10.1145/3491102.3501879},
	urldate = {2025-12-01},
	booktitle = {Proceedings of the 2022 {CHI} {Conference} on {Human} {Factors} in {Computing} {Systems}},
	publisher = {Association for Computing Machinery},
	author = {Thomas, Kurt and Kelley, Patrick Gage and Consolvo, Sunny and Samermit, Patrawat and Bursztein, Elie},
	year = {2022},
	pages = {1--15},
}

@article{zeng_content_2022,
	title = {From content moderation to visibility moderation: {A} case study of platform governance on {TikTok}},
	volume = {14},
	copyright = {© 2022 The Authors. Policy \& Internet published by Wiley Periodicals LLC on behalf of Policy Studies Organization.},
	issn = {1944-2866},
	shorttitle = {From content moderation to visibility moderation},
	url = {https://onlinelibrary.wiley.com/doi/abs/10.1002/poi3.287},
	doi = {10.1002/poi3.287},
	language = {en},
	number = {1},
	urldate = {2025-12-02},
	journal = {Policy \& Internet},
	author = {Zeng, Jing and Kaye, D. Bondy Valdovinos},
	year = {2022},
	note = {\_eprint: https://onlinelibrary.wiley.com/doi/pdf/10.1002/poi3.287},
	pages = {79--95},
}

@book{klass_continuing_1996,
	address = {Philadelphia, PA, US},
	series = {Continuing bonds:  {New} understandings of grief},
	title = {Continuing bonds:  {New} understandings of grief},
	isbn = {978-1-56032-336-5 978-1-56032-339-6},
	shorttitle = {Continuing bonds},
	publisher = {Taylor \& Francis},
	editor = {Klass, Dennis and Silverman, Phyllis R. and Nickman, Steven L.},
	year = {1996},
	note = {Pages: xxi, 361},
}

@inproceedings{massimi_death_2010,
	address = {New York, NY, USA},
	series = {{CHI} '10},
	title = {A death in the family: opportunities for designing technologies for the bereaved},
	isbn = {978-1-60558-929-9},
	shorttitle = {A death in the family},
	url = {https://dl.acm.org/doi/10.1145/1753326.1753600},
	doi = {10.1145/1753326.1753600},
	urldate = {2025-12-01},
	booktitle = {Proceedings of the {SIGCHI} {Conference} on {Human} {Factors} in {Computing} {Systems}},
	publisher = {Association for Computing Machinery},
	author = {Massimi, Michael and Baecker, Ronald M.},
	year = {2010},
	pages = {1821--1830},
}

@inproceedings{gulotta_engaging_2016,
	address = {New York, NY, USA},
	series = {{DIS} '16},
	title = {Engaging with {Death} {Online}: {An} {Analysis} of {Systems} that {Support} {Legacy}-{Making}, {Bereavement}, and {Remembrance}},
	isbn = {978-1-4503-4031-1},
	shorttitle = {Engaging with {Death} {Online}},
	url = {https://dl.acm.org/doi/10.1145/2901790.2901802},
	doi = {10.1145/2901790.2901802},
	urldate = {2025-12-01},
	booktitle = {Proceedings of the 2016 {ACM} {Conference} on {Designing} {Interactive} {Systems}},
	publisher = {Association for Computing Machinery},
	author = {Gulotta, Rebecca and Gerritsen, David B. and Kelliher, Aisling and Forlizzi, Jodi},
	year = {2016},
	pages = {736--748},
}

@inproceedings{albers_lets_2024,
	address = {New York, NY, USA},
	series = {{CHI} '24},
	title = {Let’s {Talk} {About} {Death}: {Existential} {Conversations} with {Chatbots}},
	isbn = {979-8-4007-0330-0},
	shorttitle = {Let’s {Talk} {About} {Death}},
	url = {https://dl.acm.org/doi/10.1145/3613904.3642421},
	doi = {10.1145/3613904.3642421},
	urldate = {2025-12-01},
	booktitle = {Proceedings of the 2024 {CHI} {Conference} on {Human} {Factors} in {Computing} {Systems}},
	publisher = {Association for Computing Machinery},
	author = {Albers, Ruben and Hassenzahl, Marc},
	year = {2024},
	pages = {1--14},
}

@inproceedings{albers_dying_2023,
	address = {New York, NY, USA},
	series = {{CHI} '23},
	title = {Dying, {Death}, and the {Afterlife} in {Human}-{Computer} {Interaction}. {A} {Scoping} {Review}.},
	isbn = {978-1-4503-9421-5},
	url = {https://dl.acm.org/doi/10.1145/3544548.3581199},
	doi = {10.1145/3544548.3581199},
	urldate = {2025-12-01},
	booktitle = {Proceedings of the 2023 {CHI} {Conference} on {Human} {Factors} in {Computing} {Systems}},
	publisher = {Association for Computing Machinery},
	author = {Albers, Ruben and Sadeghian, Shadan and Laschke, Matthias and Hassenzahl, Marc},
	year = {2023},
	pages = {1--16},
}

@inproceedings{mori_design_2012,
	address = {New York, NY, USA},
	series = {{OzCHI} '12},
	title = {Design considerations for after death: comparing the affordances of three online platforms},
	isbn = {978-1-4503-1438-1},
	shorttitle = {Design considerations for after death},
	url = {https://dl.acm.org/doi/10.1145/2414536.2414599},
	doi = {10.1145/2414536.2414599},
	urldate = {2025-12-01},
	booktitle = {Proceedings of the 24th {Australian} {Computer}-{Human} {Interaction} {Conference}},
	publisher = {Association for Computing Machinery},
	author = {Mori, Joji and Gibbs, Martin and Arnold, Michael and Nansen, Bjorn and Kohn, Tamara},
	year = {2012},
	pages = {395--404},
}

@inproceedings{massimi_dealing_2011,
	address = {New York, NY, USA},
	series = {{CHI} '11},
	title = {Dealing with death in design: developing systems for the bereaved},
	isbn = {978-1-4503-0228-9},
	shorttitle = {Dealing with death in design},
	url = {https://dl.acm.org/doi/10.1145/1978942.1979092},
	doi = {10.1145/1978942.1979092},
	urldate = {2025-12-01},
	booktitle = {Proceedings of the {SIGCHI} {Conference} on {Human} {Factors} in {Computing} {Systems}},
	publisher = {Association for Computing Machinery},
	author = {Massimi, Michael and Baecker, Ronald M.},
	year = {2011},
	pages = {1001--1010},
}

@article{robinson_online_2019,
	title = {Do online support groups for grief benefit the bereaved? {Systematic} review of the quantitative and qualitative literature},
	volume = {100},
	issn = {0747-5632},
	shorttitle = {Do online support groups for grief benefit the bereaved?},
	url = {https://www.sciencedirect.com/science/article/pii/S0747563219302377},
	doi = {10.1016/j.chb.2019.06.011},
	urldate = {2025-12-01},
	journal = {Computers in Human Behavior},
	author = {Robinson, Ceit and Pond, Dr Rachael},
	month = nov,
	year = {2019},
	pages = {48--59},
}

@article{marwick_i_2011,
	title = {I tweet honestly, {I} tweet passionately: {Twitter} users, context collapse, and the imagined audience},
	volume = {13},
	issn = {1461-4448},
	shorttitle = {I tweet honestly, {I} tweet passionately},
	url = {https://doi.org/10.1177/1461444810365313},
	doi = {10.1177/1461444810365313},
	language = {EN},
	number = {1},
	urldate = {2025-12-01},
	journal = {New Media \& Society},
	author = {Marwick, Alice E. and boyd, danah},
	month = feb,
	year = {2011},
	note = {Publisher: SAGE Publications},
	pages = {114--133},
}

@article{christopherson_positive_2007,
	series = {Including the {Special} {Issue}: {Education} and {Pedagogy} with {Learning} {Objects} and {Learning} {Designs}},
	title = {The positive and negative implications of anonymity in {Internet} social interactions: “{On} the {Internet}, {Nobody} {Knows} {You}’re a {Dog}”},
	volume = {23},
	issn = {0747-5632},
	shorttitle = {The positive and negative implications of anonymity in {Internet} social interactions},
	url = {https://www.sciencedirect.com/science/article/pii/S0747563206001221},
	doi = {10.1016/j.chb.2006.09.001},
	number = {6},
	urldate = {2025-11-30},
	journal = {Computers in Human Behavior},
	author = {Christopherson, Kimberly M.},
	month = nov,
	year = {2007},
	pages = {3038--3056},
}

@inproceedings{milton_i_2023,
	address = {New York, NY, USA},
	series = {{CHI} '23},
	title = {“{I} {See} {Me} {Here}”: {Mental} {Health} {Content}, {Community}, and {Algorithmic} {Curation} on {TikTok}},
	isbn = {978-1-4503-9421-5},
	shorttitle = {“{I} {See} {Me} {Here}”},
	url = {https://dl.acm.org/doi/10.1145/3544548.3581489},
	doi = {10.1145/3544548.3581489},
	urldate = {2025-11-30},
	booktitle = {Proceedings of the 2023 {CHI} {Conference} on {Human} {Factors} in {Computing} {Systems}},
	publisher = {Association for Computing Machinery},
	author = {Milton, Ashlee and Ajmani, Leah and DeVito, Michael Ann and Chancellor, Stevie},
	year = {2023},
	pages = {1--17},
}

@article{ivana_stamenkovic_motivation_nodate,
	title = {The {Motivation} for {Using} the {Social} {Media} {Platform} {TikTok} from the {Perspective} of the {Uses} and {Gratifications} {Theory} {\textbar} {Applied} {Media} {Studies} {Journal}},
	url = {https://msae.rs/index.php/home/article/view/84},
	urldate = {2025-11-30},
	author = {{Ivana Stamenković} and {Marta Mitrović}},
}

@article{schafer_i_2025,
	title = {'{I} {Blow} {Up}': {Understanding} {TikTok} {Users}' {Reactions} to {Sudden} {Social} {Media} {Attention}},
	volume = {9},
	shorttitle = {'{I} {Blow} {Up}'},
	url = {https://dl.acm.org/doi/10.1145/3711001},
	doi = {10.1145/3711001},
	number = {2},
	urldate = {2025-11-30},
	journal = {Proc. ACM Hum.-Comput. Interact.},
	author = {Schafer, Joseph S. and Denton, Annie and Seelhoff, Chloe and Vo, Jordyn and Garcia, Lance and Madan, Isha and Mudbhary, Alisha and Tang, Ruijingya and Starbird, Kate},
	year = {2025},
	pages = {CSCW103:1--CSCW103:31},
}

@article{devito_how_2022,
	title = {How {Transfeminine} {TikTok} {Creators} {Navigate} the {Algorithmic} {Trap} of {Visibility} {Via} {Folk} {Theorization}},
	volume = {6},
	url = {https://dl.acm.org/doi/10.1145/3555105},
	doi = {10.1145/3555105},
	number = {CSCW2},
	urldate = {2025-11-30},
	journal = {Proc. ACM Hum.-Comput. Interact.},
	author = {DeVito, Michael Ann},
	year = {2022},
	pages = {380:1--380:31},
}

@inproceedings{moolenijzer_they_2023,
	address = {New York, NY, USA},
	series = {{MobileHCI} '23 {Companion}},
	title = {“{They} know that it works because we are looking for ourselves” – {LGBTQ}+ {TikTok} {Users}' {Perceptions} and {Experiences} of {Queerbaiting}},
	isbn = {978-1-4503-9924-1},
	url = {https://dl.acm.org/doi/10.1145/3565066.3608705},
	doi = {10.1145/3565066.3608705},
	urldate = {2025-11-30},
	booktitle = {Proceedings of the 25th {International} {Conference} on {Mobile} {Human}-{Computer} {Interaction}},
	publisher = {Association for Computing Machinery},
	author = {Moolenijzer, Nicolaas B and Dew, Kristin},
	year = {2023},
	pages = {1--6},
}

@inproceedings{stephenson_sharenting_2024,
	address = {New York, NY, USA},
	series = {{CHI} '24},
	title = {Sharenting on {TikTok}: {Exploring} {Parental} {Sharing} {Behaviors} and the {Discourse} {Around} {Children}’s {Online} {Privacy}},
	isbn = {979-8-4007-0330-0},
	shorttitle = {Sharenting on {TikTok}},
	url = {https://dl.acm.org/doi/10.1145/3613904.3642447},
	doi = {10.1145/3613904.3642447},
	urldate = {2025-11-30},
	booktitle = {Proceedings of the 2024 {CHI} {Conference} on {Human} {Factors} in {Computing} {Systems}},
	publisher = {Association for Computing Machinery},
	author = {Stephenson, Sophie and Page, Christopher Nathaniel and Wei, Miranda and Kapadia, Apu and Roesner, Franziska},
	year = {2024},
	pages = {1--17},
}

@article{lee_algorithmic_2022,
	title = {The {Algorithmic} {Crystal}: {Conceptualizing} the {Self} through {Algorithmic} {Personalization} on {TikTok}},
	volume = {6},
	shorttitle = {The {Algorithmic} {Crystal}},
	url = {https://dl.acm.org/doi/10.1145/3555601},
	doi = {10.1145/3555601},
	number = {CSCW2},
	urldate = {2025-11-30},
	journal = {Proc. ACM Hum.-Comput. Interact.},
	author = {Lee, Angela Y. and Mieczkowski, Hannah and Ellison, Nicole B. and Hancock, Jeffrey T.},
	year = {2022},
	pages = {543:1--543:22},
}

@inproceedings{le_compte_its_2021,
	address = {New York, NY, USA},
	series = {{CSCW} '21 {Companion}},
	title = {“{It}’s {Viral}!” - {A} {Study} of the {Behaviors}, {Practices}, and {Motivations} of {TikTok} {Users} and {Social} {Activism}},
	isbn = {978-1-4503-8479-7},
	url = {https://dl.acm.org/doi/10.1145/3462204.3481741},
	doi = {10.1145/3462204.3481741},
	urldate = {2025-11-30},
	booktitle = {Companion {Publication} of the 2021 {Conference} on {Computer} {Supported} {Cooperative} {Work} and {Social} {Computing}},
	publisher = {Association for Computing Machinery},
	author = {Le Compte, Daniel and Klug, Daniel},
	year = {2021},
	pages = {108--111},
}

@inproceedings{nguyen_tiktok_2023,
	address = {New York, NY, USA},
	series = {{LAK2023}},
	title = {{TikTok} as {Learning} {Analytics} {Data}: {Framing} {Climate} {Change} and {Data} {Practices}},
	isbn = {978-1-4503-9865-7},
	shorttitle = {{TikTok} as {Learning} {Analytics} {Data}},
	url = {https://dl.acm.org/doi/10.1145/3576050.3576055},
	doi = {10.1145/3576050.3576055},
	urldate = {2025-11-30},
	booktitle = {{LAK23}: 13th {International} {Learning} {Analytics} and {Knowledge} {Conference}},
	publisher = {Association for Computing Machinery},
	author = {Nguyen, Ha},
	year = {2023},
	pages = {33--43},
}

@inproceedings{boeker_empirical_2022,
	address = {New York, NY, USA},
	series = {{WWW} '22},
	title = {An {Empirical} {Investigation} of {Personalization} {Factors} on {TikTok}},
	isbn = {978-1-4503-9096-5},
	url = {https://dl.acm.org/doi/10.1145/3485447.3512102},
	doi = {10.1145/3485447.3512102},
	urldate = {2025-11-30},
	booktitle = {Proceedings of the {ACM} {Web} {Conference} 2022},
	publisher = {Association for Computing Machinery},
	author = {Boeker, Maximilian and Urman, Aleksandra},
	year = {2022},
	pages = {2298--2309},
}

@inproceedings{wang_weaving_2023,
	address = {New York, NY, USA},
	series = {{CSCW} '23 {Companion}},
	title = {Weaving {Autistic} {Voices} on {TikTok}: {Utilizing} {Co}-{Hashtag} {Networks} for {Netnography}},
	isbn = {979-8-4007-0129-0},
	shorttitle = {Weaving {Autistic} {Voices} on {TikTok}},
	url = {https://dl.acm.org/doi/10.1145/3584931.3606995},
	doi = {10.1145/3584931.3606995},
	urldate = {2025-11-30},
	booktitle = {Companion {Publication} of the 2023 {Conference} on {Computer} {Supported} {Cooperative} {Work} and {Social} {Computing}},
	publisher = {Association for Computing Machinery},
	author = {Wang, Yihe and Ringland, Kathryn E.},
	year = {2023},
	pages = {254--258},
}

@inproceedings{klug_trick_2021,
	address = {New York, NY, USA},
	series = {{WebSci} '21},
	title = {Trick and {Please}. {A} {Mixed}-{Method} {Study} {On} {User} {Assumptions} {About} the {TikTok} {Algorithm}},
	isbn = {978-1-4503-8330-1},
	url = {https://dl.acm.org/doi/10.1145/3447535.3462512},
	doi = {10.1145/3447535.3462512},
	urldate = {2025-11-30},
	booktitle = {Proceedings of the 13th {ACM} {Web} {Science} {Conference} 2021},
	publisher = {Association for Computing Machinery},
	author = {Klug, Daniel and Qin, Yiluo and Evans, Morgan and Kaufman, Geoff},
	year = {2021},
	pages = {84--92},
}

@inproceedings{de_los_santos_tiktok_2022,
	address = {New York, NY, USA},
	series = {{MUM} '21},
	title = {The {TikTok} {Tradeoff}: {Compelling} {Algorithmic} {Content} at the {Expense} of {Personal} {Privacy}},
	isbn = {978-1-4503-8643-2},
	shorttitle = {The {TikTok} {Tradeoff}},
	url = {https://dl.acm.org/doi/10.1145/3490632.3497864},
	doi = {10.1145/3490632.3497864},
	urldate = {2025-11-30},
	booktitle = {Proceedings of the 20th {International} {Conference} on {Mobile} and {Ubiquitous} {Multimedia}},
	publisher = {Association for Computing Machinery},
	author = {De Los Santos, Maya and Klug, Daniel},
	year = {2022},
	pages = {226--229},
}

@inproceedings{biggs_tiktok_2023,
	address = {New York, NY, USA},
	series = {{DIS} '23},
	title = {{TikTok} as a {Stage}: {Performing} {Rural} \#farmqueer {Utopias} on {TikTok}},
	isbn = {978-1-4503-9893-0},
	shorttitle = {{TikTok} as a {Stage}},
	url = {https://dl.acm.org/doi/10.1145/3563657.3596038},
	doi = {10.1145/3563657.3596038},
	urldate = {2025-11-30},
	booktitle = {Proceedings of the 2023 {ACM} {Designing} {Interactive} {Systems} {Conference}},
	publisher = {Association for Computing Machinery},
	author = {Biggs, Heidi and Marcotte, Alexa and Bardzell, Shaowen},
	year = {2023},
	pages = {946--956},
}

@inproceedings{losh_are_2023,
	address = {New York, NY, USA},
	series = {{HT} '23},
	title = {Are {You} the {Main} {Character}? {Visibility} {Labor} and {Attributional} {Practices} on {TikTok}},
	isbn = {979-8-4007-0232-7},
	shorttitle = {Are {You} the {Main} {Character}?},
	url = {https://dl.acm.org/doi/10.1145/3603163.3609049},
	doi = {10.1145/3603163.3609049},
	urldate = {2025-11-30},
	booktitle = {Proceedings of the 34th {ACM} {Conference} on {Hypertext} and {Social} {Media}},
	publisher = {Association for Computing Machinery},
	author = {Losh, Elizabeth},
	year = {2023},
	pages = {1--5},
}

@inproceedings{pera_shifting_2024,
	address = {New York, NY, USA},
	series = {{WEBSCI} '24},
	title = {Shifting {Climates}: {Climate} {Change} {Communication} from {YouTube} to {TikTok}},
	isbn = {979-8-4007-0334-8},
	shorttitle = {Shifting {Climates}},
	url = {https://dl.acm.org/doi/10.1145/3614419.3644024},
	doi = {10.1145/3614419.3644024},
	urldate = {2025-11-30},
	booktitle = {Proceedings of the 16th {ACM} {Web} {Science} {Conference}},
	publisher = {Association for Computing Machinery},
	author = {Pera, Arianna and Aiello, Luca Maria},
	year = {2024},
	pages = {376--381},
}

@article{simpson_for_2021,
	title = {For {You}, or {For}"{You}"? {Everyday} {LGBTQ}+ {Encounters} with {TikTok}},
	volume = {4},
	shorttitle = {For {You}, or {For}"{You}"?},
	url = {https://dl.acm.org/doi/10.1145/3432951},
	doi = {10.1145/3432951},
	number = {CSCW3},
	urldate = {2025-11-30},
	journal = {Proc. ACM Hum.-Comput. Interact.},
	author = {Simpson, Ellen and Semaan, Bryan},
	year = {2021},
	pages = {252:1--252:34},
}

@inproceedings{al-megren_dementia_2021,
	address = {New York, NY, USA},
	series = {{MobileHCI} '21},
	title = {”{The} {Dementia} {Diva} {Strikes} {Again}!”: {A} {Thematic} {Analysis} of {How} {Informal} {Carers} of {Persons} with {Dementia} {Use} {TikTok}},
	isbn = {978-1-4503-8329-5},
	shorttitle = {”{The} {Dementia} {Diva} {Strikes} {Again}!”},
	url = {https://dl.acm.org/doi/10.1145/3447527.3474857},
	doi = {10.1145/3447527.3474857},
	urldate = {2025-11-30},
	booktitle = {Adjunct {Publication} of the 23rd {International} {Conference} on {Mobile} {Human}-{Computer} {Interaction}},
	publisher = {Association for Computing Machinery},
	author = {Al-Megren, Shiroq and Majrashi, Khalid and Allwihan, Ragad Mohammad},
	year = {2021},
	pages = {1--6},
}

@inproceedings{simpson_rethinking_2023,
	address = {New York, NY, USA},
	series = {{CHI} '23},
	title = {Rethinking {Creative} {Labor}: {A} {Sociotechnical} {Examination} of {Creativity} \& {Creative} {Work} on {TikTok}},
	isbn = {978-1-4503-9421-5},
	shorttitle = {Rethinking {Creative} {Labor}},
	url = {https://dl.acm.org/doi/10.1145/3544548.3580649},
	doi = {10.1145/3544548.3580649},
	urldate = {2025-11-30},
	booktitle = {Proceedings of the 2023 {CHI} {Conference} on {Human} {Factors} in {Computing} {Systems}},
	publisher = {Association for Computing Machinery},
	author = {Simpson, Ellen and Semaan, Bryan},
	year = {2023},
	pages = {1--16},
}

@article{lutz_were_2024,
	title = {"{We}'re not all construction workers": {Algorithmic} {Compression} of {Latinidad} on {TikTok}},
	volume = {8},
	shorttitle = {"{We}'re not all construction workers"},
	url = {https://dl.acm.org/doi/10.1145/3687019},
	doi = {10.1145/3687019},
	number = {CSCW2},
	urldate = {2025-11-30},
	journal = {Proc. ACM Hum.-Comput. Interact.},
	author = {Lutz, Nina and Aragon, Cecilia},
	year = {2024},
	pages = {480:1--480:31},
}

@inproceedings{sharevski_debunk-it-yourself_2025,
	address = {New York, NY, USA},
	series = {{NSPW} '24},
	title = {'{Debunk}-{It}-{Yourself}': {Health} {Professionals} {Strategies} for {Responding} to {Misinformation} on {TikTok}},
	isbn = {979-8-4007-1128-2},
	shorttitle = {'{Debunk}-{It}-{Yourself}'},
	url = {https://dl.acm.org/doi/10.1145/3703465.3703469},
	doi = {10.1145/3703465.3703469},
	urldate = {2025-11-30},
	booktitle = {Proceedings of the {New} {Security} {Paradigms} {Workshop}},
	publisher = {Association for Computing Machinery},
	author = {Sharevski, Filipo and Vander Loop, Jennifer and Jachim, Peter and Devine, Amy and Das, Sanchari},
	year = {2025},
	pages = {35--55},
}

@article{pehlivanova_support_2025,
	title = {Support needs after a near-death experience: {A} quantitative study with experiencers},
	issn = {2326-5531},
	shorttitle = {Support needs after a near-death experience},
	doi = {10.1037/cns0000439},
	journal = {Psychology of Consciousness: Theory, Research, and Practice},
	author = {Pehlivanova, Marieta and McNally, Katherine C. and Funk, Sabina and Greyson, Bruce},
	year = {2025},
	note = {Place: US
Publisher: Educational Publishing Foundation},
}

@article{jr_experience_1972,
	title = {The {Experience} of {Dying}†},
	copyright = {Copyright Taylor \& Francis},
	issn = {0033-2747},
	url = {https://www.tandfonline.com/doi/abs/10.1080/00332747.1972.11023710},
	language = {EN},
	urldate = {2025-11-21},
	journal = {Psychiatry},
	author = {Jr, Russell Noyes},
	month = may,
	year = {1972},
	note = {Publisher: Routledge},
}

@article{huang_gender_2018,
	title = {Gender {Differences} in {Motivations} for {Identity} {Reconstruction} on {Social} {Network} {Sites}},
	volume = {34},
	issn = {1044-7318},
	url = {https://doi.org/10.1080/10447318.2017.1383061},
	doi = {10.1080/10447318.2017.1383061},
	number = {7},
	urldate = {2025-09-11},
	journal = {International Journal of Human–Computer Interaction},
	author = {Huang, Jiao and Kumar, Sameer and Hu, Chuan},
	month = jul,
	year = {2018},
	note = {Publisher: Taylor \& Francis
\_eprint: https://doi.org/10.1080/10447318.2017.1383061},
	pages = {591--602},
}

@article{kim_examining_2011,
	series = {2009 {Fifth} {International} {Conference} on {Intelligent} {Computing}},
	title = {Examining knowledge contribution from the perspective of an online identity in blogging communities},
	volume = {27},
	issn = {0747-5632},
	url = {https://www.sciencedirect.com/science/article/pii/S0747563211000616},
	doi = {10.1016/j.chb.2011.03.003},
	number = {5},
	urldate = {2025-09-11},
	journal = {Computers in Human Behavior},
	author = {Kim, Hee-Woong and Zheng, Jun Raymond and Gupta, Sumeet},
	month = sep,
	year = {2011},
	pages = {1760--1770},
}

@article{suler_online_2004,
	title = {The {Online} {Disinhibition} {Effect}},
	volume = {7},
	issn = {1094-9313},
	url = {https://www.liebertpub.com/doi/abs/10.1089/1094931041291295},
	doi = {10.1089/1094931041291295},
	number = {3},
	urldate = {2025-09-11},
	journal = {CyberPsychology \& Behavior},
	author = {Suler, John},
	month = jun,
	year = {2004},
	note = {Publisher: Mary Ann Liebert, Inc., publishers},
	pages = {321--326},
}

@article{barak_fostering_2008,
	series = {Including the {Special} {Issue}: {Internet} {Empowerment}},
	title = {Fostering empowerment in online support groups},
	volume = {24},
	issn = {0747-5632},
	url = {https://www.sciencedirect.com/science/article/pii/S0747563208000198},
	doi = {10.1016/j.chb.2008.02.004},
	number = {5},
	urldate = {2025-09-11},
	journal = {Computers in Human Behavior},
	author = {Barak, Azy and Boniel-Nissim, Meyran and Suler, John},
	month = sep,
	year = {2008},
	pages = {1867--1883},
}

@article{lou_contributing_2013,
	title = {Contributing high quantity and quality knowledge to online {Q}\&{A} communities},
	volume = {64},
	copyright = {© 2012 ASIS\&T},
	issn = {1532-2890},
	url = {https://onlinelibrary.wiley.com/doi/abs/10.1002/asi.22750},
	doi = {10.1002/asi.22750},
	language = {en},
	number = {2},
	urldate = {2025-09-11},
	journal = {Journal of the American Society for Information Science and Technology},
	author = {Lou, Jie and Fang, Yulin and Lim, Kai H. and Peng, Jerry Zeyu},
	year = {2013},
	note = {\_eprint: https://onlinelibrary.wiley.com/doi/pdf/10.1002/asi.22750},
	pages = {356--371},
}

@article{lin_why_2013,
	title = {Why people share knowledge in virtual communities?: {The} use of {Yahoo}! {Kimo} {Knowledge}+ as an example},
	volume = {23},
	issn = {1066-2243},
	shorttitle = {Why people share knowledge in virtual communities?},
	url = {https://doi.org/10.1108/10662241311313295},
	doi = {10.1108/10662241311313295},
	number = {2},
	urldate = {2025-09-11},
	journal = {Internet Research},
	author = {Lin, Fu‐ren and Huang, Hui‐yi},
	month = mar,
	year = {2013},
	pages = {133--159},
}

@article{hsu_acceptance_2008,
	title = {Acceptance of blog usage: {The} roles of technology acceptance, social influence and knowledge sharing motivation},
	volume = {45},
	issn = {0378-7206},
	shorttitle = {Acceptance of blog usage},
	url = {https://www.sciencedirect.com/science/article/pii/S0378720607001255},
	doi = {10.1016/j.im.2007.11.001},
	number = {1},
	urldate = {2025-09-11},
	journal = {Information \& Management},
	author = {Hsu, Chin-Lung and Lin, Judy Chuan-Chuan},
	month = jan,
	year = {2008},
	pages = {65--74},
}

@article{miceli_emotional_2006,
	title = {Emotional and {Non}-{Emotional} {Persuasion}},
	volume = {20},
	issn = {0883-9514},
	url = {https://doi.org/10.1080/08839510600938193},
	doi = {10.1080/08839510600938193},
	number = {10},
	urldate = {2025-09-11},
	journal = {Applied Artificial Intelligence},
	author = {Miceli, Maria and Rosis, Fiorella de and Poggi, Isabella},
	month = jun,
	year = {2006},
	note = {Publisher: Taylor \& Francis
\_eprint: https://doi.org/10.1080/08839510600938193},
	pages = {849--879},
}

@incollection{jorgensen_chapter_1996,
	address = {San Diego},
	title = {Chapter 15 - {Affect}, persuasion, and communication processes},
	isbn = {978-0-12-057770-5},
	url = {https://www.sciencedirect.com/science/article/pii/B9780120577705500175},
	urldate = {2025-09-11},
	booktitle = {Handbook of {Communication} and {Emotion}},
	publisher = {Academic Press},
	author = {Jorgensen, Peter F.},
	editor = {Andersen, Peter A. and Guerrero, Laura K.},
	month = jan,
	year = {1996},
	doi = {10.1016/B978-012057770-5/50017-5},
	pages = {403--422},
}

@article{olson_distance_2000,
	title = {Distance {Matters}},
	volume = {15},
	issn = {0737-0024},
	url = {https://doi.org/10.1207/S15327051HCI1523_4},
	doi = {10.1207/S15327051HCI1523_4},
	number = {2-3},
	urldate = {2025-09-11},
	journal = {Human–Computer Interaction},
	author = {Olson, Gary M. and Olson, Judith S.},
	month = sep,
	year = {2000},
	note = {Publisher: Taylor \& Francis
\_eprint: https://doi.org/10.1207/S15327051HCI1523\_4},
	pages = {139--178},
}

@article{hogan_presentation_2010,
	title = {The {Presentation} of {Self} in the {Age} of {Social} {Media}: {Distinguishing} {Performances} and {Exhibitions} {Online}},
	volume = {30},
	issn = {0270-4676},
	shorttitle = {The {Presentation} of {Self} in the {Age} of {Social} {Media}},
	url = {https://doi.org/10.1177/0270467610385893},
	doi = {10.1177/0270467610385893},
	language = {EN},
	number = {6},
	urldate = {2025-09-11},
	journal = {Bulletin of Science, Technology \& Society},
	author = {Hogan, Bernie},
	month = dec,
	year = {2010},
	note = {Publisher: SAGE Publications Inc},
	pages = {377--386},
}

@book{oudshoorn_how_2005,
	title = {How {Users} {Matter}: {The} {Co}-{Construction} of {Users} and {Technology}},
	isbn = {978-0-262-65109-7},
	shorttitle = {How {Users} {Matter}},
	language = {en},
	publisher = {MIT Press},
	author = {Oudshoorn, Nelly and Pinch, Trevor},
	month = aug,
	year = {2005},
	note = {Google-Books-ID: hr5NEAAAQBAJ},
}

@inproceedings{baglione_modern_2018,
	address = {New York, NY, USA},
	series = {{CHI} '18},
	title = {Modern {Bereavement}: {A} {Model} for {Complicated} {Grief} in the {Digital} {Age}},
	isbn = {978-1-4503-5620-6},
	shorttitle = {Modern {Bereavement}},
	url = {https://dl.acm.org/doi/10.1145/3173574.3173990},
	doi = {10.1145/3173574.3173990},
	urldate = {2025-09-08},
	booktitle = {Proceedings of the 2018 {CHI} {Conference} on {Human} {Factors} in {Computing} {Systems}},
	publisher = {Association for Computing Machinery},
	author = {Baglione, Anna N. and Girard, Maxine M. and Price, Meagan and Clawson, James and Shih, Patrick C.},
	year = {2018},
	pages = {1--12},
}

@inproceedings{andalibi_self-disclosure_2017,
	address = {New York, NY, USA},
	series = {{CHI} {EA} '17},
	title = {Self-disclosure and {Response} {Behaviors} in {Socially} {Stigmatized} {Contexts} on {Social} {Media}: {The} {Case} of {Miscarriage}},
	isbn = {978-1-4503-4656-6},
	shorttitle = {Self-disclosure and {Response} {Behaviors} in {Socially} {Stigmatized} {Contexts} on {Social} {Media}},
	url = {https://dl.acm.org/doi/10.1145/3027063.3027137},
	doi = {10.1145/3027063.3027137},
	urldate = {2025-09-08},
	booktitle = {Proceedings of the 2017 {CHI} {Conference} {Extended} {Abstracts} on {Human} {Factors} in {Computing} {Systems}},
	publisher = {Association for Computing Machinery},
	author = {Andalibi, Nazanin},
	year = {2017},
	pages = {248--253},
}

@article{fairbairn_good_2002,
	title = {A {Good} {Death}: {On} the {Value} of {Death} and {Dying}},
	volume = {3},
	issn = {1466-769X},
	shorttitle = {A {Good} {Death}},
	url = {https://onlinelibrary.wiley.com/doi/abs/10.1046/j.1466-769X.2002.t01-2-00103.x},
	doi = {10.1046/j.1466-769X.2002.t01-2-00103.x},
	language = {en},
	number = {3},
	urldate = {2025-09-06},
	journal = {Nursing Philosophy},
	author = {Fairbairn, Gavin},
	year = {2002},
	note = {\_eprint: https://onlinelibrary.wiley.com/doi/pdf/10.1046/j.1466-769X.2002.t01-2-00103.x},
	pages = {274--275},
}

@book{glaser_awareness_2017,
	address = {New York},
	title = {Awareness of {Dying}},
	isbn = {978-1-351-32792-3},
	publisher = {Routledge},
	author = {Glaser, Barney G. and Strauss, Anselm L.},
	month = jul,
	year = {2017},
	doi = {10.4324/9781351327923},
}

@book{kellehear_social_2007,
	address = {Cambridge},
	title = {A {Social} {History} of {Dying}},
	isbn = {978-0-521-69429-2},
	url = {https://www.cambridge.org/core/books/social-history-of-dying/557FA3B9285436ADBA5C858900B83661},
	urldate = {2025-09-06},
	publisher = {Cambridge University Press},
	author = {Kellehear, Allan},
	year = {2007},
	doi = {10.1017/CBO9780511481352},
}

@article{meier_defining_2016,
	title = {Defining a {Good} {Death} ({Successful} {Dying}): {Literature} {Review} and a {Call} for {Research} and {Public} {Dialogue}},
	volume = {24},
	issn = {1545-7214},
	shorttitle = {Defining a {Good} {Death} ({Successful} {Dying})},
	doi = {10.1016/j.jagp.2016.01.135},
	language = {eng},
	number = {4},
	journal = {The American Journal of Geriatric Psychiatry: Official Journal of the American Association for Geriatric Psychiatry},
	author = {Meier, Emily A. and Gallegos, Jarred V. and Thomas, Lori P. Montross and Depp, Colin A. and Irwin, Scott A. and Jeste, Dilip V.},
	month = apr,
	year = {2016},
	pmid = {26976293},
	pmcid = {PMC4828197},
	pages = {261--271},
}

@inproceedings{massimi_dying_2009,
	address = {New York, NY, USA},
	series = {{CHI} {EA} '09},
	title = {Dying, death, and mortality: towards thanatosensitivity in {HCI}},
	isbn = {978-1-60558-247-4},
	shorttitle = {Dying, death, and mortality},
	url = {https://doi.org/10.1145/1520340.1520349},
	doi = {10.1145/1520340.1520349},
	urldate = {2025-09-06},
	booktitle = {{CHI} '09 {Extended} {Abstracts} on {Human} {Factors} in {Computing} {Systems}},
	publisher = {Association for Computing Machinery},
	author = {Massimi, Michael and Charise, Andrea},
	year = {2009},
	pages = {2459--2468},
}

@inproceedings{bahng_reflexive_2020,
	address = {New York, NY, USA},
	series = {{CHI} '20},
	title = {Reflexive {VR} {Storytelling} {Design} {Beyond} {Immersion}: {Facilitating} {Self}-{Reflection} on {Death} and {Loneliness}},
	isbn = {978-1-4503-6708-0},
	shorttitle = {Reflexive {VR} {Storytelling} {Design} {Beyond} {Immersion}},
	url = {https://doi.org/10.1145/3313831.3376582},
	doi = {10.1145/3313831.3376582},
	urldate = {2025-09-06},
	booktitle = {Proceedings of the 2020 {CHI} {Conference} on {Human} {Factors} in {Computing} {Systems}},
	publisher = {Association for Computing Machinery},
	author = {Bahng, Sojung and Kelly, Ryan M. and McCormack, Jon},
	year = {2020},
	pages = {1--13},
}

@inproceedings{eum_how_2021,
	address = {New York, NY, USA},
	series = {{CHI} {EA} '21},
	title = {How the {Death}-themed {Game} {Spiritfarer} {Can} {Help} {Players} {Cope} with the {Loss} of a {Loved} {One}},
	isbn = {978-1-4503-8095-9},
	url = {https://doi.org/10.1145/3411763.3451608},
	doi = {10.1145/3411763.3451608},
	urldate = {2025-09-06},
	booktitle = {Extended {Abstracts} of the 2021 {CHI} {Conference} on {Human} {Factors} in {Computing} {Systems}},
	publisher = {Association for Computing Machinery},
	author = {Eum, Karam and Erb, Valérie and Lin, Subin and Wang, Sungpil and Doh, Young Yim},
	year = {2021},
	pages = {1--6},
}

@inproceedings{chen_what_2021,
	address = {New York, NY, USA},
	series = {{CHI} '21},
	title = {What {Happens} {After} {Death}? {Using} a {Design} {Workbook} to {Understand} {User} {Expectations} for {Preparing} their {Data}},
	isbn = {978-1-4503-8096-6},
	shorttitle = {What {Happens} {After} {Death}?},
	url = {https://doi.org/10.1145/3411764.3445359},
	doi = {10.1145/3411764.3445359},
	urldate = {2025-09-06},
	booktitle = {Proceedings of the 2021 {CHI} {Conference} on {Human} {Factors} in {Computing} {Systems}},
	publisher = {Association for Computing Machinery},
	author = {Chen, Janet X. and Vitale, Francesco and McGrenere, Joanna},
	year = {2021},
	pages = {1--13},
}

@inproceedings{luo_between_2025,
	address = {New York, NY, USA},
	series = {{CHI} {EA} '25},
	title = {Between: {Easing} the {Fear} of {Death} {Through} {Peaceful} {Play}},
	isbn = {979-8-4007-1395-8},
	shorttitle = {Between},
	url = {https://doi.org/10.1145/3706599.3720320},
	doi = {10.1145/3706599.3720320},
	urldate = {2025-09-06},
	booktitle = {Proceedings of the {Extended} {Abstracts} of the {CHI} {Conference} on {Human} {Factors} in {Computing} {Systems}},
	publisher = {Association for Computing Machinery},
	author = {Luo, Bingxiao and Hämäläinen, Perttu and Rautalahti, Heidi},
	year = {2025},
	pages = {1--5},
}

@article{zeng_whatieatinaday_2025,
	title = {\#{WhatIEatinaDay}: {The} {Quality}, {Accuracy}, and {Engagement} of {Nutrition} {Content} on {TikTok}},
	volume = {17},
	copyright = {http://creativecommons.org/licenses/by/3.0/},
	issn = {2072-6643},
	shorttitle = {\#{WhatIEatinaDay}},
	url = {https://www.mdpi.com/2072-6643/17/5/781},
	doi = {10.3390/nu17050781},
	language = {en},
	number = {5},
	urldate = {2025-05-10},
	journal = {Nutrients},
	author = {Zeng, Michelle and Grgurevic, Jacqueline and Diyab, Rayan and Roy, Rajshri},
	month = jan,
	year = {2025},
	note = {Number: 5
Publisher: Multidisciplinary Digital Publishing Institute},
	pages = {781},
}

@misc{baumann_dynamics_2025,
	title = {Dynamics of {Algorithmic} {Content} {Amplification} on {TikTok}},
	url = {http://arxiv.org/abs/2503.20231},
	doi = {10.48550/arXiv.2503.20231},
	urldate = {2025-05-10},
	publisher = {arXiv},
	author = {Baumann, Fabian and Arora, Nipun and Rahwan, Iyad and Czaplicka, Agnieszka},
	month = mar,
	year = {2025},
	note = {arXiv:2503.20231 [physics]},
}

@article{onwuegbuzie_validity_2007,
	title = {Validity and {Qualitative} {Research}: {An} {Oxymoron}?},
	volume = {41},
	issn = {1573-7845},
	shorttitle = {Validity and {Qualitative} {Research}},
	url = {https://doi.org/10.1007/s11135-006-9000-3},
	doi = {10.1007/s11135-006-9000-3},
	language = {en},
	number = {2},
	urldate = {2025-05-10},
	journal = {Quality \& Quantity},
	author = {Onwuegbuzie, Anthony J. and Leech, Nancy L.},
	month = apr,
	year = {2007},
	pages = {233--249},
}

@article{chiu_last_2021,
	series = {{ACM} {Conferences}},
	title = {To {Last} {Long} or to {Fade} {Away}: {Investigating} {Users}' {Instagram} {Post} and {Story} {Practices}},
	issn = {9781450384797},
	shorttitle = {To {Last} {Long} or to {Fade} {Away}},
	url = {https://dl.acm.org/doi/abs/10.1145/3462204.3481778},
	doi = {10.1145/3462204.3481778},
	urldate = {2025-05-10},
	journal = {Companion Publication of the 2021 Conference on Computer Supported Cooperative Work and Social Computing},
	author = {Chiu, Hsuen Chi and {View Profile} and Yuan, Chien Wen (Tina) and {View Profile}},
	month = oct,
	year = {2021},
	pages = {32--35},
}

@inproceedings{zhang_pragmatic_2023,
	address = {New York, NY, USA},
	series = {{CSCW} '23 {Companion}},
	title = {\#{Pragmatic} or \#{Clinical}: {Analyzing} {TikTok} {Mental} {Health} {Videos}},
	isbn = {9798400701290},
	shorttitle = {\#{Pragmatic} or \#{Clinical}},
	url = {https://dl.acm.org/doi/10.1145/3584931.3607013},
	doi = {10.1145/3584931.3607013},
	urldate = {2025-05-09},
	booktitle = {Companion {Publication} of the 2023 {Conference} on {Computer} {Supported} {Cooperative} {Work} and {Social} {Computing}},
	publisher = {Association for Computing Machinery},
	author = {Zhang, Alice Qian and Milton, Ashlee and Chancellor, Stevie},
	year = {2023},
	pages = {149--153},
}

@article{giaxoglou_mediatization_2018,
	title = {Mediatization of {Emotion} on {Social} {Media}: {Forms} and {Norms} in {Digital} {Mourning} {Practices}},
	copyright = {© The Author(s) 2018},
	shorttitle = {Mediatization of {Emotion} on {Social} {Media}},
	url = {https://journals.sagepub.com/doi/full/10.1177/2056305117744393},
	doi = {10.1177/2056305117744393},
	language = {EN},
	urldate = {2025-04-30},
	journal = {Social Media + Society},
	author = {Giaxoglou, Korina and Döveling, Katrin},
	month = jan,
	year = {2018},
	note = {Publisher: SAGE PublicationsSage UK: London, England},
}

@article{walter_new_2015,
	title = {New mourners, old mourners: online memorial culture as a chapter in the history of mourning},
	copyright = {© 2014 Taylor \& Francis},
	issn = {1361-4568},
	shorttitle = {New mourners, old mourners},
	url = {https://www.tandfonline.com/doi/abs/10.1080/13614568.2014.983555},
	language = {EN},
	urldate = {2025-04-30},
	journal = {New Review of Hypermedia and Multimedia},
	author = {Walter, Tony},
	month = apr,
	year = {2015},
	note = {Publisher: Taylor \& Francis},
}

@phdthesis{roland_affective_2017,
	title = {The affective resonance of personal narratives : creating a deeper experience of identity, empathy and historical understanding : a thesis presented in partial fulfilment of the requirements for the degree of {Masters} of {Arts} in {Museum} {Studies}, {Massey} {University}, {Palmerston} {North}, {Aotearoa}, {New} {Zealand}},
	shorttitle = {The affective resonance of personal narratives},
	url = {http://hdl.handle.net/10179/12991},
	language = {en},
	urldate = {2025-04-30},
	school = {Massey University},
	author = {Roland, Zoë Gabrielle},
	year = {2017},
}

@article{lagerkvist_grand_2017,
	title = {The grand interruption: death online and mediated lifelines of shared vulnerability},
	copyright = {© 2017 Informa UK Limited, trading as Taylor \& Francis Group},
	issn = {1468-0777},
	shorttitle = {The grand interruption},
	url = {https://www.tandfonline.com/doi/abs/10.1080/14680777.2017.1326554},
	language = {EN},
	urldate = {2025-04-30},
	journal = {Feminist Media Studies},
	author = {Lagerkvist, Amanda and Andersson, Yvonne},
	month = jul,
	year = {2017},
	note = {Publisher: Routledge},
}

@article{trepte_reciprocal_2013,
	title = {The reciprocal effects of social network site use and the disposition for self-disclosure: {A} longitudinal study},
	volume = {29},
	issn = {0747-5632},
	shorttitle = {The reciprocal effects of social network site use and the disposition for self-disclosure},
	url = {https://www.sciencedirect.com/science/article/pii/S0747563212002701},
	doi = {10.1016/j.chb.2012.10.002},
	number = {3},
	urldate = {2025-04-30},
	journal = {Computers in Human Behavior},
	author = {Trepte, Sabine and Reinecke, Leonard},
	month = may,
	year = {2013},
	pages = {1102--1112},
}

@article{sprecher_taking_2013,
	title = {Taking turns: {Reciprocal} self-disclosure promotes liking in initial interactions},
	volume = {49},
	issn = {0022-1031},
	shorttitle = {Taking turns},
	url = {https://www.sciencedirect.com/science/article/pii/S002210311300070X},
	doi = {10.1016/j.jesp.2013.03.017},
	number = {5},
	urldate = {2025-04-30},
	journal = {Journal of Experimental Social Psychology},
	author = {Sprecher, Susan and Treger, Stanislav and Wondra, Joshua D. and Hilaire, Nicole and Wallpe, Kevin},
	month = sep,
	year = {2013},
	pages = {860--866},
}

@article{osler_taking_2024,
	title = {Taking empathy online},
	volume = {67},
	issn = {0020-174X},
	url = {https://www.tandfonline.com/doi/full/10.1080/0020174X.2021.1899045},
	doi = {10.1080/0020174X.2021.1899045},
	number = {1},
	urldate = {2025-04-30},
	journal = {Inquiry},
	author = {Osler, Lucy},
	month = jan,
	year = {2024},
	note = {Publisher: Routledge},
	pages = {302--329},
}

@article{greyson_near-death_2006,
	title = {Near-{Death} {Experiences} and {Spirituality}},
	volume = {41},
	issn = {1467-9744},
	url = {https://onlinelibrary.wiley.com/doi/abs/10.1111/j.1467-9744.2005.00745.x},
	doi = {10.1111/j.1467-9744.2005.00745.x},
	language = {en},
	number = {2},
	urldate = {2025-04-30},
	journal = {Zygon®},
	author = {Greyson, Bruce},
	year = {2006},
	note = {\_eprint: https://onlinelibrary.wiley.com/doi/pdf/10.1111/j.1467-9744.2005.00745.x},
	pages = {393--414},
}

@article{dimaggio_culture_1997,
	title = {Culture and {Cognition}},
	volume = {23},
	issn = {0360-0572, 1545-2115},
	url = {https://www.annualreviews.org/content/journals/10.1146/annurev.soc.23.1.263},
	doi = {10.1146/annurev.soc.23.1.263},
	language = {en},
	number = {Volume 23, 1997},
	urldate = {2025-04-30},
	journal = {Annual Review of Sociology},
	author = {DiMaggio, Paul},
	month = aug,
	year = {1997},
	note = {Publisher: Annual Reviews},
	pages = {263--287},
}

@article{cinelli_echo_2021,
	title = {The echo chamber effect on social media},
	volume = {118},
	issn = {1091-6490},
	doi = {10.1073/pnas.2023301118},
	language = {eng},
	number = {9},
	journal = {Proceedings of the National Academy of Sciences of the United States of America},
	author = {Cinelli, Matteo and De Francisci Morales, Gianmarco and Galeazzi, Alessandro and Quattrociocchi, Walter and Starnini, Michele},
	month = mar,
	year = {2021},
	pmid = {33622786},
	pmcid = {PMC7936330},
	pages = {e2023301118},
}

@article{moller_online_2021,
	title = {Online social environments and their impact on video viewers: {The} effects of user comments on entertainment experiences and knowledge gain during political satire consumption},
	copyright = {© The Author(s) 2021},
	shorttitle = {Online social environments and their impact on video viewers},
	url = {https://journals.sagepub.com/doi/full/10.1177/14614448211015984},
	doi = {10.1177/14614448211015984},
	language = {EN},
	urldate = {2025-04-30},
	journal = {New Media \& Society},
	author = {Möller, Anne Marthe and Boukes, Mark},
	month = jun,
	year = {2021},
	note = {Publisher: SAGE PublicationsSage UK: London, England},
}

@article{byun_effect_2022,
	title = {The effect of {YouTube} comment interaction on video engagement: focusing on interactivity centralization and creators' interactivity},
	volume = {47},
	issn = {1468-4527},
	shorttitle = {The effect of {YouTube} comment interaction on video engagement},
	url = {https://www.emerald.com/insight/content/doi/10.1108/oir-04-2022-0217/full/html},
	doi = {10.1108/OIR-04-2022-0217},
	language = {en},
	number = {6},
	urldate = {2025-04-30},
	journal = {Online Information Review},
	author = {Byun, Unji and Jang, Moonkyoung and Baek, Hyunmi},
	month = nov,
	year = {2022},
	note = {Publisher: Emerald Publishing Limited},
	pages = {1083--1097},
}

@book{james_varieties_2003,
	address = {London},
	edition = {2},
	title = {The {Varieties} of {Religious} {Experience}: {A} {Study} in {Human} {Nature}},
	isbn = {978-0-203-39847-0},
	shorttitle = {The {Varieties} of {Religious} {Experience}},
	publisher = {Routledge},
	author = {James, William},
	month = sep,
	year = {2003},
	doi = {10.4324/9780203398470},
}

@article{carr_reason_2004,
	title = {Reason, {Meaning} and {Truth} in {Religious} {Narrative}: {Towards} an {Epistemic} {Rationale} for {Religious} and {Faith} {School} {Education}},
	volume = {17},
	issn = {0953-9468},
	shorttitle = {Reason, {Meaning} and {Truth} in {Religious} {Narrative}},
	url = {https://journals.sagepub.com/doi/abs/10.1177/095394680401700103},
	doi = {10.1177/095394680401700103},
	number = {1},
	urldate = {2025-04-30},
	journal = {Studies in Christian Ethics},
	author = {Carr, David},
	month = apr,
	year = {2004},
	note = {Publisher: SAGE Publications Ltd},
	pages = {38--53},
}

@article{yaden_varieties_2017,
	title = {The {Varieties} of {Self}-{Transcendent} {Experience}},
	volume = {21},
	issn = {1089-2680},
	url = {https://journals.sagepub.com/doi/full/10.1037/gpr0000102},
	doi = {10.1037/gpr0000102},
	number = {2},
	urldate = {2025-04-30},
	journal = {Review of General Psychology},
	author = {Yaden, David Bryce and Haidt, Jonathan and Hood, Ralph W. and Vago, David R. and Newberg, Andrew B.},
	month = jun,
	year = {2017},
	note = {Publisher: SAGE Publications Inc},
	pages = {143--160},
}

@article{hashemi_explanation_2023,
	title = {Explanation of near-death experiences: a systematic analysis of case reports and qualitative research},
	volume = {14},
	issn = {1664-1078},
	shorttitle = {Explanation of near-death experiences},
	url = {https://www.frontiersin.orghttps://www.frontiersin.org/journals/psychology/articles/10.3389/fpsyg.2023.1048929/full},
	doi = {10.3389/fpsyg.2023.1048929},
	language = {English},
	urldate = {2025-04-24},
	journal = {Frontiers in Psychology},
	author = {Hashemi, Amirhossein and Oroojan, Ali Akbar and Rassouli, Maryam and Ashrafizadeh, Hadis},
	month = apr,
	year = {2023},
	note = {Publisher: Frontiers},
}

@inproceedings{lingel_city_2014,
	address = {New York, NY, USA},
	series = {{CSCW} '14},
	title = {City, self, network: transnational migrants and online identity work},
	isbn = {978-1-4503-2540-0},
	shorttitle = {City, self, network},
	url = {https://doi.org/10.1145/2531602.2531693},
	doi = {10.1145/2531602.2531693},
	urldate = {2025-04-23},
	booktitle = {Proceedings of the 17th {ACM} conference on {Computer} supported cooperative work \& social computing},
	publisher = {Association for Computing Machinery},
	author = {Lingel, Jessica and Naaman, Mor and boyd, danah m.},
	year = {2014},
	pages = {1502--1510},
}

@inproceedings{haque_perceptions_2023,
	address = {New York, NY, USA},
	series = {{CSCW} '23 {Companion}},
	title = {Perceptions of {Mental} {Health} {Crisis} among {U}.{S}. {Military} {Veteran} {Peer} {Mentors} and {Potential} of {Mobile}-{Based} {Peer}-{Support} {Interventions}},
	isbn = {9798400701290},
	url = {https://dl.acm.org/doi/10.1145/3584931.3607009},
	doi = {10.1145/3584931.3607009},
	urldate = {2025-04-23},
	booktitle = {Companion {Publication} of the 2023 {Conference} on {Computer} {Supported} {Cooperative} {Work} and {Social} {Computing}},
	publisher = {Association for Computing Machinery},
	author = {Haque, Md Romael and Franco, Zeno and Hossain, Md Fitrat and Frydrychowicz, Wylie and Madiraju, Praveen and Baker, Natalie D and Hooyer, Katinka and Ahamed, Sheikh Iqbal and Winstead, Otis and Curry, Robert and Rubya, Sabirat},
	year = {2023},
	pages = {33--38},
}

@inproceedings{wang_modeling_2016,
	address = {New York, NY, USA},
	series = {{CSCW} '16},
	title = {Modeling {Self}-{Disclosure} in {Social} {Networking} {Sites}},
	isbn = {978-1-4503-3592-8},
	url = {https://dl.acm.org/doi/10.1145/2818048.2820010},
	doi = {10.1145/2818048.2820010},
	urldate = {2025-04-23},
	booktitle = {Proceedings of the 19th {ACM} {Conference} on {Computer}-{Supported} {Cooperative} {Work} \& {Social} {Computing}},
	publisher = {Association for Computing Machinery},
	author = {Wang, Yi-Chia and Burke, Moira and Kraut, Robert},
	year = {2016},
	pages = {74--85},
}

@inproceedings{punuru_cultural_2020,
	address = {New York, NY, USA},
	series = {{CSCW} '20 {Companion}},
	title = {Cultural {Norms} and {Interpersonal} {Relationships}: {Comparing} {Disclosure} {Behaviors} on {Twitter}},
	isbn = {978-1-4503-8059-1},
	shorttitle = {Cultural {Norms} and {Interpersonal} {Relationships}},
	url = {https://doi.org/10.1145/3406865.3418341},
	doi = {10.1145/3406865.3418341},
	urldate = {2025-04-23},
	booktitle = {Companion {Publication} of the 2020 {Conference} on {Computer} {Supported} {Cooperative} {Work} and {Social} {Computing}},
	publisher = {Association for Computing Machinery},
	author = {Punuru, Anju and Cheng, Tyng-Wen and Ghosh, Isha and Page, Xinru and Mondal, Mainack},
	year = {2020},
	pages = {371--375},
}

@inproceedings{semaan_military_2017,
	address = {New York, NY, USA},
	series = {{CSCW} '17},
	title = {Military {Masculinity} and the {Travails} of {Transitioning}: {Disclosure} in {Social} {Media}},
	isbn = {978-1-4503-4335-0},
	shorttitle = {Military {Masculinity} and the {Travails} of {Transitioning}},
	url = {https://doi.org/10.1145/2998181.2998221},
	doi = {10.1145/2998181.2998221},
	urldate = {2025-04-23},
	booktitle = {Proceedings of the 2017 {ACM} {Conference} on {Computer} {Supported} {Cooperative} {Work} and {Social} {Computing}},
	publisher = {Association for Computing Machinery},
	author = {Semaan, Bryan and Britton, Lauren M. and Dosono, Bryan},
	year = {2017},
	pages = {387--403},
}

@article{bazarova_contents_2012,
	series = {{ACM} {Conferences}},
	title = {Contents and contexts},
	issn = {9781450310864},
	url = {https://dl.acm.org/doi/10.1145/2145204.2145262},
	doi = {10.1145/2145204.2145262},
	urldate = {2025-04-24},
	journal = {Proceedings of the ACM 2012 conference on Computer Supported Cooperative Work},
	author = {Bazarova, Natalya N. and {View Profile}},
	month = feb,
	year = {2012},
	pages = {369--372},
}

@inproceedings{bak_examining_2024,
	address = {New York, NY, USA},
	series = {{CSCW} {Companion} '24},
	title = {Examining the {Social} {Dynamics} on {Health} {Behavior} {Promotion} in {Online} {Peer} {Discussions} about {Breast} {Cancer} {Screening}: {Understanding} {Peer} {Discussion} for {Breast} {Cancer} {Screening} {Promotion} {Using} the {Health} {Action} {Process} {Approach}},
	isbn = {9798400711145},
	shorttitle = {Examining the {Social} {Dynamics} on {Health} {Behavior} {Promotion} in {Online} {Peer} {Discussions} about {Breast} {Cancer} {Screening}},
	url = {https://dl.acm.org/doi/10.1145/3678884.3681893},
	doi = {10.1145/3678884.3681893},
	urldate = {2025-04-23},
	booktitle = {Companion {Publication} of the 2024 {Conference} on {Computer}-{Supported} {Cooperative} {Work} and {Social} {Computing}},
	publisher = {Association for Computing Machinery},
	author = {Bak, Michelle and Chin, Jessie},
	year = {2024},
	pages = {472--477},
}

@inproceedings{mckim_investigating_2021,
	address = {New York, NY, USA},
	series = {{CSCW} '21 {Companion}},
	title = {Investigating {Drug} {Addiction} {Discourse} on {YouTube}},
	isbn = {978-1-4503-8479-7},
	url = {https://doi.org/10.1145/3462204.3481762},
	doi = {10.1145/3462204.3481762},
	urldate = {2025-04-23},
	booktitle = {Companion {Publication} of the 2021 {Conference} on {Computer} {Supported} {Cooperative} {Work} and {Social} {Computing}},
	publisher = {Association for Computing Machinery},
	author = {McKim, Katherine G. and Mai, Cat and Hess, Danielle and Niu, Shuo},
	year = {2021},
	pages = {130--134},
}

@article{de_choudhury_gender_2017,
	series = {{ACM} {Conferences}},
	title = {Gender and {Cross}-{Cultural} {Differences} in {Social} {Media} {Disclosures} of {Mental} {Illness}},
	issn = {9781450343350},
	url = {https://dl.acm.org/doi/10.1145/2998181.2998220},
	doi = {10.1145/2998181.2998220},
	urldate = {2025-04-24},
	journal = {Proceedings of the 2017 ACM Conference on Computer Supported Cooperative Work and Social Computing},
	author = {De Choudhury, Munmun and {View Profile} and Sharma, Sanket S. and {View Profile} and Logar, Tomaz and {View Profile} and Eekhout, Wouter and {View Profile} and Nielsen, René Clausen and {View Profile}},
	month = feb,
	year = {2017},
	pages = {353--369},
}

@inproceedings{ammari_thanks_2016,
	address = {New York, NY, USA},
	series = {{CSCW} '16},
	title = {“{Thanks} for your interest in our {Facebook} group, but it's only for dads”: {Social} {Roles} of {Stay}-at-{Home} {Dads}},
	isbn = {978-1-4503-3592-8},
	shorttitle = {“{Thanks} for your interest in our {Facebook} group, but it's only for dads”},
	url = {https://dl.acm.org/doi/10.1145/2818048.2819927},
	doi = {10.1145/2818048.2819927},
	urldate = {2025-04-23},
	booktitle = {Proceedings of the 19th {ACM} {Conference} on {Computer}-{Supported} {Cooperative} {Work} \& {Social} {Computing}},
	publisher = {Association for Computing Machinery},
	author = {Ammari, Tawfiq and Schoenebeck, Sarita},
	year = {2016},
	pages = {1363--1375},
}

@article{pinch_its_2021,
	series = {{ACM} {Conferences}},
	title = {?{It}'s not exactly prominent or direct, but it's there?: {Understanding} {Strategies} for {Sensitive} {Disclosure} {Online}},
	issn = {9781450384797},
	shorttitle = {?},
	url = {https://dl.acm.org/doi/10.1145/3462204.3481740},
	doi = {10.1145/3462204.3481740},
	urldate = {2025-04-24},
	journal = {Companion Publication of the 2021 Conference on Computer Supported Cooperative Work and Social Computing},
	author = {Pinch, Annika and {View Profile} and Birnholtz, Jeremy and {View Profile} and Kraus, Ashley and {View Profile} and Macapagal, Kathryn and {View Profile} and A. Moskowitz, David and {View Profile}},
	month = oct,
	year = {2021},
	pages = {149--152},
}

@inproceedings{andalibi_sensitive_2017,
	address = {New York, NY, USA},
	series = {{CSCW} '17},
	title = {Sensitive {Self}-disclosures, {Responses}, and {Social} {Support} on {Instagram}: {The} {Case} of \#{Depression}},
	isbn = {978-1-4503-4335-0},
	shorttitle = {Sensitive {Self}-disclosures, {Responses}, and {Social} {Support} on {Instagram}},
	url = {https://dl.acm.org/doi/10.1145/2998181.2998243},
	doi = {10.1145/2998181.2998243},
	urldate = {2025-04-23},
	booktitle = {Proceedings of the 2017 {ACM} {Conference} on {Computer} {Supported} {Cooperative} {Work} and {Social} {Computing}},
	publisher = {Association for Computing Machinery},
	author = {Andalibi, Nazanin and Ozturk, Pinar and Forte, Andrea},
	year = {2017},
	pages = {1485--1500},
}

@inproceedings{lu_when_2015,
	address = {New York, NY, USA},
	series = {{CSCW}'15 {Companion}},
	title = {When to {Break} the {Ice}: {Self}-disclosure {Strategies} for {Newcomers} in {Online} {Communities}},
	isbn = {978-1-4503-2946-0},
	shorttitle = {When to {Break} the {Ice}},
	url = {https://doi.org/10.1145/2685553.2698997},
	doi = {10.1145/2685553.2698997},
	urldate = {2025-04-23},
	booktitle = {Proceedings of the 18th {ACM} {Conference} {Companion} on {Computer} {Supported} {Cooperative} {Work} \& {Social} {Computing}},
	publisher = {Association for Computing Machinery},
	author = {Lu, Di and Farzan, Rosta},
	year = {2015},
	pages = {163--166},
}

@article{ma_self-disclosure_2017,
	series = {{ACM} {Conferences}},
	title = {Self-{Disclosure} and {Perceived} {Trustworthiness} of {Airbnb} {Host} {Profiles}},
	issn = {9781450343350},
	url = {https://dl.acm.org/doi/10.1145/2998181.2998269},
	doi = {10.1145/2998181.2998269},
	urldate = {2025-04-24},
	journal = {Proceedings of the 2017 ACM Conference on Computer Supported Cooperative Work and Social Computing},
	author = {Ma, Xiao and {View Profile} and Hancock, Jeffrey T. and {View Profile} and Lim Mingjie, Kenneth and {View Profile} and Naaman, Mor and {View Profile}},
	month = feb,
	year = {2017},
	pages = {2397--2409},
}

@article{tang_tale_2011,
	series = {{ACM} {Conferences}},
	title = {A tale of two languages},
	issn = {9781450305563},
	url = {https://dl.acm.org/doi/10.1145/1958824.1958884},
	doi = {10.1145/1958824.1958884},
	urldate = {2025-04-24},
	journal = {Proceedings of the ACM 2011 conference on Computer supported cooperative work},
	author = {Tang, Dai and {View Profile} and Chou, Tina and {View Profile} and Drucker, Naomi and {View Profile} and Robertson, Adi and {View Profile} and Smith, William C. and {View Profile} and Hancock, Jeffery T. and {View Profile}},
	month = mar,
	year = {2011},
	pages = {387--390},
}

@article{greyson_defining_1999,
	title = {Defining near-death experiences},
	copyright = {Copyright Taylor and Francis Group, LLC},
	url = {https://www.tandfonline.com/doi/abs/10.1080/713685958},
	doi = {10.1080/713685958},
	language = {EN},
	urldate = {2025-04-24},
	journal = {Mortality},
	author = {Greyson, Bruce},
	month = jan,
	year = {1999},
	note = {Publisher: Taylor \& Francis Group},
}

@article{martial_near-death_2020,
	title = {The {Near}-{Death} {Experience} {Content} ({NDE}-{C}) scale: {Development} and psychometric validation},
	volume = {86},
	issn = {1090-2376},
	shorttitle = {The {Near}-{Death} {Experience} {Content} ({NDE}-{C}) scale},
	doi = {10.1016/j.concog.2020.103049},
	language = {eng},
	journal = {Consciousness and Cognition},
	author = {Martial, Charlotte and Simon, Jessica and Puttaert, Ninon and Gosseries, Olivia and Charland-Verville, Vanessa and Nyssen, Anne-Sophie and Greyson, Bruce and Laureys, Steven and Cassol, Héléna},
	month = nov,
	year = {2020},
	pmid = {33227590},
	pages = {103049},
}

@article{greyson_near-death_1983,
	title = {The near-death experience scale. {Construction}, reliability, and validity},
	volume = {171},
	issn = {0022-3018},
	doi = {10.1097/00005053-198306000-00007},
	language = {eng},
	number = {6},
	journal = {The Journal of Nervous and Mental Disease},
	author = {Greyson, B.},
	month = jun,
	year = {1983},
	pmid = {6854303},
	pages = {369--375},
}

@article{morris_nature_2003,
	title = {The {Nature} and {Meaning} of the {Near}-{Death} {Experience} for {Patients} and {Critical} {Care} {Nurses}},
	volume = {21},
	issn = {1573-3661},
	url = {https://doi.org/10.1023/A:1021235922301},
	doi = {10.1023/A:1021235922301},
	language = {en},
	number = {3},
	urldate = {2025-04-24},
	journal = {Journal of Near-Death Studies},
	author = {Morris, Linda L. and Knafl, Kathleen},
	month = mar,
	year = {2003},
	pages = {139--167},
}

@phdthesis{sutherland_very_1991,
	type = {Thesis},
	title = {A very different way : a sociological investigation of life after a near-death experience},
	shorttitle = {A very different way},
	url = {http://hdl.handle.net/1959.4/65900},
	language = {English},
	urldate = {2025-04-24},
	school = {UNSW Sydney},
	author = {Sutherland, Cherie},
	year = {1991},
	doi = {10.26190/unsworks/10859},
}

@incollection{foster_practical_2009,
	address = {Santa Barbara, CA, US},
	title = {Practical applications of research on near-death experiences},
	isbn = {978-0-313-35864-7 978-0-313-35865-4},
	booktitle = {The handbook of near-death experiences: {Thirty} years of investigation},
	publisher = {Praeger/ABC-CLIO},
	author = {Foster, Ryan D. and James, Debbie and Holden, Janice Miner},
	year = {2009},
	pages = {235--258},
}

@incollection{bush_distressing_2009,
	address = {Santa Barbara, CA, US},
	title = {Distressing {Western} near-death experiences: {Finding} a way through the abyss},
	isbn = {978-0-313-35864-7 978-0-313-35865-4},
	shorttitle = {Distressing {Western} near-death experiences},
	booktitle = {The handbook of near-death experiences: {Thirty} years of investigation},
	publisher = {Praeger/ABC-CLIO},
	author = {Bush, Nancy Evans},
	year = {2009},
	pages = {63--86},
}

@article{palmieri_reality_2014,
	title = {"{Reality}" of near-death-experience memories: evidence from a psychodynamic and electrophysiological integrated study},
	volume = {8},
	issn = {1662-5161},
	shorttitle = {"{Reality}" of near-death-experience memories},
	doi = {10.3389/fnhum.2014.00429},
	language = {eng},
	journal = {Frontiers in Human Neuroscience},
	author = {Palmieri, Arianna and Calvo, Vincenzo and Kleinbub, Johann R. and Meconi, Federica and Marangoni, Matteo and Barilaro, Paolo and Broggio, Alice and Sambin, Marco and Sessa, Paola},
	year = {2014},
	pmid = {24994974},
	pmcid = {PMC4063168},
	pages = {429},
}

@article{thonnard_characteristics_2013,
	title = {Characteristics of near-death experiences memories as compared to real and imagined events memories},
	volume = {8},
	issn = {1932-6203},
	doi = {10.1371/journal.pone.0057620},
	language = {eng},
	number = {3},
	journal = {PloS One},
	author = {Thonnard, Marie and Charland-Verville, Vanessa and Brédart, Serge and Dehon, Hedwige and Ledoux, Didier and Laureys, Steven and Vanhaudenhuyse, Audrey},
	year = {2013},
	pmid = {23544039},
	pmcid = {PMC3609762},
	pages = {e57620},
}

@incollection{greyson_near-death_2000,
	address = {Washington, DC, US},
	title = {Near-death experiences},
	isbn = {978-1-55798-625-2},
	booktitle = {Varieties of anomalous experience:  {Examining} the scientific evidence},
	publisher = {American Psychological Association},
	author = {Greyson, Bruce},
	year = {2000},
	doi = {10.1037/10371-010},
	pages = {315--352},
}

@book{moody_life_1976,
	address = {New York},
	title = {Life after life: the investigation of a phenomenon--survival of bodily death},
	shorttitle = {Life after life},
	publisher = {Bantam Books},
	author = {Moody, Raymond A. and Hill, Dick},
	year = {1976},
	note = {Open Library ID: OL17976439M},
}

@article{khanna_daily_2014,
	title = {Daily spiritual experiences before and after near-death experiences},
	volume = {6},
	issn = {1943-1562},
	doi = {10.1037/a0037258},
	number = {4},
	journal = {Psychology of Religion and Spirituality},
	author = {Khanna, Surbhi and Greyson, Bruce},
	year = {2014},
	note = {Place: US
Publisher: Educational Publishing Foundation},
	pages = {302--309},
}

@article{pennachio_near-death_1986,
	title = {Near-death experience as mystical experience},
	volume = {25},
	issn = {1573-6571},
	url = {https://doi.org/10.1007/BF01533055},
	doi = {10.1007/BF01533055},
	language = {en},
	number = {1},
	urldate = {2025-04-24},
	journal = {Journal of Religion and Health},
	author = {Pennachio, John},
	month = mar,
	year = {1986},
	pages = {64--72},
}

@article{pahnke_psychedelic_1969,
	title = {The {Psychedelic} {Mystical} {Experience} in the {Human} {Encounter} with {Death}},
	volume = {62},
	issn = {1475-4517, 0017-8160},
	url = {https://www.cambridge.org/core/journals/harvard-theological-review/article/abs/psychedelic-mystical-experience-in-the-human-encounter-with-death/3EA09A9F82B0044C841C75F304046DC2},
	doi = {10.1017/S0017816000027577},
	language = {en},
	number = {1},
	urldate = {2025-04-24},
	journal = {Harvard Theological Review},
	author = {Pahnke, Walter N.},
	month = jan,
	year = {1969},
	pages = {1--21},
}

@article{greyson_congruence_2014,
	title = {Congruence {Between} {Near}-{Death} and {Mystical} {Experience}},
	copyright = {Copyright Taylor \& Francis},
	issn = {1050-8619},
	url = {https://www.tandfonline.com/doi/abs/10.1080/10508619.2013.845005},
	language = {EN},
	urldate = {2025-04-24},
	journal = {The International Journal for the Psychology of Religion},
	author = {Greyson, Bruce},
	month = oct,
	year = {2014},
	note = {Publisher: Routledge},
}

@article{greyson_western_2015,
	title = {Western {Scientific} {Approaches} to {Near}-{Death} {Experiences}},
	volume = {4},
	copyright = {http://creativecommons.org/licenses/by/3.0/},
	issn = {2076-0787},
	url = {https://www.mdpi.com/2076-0787/4/4/775},
	doi = {10.3390/h4040775},
	language = {en},
	number = {4},
	urldate = {2025-04-24},
	journal = {Humanities},
	author = {Greyson, Bruce},
	month = dec,
	year = {2015},
	note = {Number: 4
Publisher: Multidisciplinary Digital Publishing Institute},
	pages = {775--796},
}

@article{greyson_implications_2010,
	title = {Implications of near-death experiences for a postmaterialist psychology},
	volume = {2},
	issn = {1943-1562},
	doi = {10.1037/a0018548},
	number = {1},
	journal = {Psychology of Religion and Spirituality},
	author = {Greyson, Bruce},
	year = {2010},
	note = {Place: US
Publisher: Educational Publishing Foundation},
	pages = {37--45},
}

@misc{liu_many_2009,
	title = {Many {Americans} {Mix} {Multiple} {Faiths}},
	url = {https://www.pewresearch.org/religion/2009/12/09/many-americans-mix-multiple-faiths/},
	language = {en-US},
	urldate = {2025-04-24},
	journal = {Pew Research Center},
	author = {Liu, Joseph},
	month = dec,
	year = {2009},
}

@article{badham_religious_1997,
	title = {Religious and near-death experience in relation to belief in a future life},
	copyright = {Copyright Taylor \& Francis Group, LLC},
	url = {https://www.tandfonline.com/doi/abs/10.1080/713685847},
	doi = {10.1080/713685847},
	language = {EN},
	urldate = {2025-04-24},
	journal = {Mortality},
	author = {Badham, Paul},
	month = jan,
	year = {1997},
	note = {Publisher: Taylor \& Francis Group},
}

@article{mclaughlin_near-death_1984,
	title = {Near-death experiences and religion: {A} further investigation},
	volume = {23},
	issn = {1573-6571},
	shorttitle = {Near-death experiences and religion},
	url = {https://doi.org/10.1007/BF00996157},
	doi = {10.1007/BF00996157},
	language = {en},
	number = {2},
	urldate = {2025-04-24},
	journal = {Journal of Religion and Health},
	author = {McLaughlin, Steven A. and Malony, H. Newton},
	month = jun,
	year = {1984},
	pages = {149--159},
}

@article{spilka_death_1977,
	title = {Death and personal faith: {A} psychometric investigation},
	volume = {16},
	issn = {1468-5906},
	shorttitle = {Death and personal faith},
	doi = {10.2307/1385748},
	number = {2},
	journal = {Journal for the Scientific Study of Religion},
	author = {Spilka, Bernard and Stout, Larry and Minton, Barbara and Sizemore, Douglas},
	year = {1977},
	note = {Place: United Kingdom
Publisher: Blackwell Publishing},
	pages = {169--178},
}

@article{martin_relationship_1965,
	title = {The relationship between religious behavior and concern about death},
	volume = {65},
	issn = {1940-1183},
	doi = {10.1080/00224545.1965.9919609},
	number = {2},
	journal = {The Journal of Social Psychology},
	author = {Martin, David and Wrightsman Jr., Lawrence S.},
	year = {1965},
	note = {Place: US
Publisher: Heldref Publications},
	pages = {317--323},
}

@book{sabom_recollections_1982,
	title = {Recollections of death : a medical investigation},
	isbn = {978-0-06-014895-9},
	shorttitle = {Recollections of death},
	url = {http://archive.org/details/recollectionsofd0000sabo},
	language = {eng},
	urldate = {2025-04-24},
	publisher = {New York : Harper \& Row},
	author = {Sabom, Michael B.},
	collaborator = {{Internet Archive}},
	year = {1982},
}

@book{ring_life_1982,
	title = {Life at death : a scientific investigation of the near-death experience},
	isbn = {978-0-688-01253-3},
	shorttitle = {Life at death},
	url = {http://archive.org/details/lifeatdeathscien0000ring_e1r9},
	language = {eng},
	urldate = {2025-04-24},
	publisher = {New York : Quill},
	author = {Ring, Kenneth},
	collaborator = {{Internet Archive}},
	year = {1982},
}

@phdthesis{erika_domeika_pratte_wellbeing_2021,
	title = {Wellbeing {Impacts} and {Clinical} {Implications} of {Near}-{Death} {Experiences}},
	url = {https://pure.northampton.ac.uk/en/studentTheses/wellbeing-impacts-and-clinical-implications-of-near-death-experie},
	language = {en},
	urldate = {2025-04-24},
	school = {University of Northampto},
	author = {{Erika Domeika Pratte}},
	year = {2021},
}

@article{walther_attraction_2002,
	title = {Attraction to {Computer}-{Mediated} {Social} {Support} ({Walther} \& {Boyd})},
	volume = {153188},
	url = {https://cmcresearch.org/articles/phpmailer/articles/support.html},
	urldate = {2025-04-24},
	journal = {Communication technology and society: Audience adoption and uses},
	author = {Walther, Joseph B.},
	year = {2002},
	pages = {2},
}

@article{ring_near-death_1997,
	title = {Near-{Death} and {Out}-of-{Body} {Experiences} in the {Blind}: {A} {Study} of {Apparent} {Eyeless} {Vision}},
	volume = {16},
	issn = {1573-3661},
	shorttitle = {Near-{Death} and {Out}-of-{Body} {Experiences} in the {Blind}},
	url = {https://doi.org/10.1023/A:1025010015662},
	doi = {10.1023/A:1025010015662},
	language = {en},
	number = {2},
	urldate = {2025-04-24},
	journal = {Journal of Near-Death Studies},
	author = {Ring, Kenneth and Cooper, Sharon},
	month = dec,
	year = {1997},
	pages = {101--147},
}

@book{blackmore_dying_1993,
	title = {Dying to live : near-death experiences},
	isbn = {978-0-87975-870-7},
	shorttitle = {Dying to live},
	url = {http://archive.org/details/dyingtoliveneard0000blac},
	language = {eng},
	urldate = {2025-04-24},
	publisher = {Buffalo, N.Y. : Prometheus Books},
	author = {Blackmore, Susan J.},
	collaborator = {{Internet Archive}},
	year = {1993},
}

@article{belanti_phenomenology_2008,
	title = {Phenomenology of near-death experiences: a cross-cultural perspective},
	volume = {45},
	issn = {1363-4615},
	shorttitle = {Phenomenology of near-death experiences},
	doi = {10.1177/1363461507088001},
	language = {eng},
	number = {1},
	journal = {Transcultural Psychiatry},
	author = {Belanti, John and Perera, Mahendra and Jagadheesan, Karuppiah},
	month = mar,
	year = {2008},
	pmid = {18344255},
	pages = {121--133},
}

@article{charland-verville_near-death_2015,
	title = {Near-{Death} {Experiences} in patients with locked-in syndrome: {Not} always a blissful journey},
	volume = {34},
	issn = {1053-8100},
	shorttitle = {Near-{Death} {Experiences} in patients with locked-in syndrome},
	url = {https://www.sciencedirect.com/science/article/pii/S1053810015000604},
	doi = {10.1016/j.concog.2015.03.011},
	urldate = {2025-04-24},
	journal = {Consciousness and Cognition},
	author = {Charland-Verville, Vanessa and Lugo, Zulay and Jourdan, Jean-Pierre and Donneau, Anne-Françoise and Laureys, Steven},
	month = jul,
	year = {2015},
	pages = {28--32},
}

@incollection{staudt_covering_2009,
	title = {Covering ({Up}?) {Death} : {A} {Close} {Reading} of {Time} {Magazine}'s {September} 11, 2001, {Special} {Issue}},
	shorttitle = {Covering ({Up}?},
	booktitle = {Speaking of death: {America}'s new sense of mortality},
	publisher = {Praeger},
	author = {Staudt, Christina},
	editor = {Bartalos, Michael K.},
	year = {2009},
}

@article{hoffman_disclosure_1995,
	title = {Disclosure habits after near-death experiences: {Influences}, obstacles, and listener selection},
	volume = {14},
	issn = {1573-3661},
	shorttitle = {Disclosure habits after near-death experiences},
	number = {1},
	journal = {Journal of Near-Death Studies},
	author = {Hoffman, Regina M.},
	year = {1995},
	note = {Publisher: Human Sciences Press, Inc.},
	pages = {29--48},
}

@article{hoffman_disclosure_1995-1,
	title = {Disclosure needs and motives after a near-death experience},
	volume = {13},
	issn = {1573-3661},
	number = {4},
	journal = {Journal of Near-Death Studies},
	author = {Hoffman, Regina M.},
	year = {1995},
	note = {Publisher: Human Sciences Press, Inc.},
	pages = {237--266},
}

@incollection{zingrone_pleasurable_2009,
	address = {Santa Barbara, CA, US},
	title = {Pleasurable {Western} adult near-death experiences: {Features}, circumstances, and incidence},
	isbn = {978-0-313-35864-7 978-0-313-35865-4},
	shorttitle = {Pleasurable {Western} adult near-death experiences},
	booktitle = {The handbook of near-death experiences: {Thirty} years of investigation},
	publisher = {Praeger/ABC-CLIO},
	author = {Zingrone, Nancy L. and Alvarado, Carlos S.},
	year = {2009},
	pages = {17--40},
}

@article{moody_semantic_1975,
	title = {A {Semantic} {Organization} of the {Portuguese} {Subjunctive}},
	volume = {58},
	issn = {0018-2133},
	url = {https://www.jstor.org/stable/339638},
	doi = {10.2307/339638},
	number = {3},
	urldate = {2025-04-24},
	journal = {Hispania},
	author = {Moody, Raymond},
	year = {1975},
	note = {Publisher: American Association of Teachers of Spanish and Portuguese},
	pages = {502--517},
}

@inproceedings{zhou_eternagram_2024,
	address = {New York, NY, USA},
	series = {{CHI} '24},
	title = {Eternagram: {Probing} {Player} {Attitudes} {Towards} {Climate} {Change} {Using} a {ChatGPT}-driven {Text}-based {Adventure}},
	copyright = {All rights reserved},
	isbn = {979-8-4007-0330-0},
	shorttitle = {Eternagram},
	url = {https://dl.acm.org/doi/10.1145/3613904.3642850},
	doi = {10.1145/3613904.3642850},
	urldate = {2024-05-13},
	booktitle = {Proceedings of the {CHI} {Conference} on {Human} {Factors} in {Computing} {Systems}},
	publisher = {Association for Computing Machinery},
	author = {Zhou, Suifang and Hendra, Latisha Besariani and Zhang, Qinshi and Holopainen, Jussi and LC, RAY},
	month = may,
	year = {2024},
	pages = {1--23},
}

@article{zhang_image_2025,
	title = {"{An} {Image} of {Ourselves} in {Our} {Minds}": {How} {College}-educated {Online} {Dating} {Users} {Construct} {Profiles} for {Effective} {Self} {Presentation}},
	volume = {9},
	copyright = {All rights reserved},
	shorttitle = {An {Image} of {Ourselves} in {Our} {Minds}},
	url = {https://doi.org/10.1145/3710929},
	doi = {10.1145/3710929},
	number = {CSCW 031},
	journal = {Proceedings of the ACM on Human-Computer Interaction},
	author = {Zhang, Fan and Chen, Yun and Zeng, Xiaoke and Wang, Tianqi and Ling, Long and LC, RAY},
	month = oct,
	year = {2025},
	pages = {32},
}

@inproceedings{zhou_retrochat_2025,
	address = {New York, NY, USA},
	series = {C\&{C} '25},
	title = {{RetroChat}: {Designing} for the {Preservation} of {Past} {Chinese} {Online} {Social} {Experiences}},
	copyright = {All rights reserved},
	shorttitle = {{RetroChat}},
	url = {https://doi.org/10.1145/3698061.3726920},
	doi = {10.1145/3698061.3726920},
	booktitle = {Creativity and {Cognition}},
	publisher = {Association for Computing Machinery},
	author = {Zhou, Suifang and Fu, Kexue and Yi, Huamin and LC, RAY},
	month = jun,
	year = {2025},
	pages = {19},
}

@inproceedings{gong_if_2025,
	address = {New York, NY, USA},
	series = {{DIS} '25},
	title = {"{If} {I} were in {Space}": {Understanding} and {Adapting} to {Social} {Isolation} through {Designing} {Collaborative} {Storytelling}},
	copyright = {All rights reserved},
	isbn = {979-8-4007-1485-6},
	shorttitle = {"{If} {I} were in {Space}"},
	url = {https://dl.acm.org/doi/10.1145/3715336.3735846},
	doi = {10.1145/3715336.3735846},
	urldate = {2025-07-17},
	booktitle = {Proceedings of the 2025 {ACM} {Designing} {Interactive} {Systems} {Conference}},
	publisher = {Association for Computing Machinery},
	author = {Gong, Qi and Shen, Ximing and Yin, Ziyou and Li, Yaning and LC, RAY},
	month = jul,
	year = {2025},
	pages = {1455--1482},
}



\end{document}